\documentclass[iop]{emulateapj}
\usepackage{apjfonts}
\usepackage{graphicx}
\usepackage{amsmath}
\usepackage{appendix}
\usepackage{natbib}

\begin{document}

\newcommand{\thetae}{\ensuremath{\theta_{\rm E}}}
\newcommand{\re}{\ensuremath{R_{\rm E}}}
\newcommand{\dl}{\ensuremath{D_{l}}}
\newcommand{\ds}{\ensuremath{D_{s}}}
\newcommand{\vrel}{\ensuremath{v_{\rm rel}}}
\newcommand{\murel}{\ensuremath{\mu_{\rm rel}}}
\newcommand{\omegafov}{\ensuremath{\Omega_{\rm FoV}}}
\newcommand{\omegaeff}{\ensuremath{\Omega_{\rm eff}}}
\newcommand{\nmc}{\ensuremath{N_{\rm MC}}}
\newcommand{\ai}{\ensuremath{A_{I}}}
\newcommand{\ak}{\ensuremath{A_{K}}}
\newcommand{\ml}{\ensuremath{M_{l}}}
\newcommand{\mpl}{\ensuremath{M_{p}}}
\newcommand{\mplme}{\ensuremath{M_{p}/{\rm M}_{\Earth}}}
\newcommand{\mprange}{\ensuremath{-1.00 \leq \mathrm{log}(M_{p}/\mathrm{M}_{\Earth}) \leq 3.00}}
\newcommand{\arange}{\ensuremath{-0.40 \leq \mathrm{log}(a/{\rm AU}) \leq 1.20}}
\newcommand{\mprangelin}{\ensuremath{0.1 \leq M_{p}/\mathrm{M}_{\Earth} \leq 1000}}
\newcommand{\arangelin}{\ensuremath{0.4 \leq a/{\rm AU} \leq 16}}
\newcommand{\tnot}{\ensuremath{t_{\rm 0}}}
\newcommand{\unot}{\ensuremath{u_{\rm 0}}}
\newcommand{\te}{\ensuremath{t_{\rm E}}}
\newcommand{\snot}{\ensuremath{s_{\rm 0}}}
\newcommand{\nfld}{\ensuremath{N_{\rm fld}}}
\newcommand{\nev}{\ensuremath{N_{\rm ev}}}
\newcommand{\texp}{\ensuremath{t_{\rm exp}}}
\newcommand{\dchisq}{\ensuremath{\Delta\chi^{2}}}
\newcommand{\dchisqpri}{\ensuremath{\Delta\chi_{\rm pri}^{\rm 2}}}
\newcommand{\dchisqprith}{\ensuremath{\Delta\chi_{\rm pri,th}^{\rm 2}}}
\newcommand{\dchisqsli}{\ensuremath{\Delta\chi_{\rm init}^{\rm 2}}}
\newcommand{\dchisqslith}{\ensuremath{\Delta\chi_{\rm init,th}^{\rm 2}}}
\newcommand{\dchisqslf}{\ensuremath{\Delta\chi_{\rm fit}^{\rm 2}}}
\newcommand{\dchisqslfcut}{\ensuremath{\Delta\chi_{\rm fit,cut}^{\rm 2}}}
\newcommand{\dchisqslfth}{\ensuremath{\Delta\chi_{\rm fit,th}^{\rm 2}}}
\newcommand{\sigsys}{\ensuremath{\sigma_{\rm sys}}}
\newcommand{\fls}{\ensuremath{F_{s}}}
\newcommand{\flb}{\ensuremath{F_{b}}}

\title{Optimal Survey Strategies and Predicted Planet Yields for the Korean Microlensing Telescope Network}
\author{
{Calen B.~Henderson}\altaffilmark{1},
{B.~Scott Gaudi}\altaffilmark{1},
{Cheongho Han}\altaffilmark{2},
{Jan Skowron}\altaffilmark{1,3},
{Matthew T.~Penny}\altaffilmark{1},
{David Nataf}\altaffilmark{4}, and
{Andrew P.~Gould}\altaffilmark{1}
}
\altaffiltext{1}{Department of Astronomy, The Ohio State University, 140 W.\ 18th Ave., Columbus, OH 43210, USA}
\altaffiltext{2}{Department of Physics, Chungbuk National University, Cheongju 361-763, Republic of Korea}
\altaffiltext{3}{Warsaw University Observatory, Al.~Ujazdowskie 4, 00-478 Warszawa, Poland}
\altaffiltext{4}{Research School of Astronomy and Astrophysics, The Australian National University, Canberra, ACT 2611, Australia}
\email{henderson@astronomy.ohio-state.edu}


\shorttitle{Next Generation Microlensing Simulations}
\shortauthors{Henderson, et al.}

\begin{abstract}

The Korean Microlensing Telescope Network (KMTNet) will consist of three 1.6m telescopes each with a 4 deg$^{2}$ field of view (FoV) and will be dedicated to monitoring the Galactic Bulge to detect exoplanets via gravitational microlensing.
KMTNet's combination of aperture size, FoV, cadence, and longitudinal coverage will provide a unique opportunity to probe exoplanet demographics in an unbiased way.
Here we present simulations that optimize the observing strategy for, and predict the planetary yields of, KMTNet.
We find preferences for four target fields located in the central Bulge and an exposure time of ${\texp} = 120$s, leading to the detection of $\sim$2,200 microlensing events per year.
We estimate the planet detection rates for planets with mass and separation across the ranges {\mprangelin} and {\arangelin}, respectively.
Normalizing these rates to the cool-planet mass function of \citet{cassan12}, we predict KMTNet will be approximately uniformly sensitive to planets with mass $5 \leq {\mplme} \leq 1000$ and will detect $\sim$20 planets per year per dex in mass across that range.
For lower-mass planets with mass $0.1 \leq {\mplme} < 5$, we predict KMTNet will detect $\sim$10 planets per year.
We also compute the yields KMTNet will obtain for free-floating planets (FFPs) and predict KMTNet will detect $\sim$1 Earth-mass FFP per year, assuming an underlying population of one such planet per star in the Galaxy.
Lastly, we investigate the dependence of these detection rates on the number of observatories, the photometric precision limit, and optimistic assumptions regarding seeing, throughput, and flux measurement uncertainties.

\end{abstract}

\section{Introduction} \label{sec:intro}

The past twenty years have witnessed a continual acceleration of the pace of the discovery of planets orbiting other stars, resulting in an explosion in the number of known exoplanetary systems.
To date, nearly $\sim$1800 planets have been verified using five different techniques\footnote{See \url{http://exoplanets.org} and
\url{http://exoplanet.eu} for catalogs of known exoplanets with references}.
With these discoveries, first using results from high-precision Doppler surveys (e.g., \citealt{cumming99,udry03,cumming08,bonfils13}) and then using results from Kepler (e.g., \citealt{youdin11,howard12,dong13,dressing13,morton13,petigura13}), we have been able to construct the first detailed determinations of the demographics of exoplanets over a broad range of planet masses and sizes based on large samples of detections.
These results have revolutionized our view of exoplanetary systems, demonstrating a broad diversity of architectures (e.g., \citealt{mayor95,butler99,lovis06,bakos09,charbonneau09,mayor09,lissauer11,orosz12,barclay13}) and revealing the ubiquity of small planets with masses below a few times that of Earth \citep{howard12}.

As exciting as these results are, they are nevertheless painting an incomplete picture of the demographics of planetary systems.
In particular, the Doppler and transit methods are restricted to relatively close orbits of less than a few AU, particularly for low-mass planets.
However, there are substantial reasons to believe that the physics of planet formation, and thus the population of exoplanets, may be substantially different in the outer regions of planetary systems that are not currently being probed by these techniques.
In particular, in a bottom-up picture of planet formation the location of the ``snow line'' in the protoplanetary disks plays a crucial role.
The snow line demarcates the distance from the host star at which it becomes cool enough for water to form as a solid in a vacuum.  Beyond the snow line the surface density of solid material is expected to increase by a factor of two to three \citep{lissauer87}.
This reservoir of solids is crucial for planet formation, facilitating the growth of more massive protoplanets and shorter formation time scales.
In particular, under the core accretion model of giant planet formation it is thought that the majority of gas giants must form beyond the snow lines in their protoplanetary disks (\citealt{ida05,kennedy08}).
Furthermore, it is likely that the liquid water on habitable planets, thought to be a critical requirement for habitability, originated from outside the snow line and was subsequently delivered to such planets via dynamical processes (\citealt{alexander12,jacquet13}).
Thus, determining the demographics of planets beyond the snow line is integral for understanding both the formation and habitability of planets (see \citealt{raymond04} and references therein).

Aside from these more theoretically motivated arguments, it is of interest to survey the outer regions of planetary systems on purely empirical grounds.
For example, of the four giant planets in our solar system, current and near-future surveys using the Doppler or transit method will be sensitive to analogs of only Jupiter.
As a result, it is currently unknown how common systems of giant planets like our own are in the Galaxy.
More generally, protoplanetary disks are known to extend out to $\sim$200--1100 AU (see \citealt{williams11} and references therein), and thus we might expect planetary systems to extend to such distant orbits as well.
Direct imaging surveys are potentially sensitive to planets on wider orbits (e.g., \citealt{marois08,kalas08,lagrange10}), however with current instrumentation these surveys can only detect relatively massive ($\ga {\rm M}_{\rm Jup}$) planets on relatively wide ($\ga$10 AU) orbits, and then only in relatively young ($\la$ Gyr) planetary systems.

Because microlensing is intrinsically sensitive to planets with more distant orbits as well as very low-mass planets, it provides a required complement to our present array of planet detection methods, without which it is currently impossible to obtain a complete picture of exoplanet demographics (see \citealt{gaudi12} for a review).
The images created during a microlensing event have an angular separation from the star that is of order the angular Einstein ring,
\begin{equation} \label{eq:thetae}
   {\thetae} \equiv \left(\kappa {\ml} \pi_{\rm rel}\right)^{1/2},
\end{equation}
where {\ml} is the mass of the lens star, $\pi_{\rm rel}$ is the relative lens-source parallax, given by $\pi_{\rm rel} = {\rm AU}(D_{l}^{-1} - D_{s}^{-1})$, {\dl} and {\ds} are the distances to the lens and source, respectively, and $\kappa \equiv 4G/(c^{2}{\rm AU}) = 8.144~{\rm mas/M_{\odot}}$.
The presence of a planetary companion to this lens star can induce a perturbation of one of these images, resulting in a deviation in the light curve from that which is expected from an isolated star (\citealt{mao91,gould92b}).
Since the angular distance of the planet from the host star must place it near these images in order to create a significant perturbation, and because these images are separated from the host star by $\sim${\thetae}, microlensing is naturally most sensitive to planets with separations of order the physical Einstein ring radius at the lens,
\begin{equation} \label{eq:re}
   {\re} \equiv {\dl} {\thetae}.
\end{equation}
By coincidence, these distances are of order the location of the snow line for a wide range of host star masses and distances \citep{gould92b}.
Therefore, microlensing is an ideal technique for probing exoplanet demographics at and beyond the snow line.

However, there are several practical challenges associated with conducting microlensing surveys for exoplanets.
The primary events are rare (one per star per $\sim$10$^{5}$ years) and, for the most part, unpredictable.
Moreover, {\thetae} is sufficiently small that the individual images are unable to be resolved (${\thetae}\lesssim{\rm mas}$ for lens star masses and lens and source distances that are typical for microlensing events toward the Galactic Bulge), forcing microlensing searches to rely solely on the time evolution of the integral flux of the images.
Tens of millions of stars must thus be monitored on the time scales of the primary events (of order 25 days) simply to find several hundred events per year.
Furthermore, only a handful of these primary events contain planetary perturbations, and, with the important exception of high-magnification events, these image distortions are brief and unpredictable, so these primary events must be monitored at even higher cadence.

Due to the relatively small detectors that were available at the time when microlensing planet surveys were first initiated, they followed a two-tiered strategy that was first advocated by \citet{gould92b}.
Survey telescopes with bigger apertures and the largest available fields-of-view (FoVs) would monitor many tens of square degrees of high stellar density, low extinction fields toward the Galactic Bulge with cadences of once or twice per night.
These cadences were sufficient to detect and alert the primary events themselves, but insufficient to accurately characterize planetary perturbations on these events.
Networks of smaller telescopes with more readily available narrow-angle detectors would then monitor a subset of the most promising of these alerted events with the cadence and wider longitudinal coverage necessary to accurately characterize these planetary perturbations.

The first planet found by microlensing was published by \citet{bond04}, and since then a total of 29 planets orbiting 27 stars have been
 published\footnote{From \url{http://exoplanet.eu} as of 29/May/2014}, primarily using variants of this strategy, including a Jupiter/Saturn analog \citep{gaudi08, bennett10a}, a system with two Jovian-mass planets beyond the snow line \citep{han13}, and two super-Earths \citep{beaulieu06,bennett08}.
The published microlensing planet detections have masses {\mpl} from $0.01 \lesssim M_{p}/{\rm M}_{\rm Jup} \lesssim 9.4$ and semimajor axes $a$ from $0.19 \lesssim a/{\rm AU} \lesssim 8.3$.
These detections have allowed for unique constraints on the demographics of planets beyond the snow line \citep{gould10,sumi11,cassan12} that are complementary to the constraints from other methods.

Nevertheless, there are several problems that confront current microlensing surveys.
The two-stage methodology introduces biases due to its reliance on human judgment for the selection of follow-up targets.
Furthermore, the impact of microlensing exoplanet surveys has been limited by a relatively low number of detections.
It is difficult to improve on the planet yield using the current observational approach because its very design leads to the surveys missing the majority of planetary perturbations.
Thus, while microlensing has produced several interesting results, there is a strong need for more detections and for these to be obtained in an unbiased and automated fashion.

Recent technological developments have facilitated such a transition.
Large format detectors, with FoVs of a few square degrees, on moderate aperture telescopes make it possible to simultaneously image tens of millions of stars in a single pointing.
With such a system, one can dispense with the two-tier strategy and instead enter into a ``Next Generation'' observationally, whereby larger-aperture telescopes monitor a significant fraction of the Bulge with a small number of pointings, thus achieving the cadence needed to detect the primary microlensing events as well as the planetary perturbations.

Using these advances, exoplanetary microlensing has already begun an observational evolution.
There are currently three survey telescopes exclusively dedicated to monitoring the Galactic Bulge to detect exoplanetary microlensing events.
The Optical Gravitational Lens Experiment (OGLE-IV) telescope, located at Las Campanas Observatory in Chile, has a 1.3m aperture, a 1.4 deg$^{2}$ FoV, and attains field-dependent observational cadences of 15--45 minutes \citep{udalski03}.
The Microlensing Observations in Astrophysics (MOA-II) telescope resides at Mt. John University Observatory in New Zealand and has a 1.8m aperture, a 2.18 deg$^{2}$ FoV, and a field-dependent cadence of 15--45 minutes \citep{bond01,sumi03}.
The Wise observatory near Mitzpe Ramon, Israel \citep{gorbikov10} has a 1m aperture, a 1 deg$^{2}$ FoV, and a constant cadence of $\sim$30 minutes \citep{shvartzvald12}.
These three observatories work in concert to tile the Bulge, and they reduce data on daily time scales and alert follow-up networks, including MicroFUN \citep{gould06}, PLANET \citep{beaulieu06}, RoboNet \citep{tsapras09}, and MiNDSTEp \citep{dominik10}.
Together, this current observational approach detects $\sim$2000 microlensing events toward the Bulge each observing season along with of order 10 exoplanetary anomalies from both single and multi-planet systems.

The Korean Microlensing Telescope Network (KMTNet) represents the next stage of this observational transition.
KMTNet is a network of three survey telescopes to be dedicated to monitoring the Galactic Bulge during the microlensing observing season, from early February through early November.
Each telescope has a 1.6m aperture and a 4 deg$^{2}$ FoV.
With these characteristics, KMTNet will provide near-complete longitudinal coverage, and so nearly continuous observations, of the Bulge for a significant portion of the observing season and will obtain deeper photometry at a higher cadence than the current network.
KMTNet will thus significantly increase the number of known planets at planet-star distances near and beyond the snow line.

Here we present the result of simulations that optimize the observing strategy for KMTNet and predict the planet detection rates that the full KMTNet will obtain.
In \S \ref{sec:kmtnet} we detail the characteristics and implementation of KMTNet.
We explain the details, ingredients, and methodology of our simulations in \S \ref{sec:sim_over}.
We vary observational parameters in an attempt to converge on an optimal observing strategy in \S \ref{sec:obsparm_opt}.
In \S \ref{sec:sim_fid} we use said observing parameters to compute fiducial planet detection rates, including rates for free-floating planets (FFPs).
We then investigate the effects that varying different extrinsic parameters has on our fiducial detection rates in \S \ref{sec:extparm_vary}.
Finally, we discuss our results in \S \ref{sec:discussion} and identify our assumptions and how they affect our calculated detection rates.

\section{The Korean Microlensing Telescope Network} \label{sec:kmtnet}

KMTNet will use microlensing as a tool with which to probe the demographics of exoplanets near and beyond the snow line.
The full KMTNet will consist of three survey telescopes that will be dedicated exclusively to monitoring the Galactic Bulge in the Cousins $I$-band for exoplanetary microlensing events during the Galactic Bulge observing season, approximately early February through early November.
Each telescope has a 1.6m aperture, a 4.0 deg$^{2}$ FoV, and uses an equatorial mount.
Tables \ref{tab:kmt_scope_parms} and \ref{tab:kmt_cam_parms} list the parameters for the telescope and camera, respectively.
\begin{deluxetable}{ccc}
\tablecaption{KMTNet Telescope Parameters}
\tablewidth{0pt}
\tablehead{
\colhead{Clear aperture}   &
\colhead{Throughput}       &
\colhead{$f$ ratio}        \\
\colhead{[m]}              &
\colhead{[$\%$]}           &
\colhead{}
}
\startdata
1.6   &   66.2$^{a}$   &   f/3.2
\enddata
\tablenotetext{a}{For $I$-band and includes the effects of the telescope optics, the central telescope obscuration, and the $I$-band filter throughput.}
\label{tab:kmt_scope_parms}
\end{deluxetable}
\begin{deluxetable*}{ccccccccc}
\tablecaption{KMTNet Camera Parameters}
\tablewidth{0pt}
\tablehead{
\colhead{FoV}                     &
\multicolumn{2}{c}{Plate scale}   &
\colhead{Number of pixels}        &
\colhead{Wavelength range}        &
\colhead{Readout noise}           &
\colhead{Full well depth}         &
\colhead{$t_{\rm over}$}          &
\colhead{QE}                      \\
\colhead{[deg$^{2}$]}             &
\colhead{[$\arcsec/$pixel]}       &
\colhead{[$\arcsec/$mm]}          &
\colhead{[$10^{6}$]}              &
\colhead{[nm]}                    &
\colhead{[electrons rms]}         &
\colhead{[electrons]}             &
\colhead{[s]}                     &
\colhead{[$\%$]}
}
\startdata
4   &   0.40   &   40   &   340   &   400-1000   &   5   &   80,000$^{a}$   &   30$^{b}$   &   70$^{c}$
\enddata
\tablenotetext{a}{For $<$3$\%$ nonlinearity.}
\tablenotetext{b}{Encompasses readout time as well as telescope slew and settle time.}
\tablenotetext{c}{For the detector in Cousins $I$-band.}
\label{tab:kmt_cam_parms}
\end{deluxetable*}

The goal of the network will be to conduct a uniform survey that has fewer selection biases and higher detection rates than the current surveys.
\begin{figure}
\centerline{
\includegraphics[width=9cm]{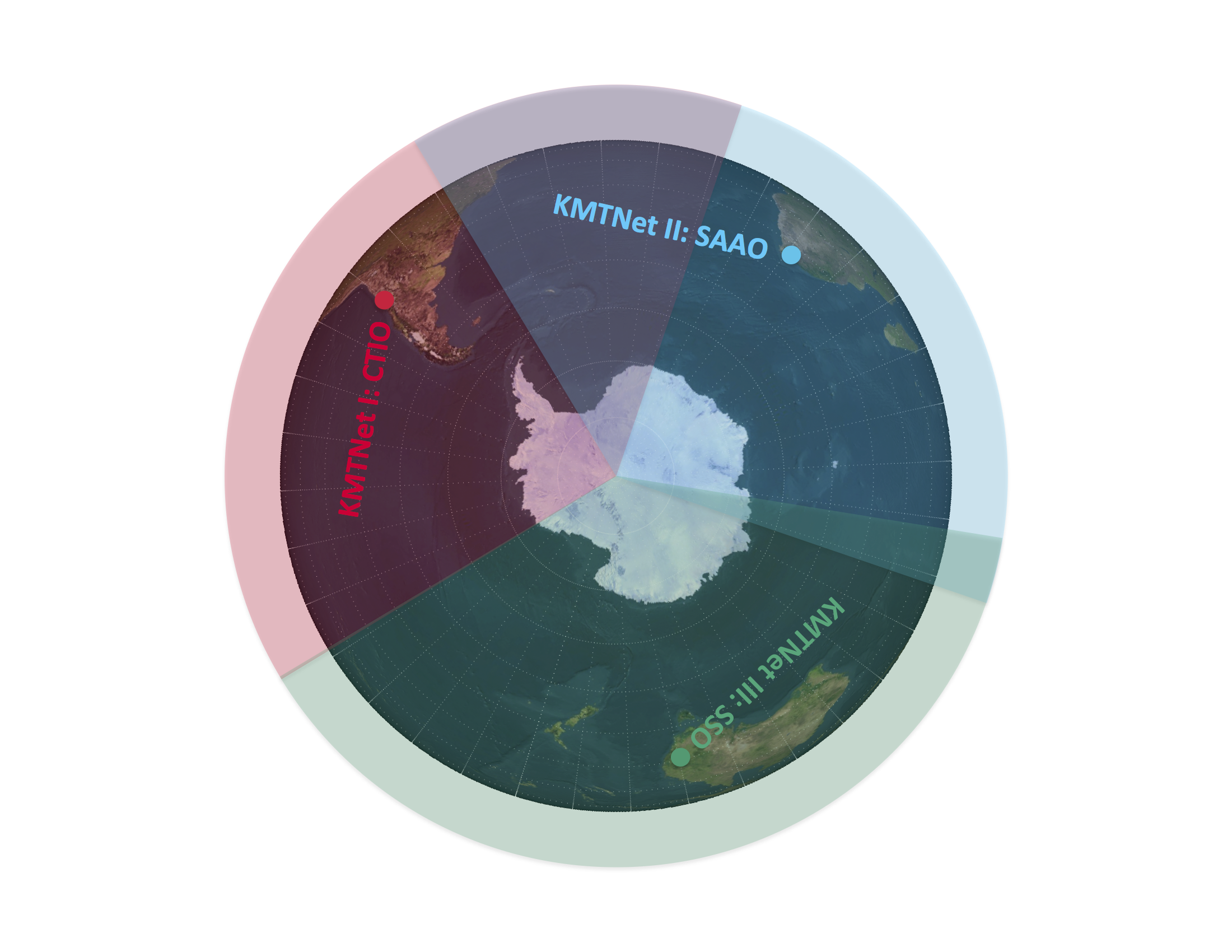}
}
\caption{
\footnotesize{
Location and chronology of the KMTNet sites as viewed from the south pole.
The corresponding wedges indicate the fraction of the night during which the Bulge will be visible (airmass $<2.0$ and nautical twilight) from each site during the peak of the Bulge observing season ($\sim$10/June).
The overlapping regions show when the Bulge will be visible from two observatories simultaneously on that date and correspond to the two bottom curves in the left panel of Figure \ref{fig:bulge_visobs}.
This image was generated in part by {\tt xplanet}.
}
}
\label{fig:earth_kmtnet}
\end{figure}
\begin{figure*}
\centerline{
\includegraphics[width=9cm]{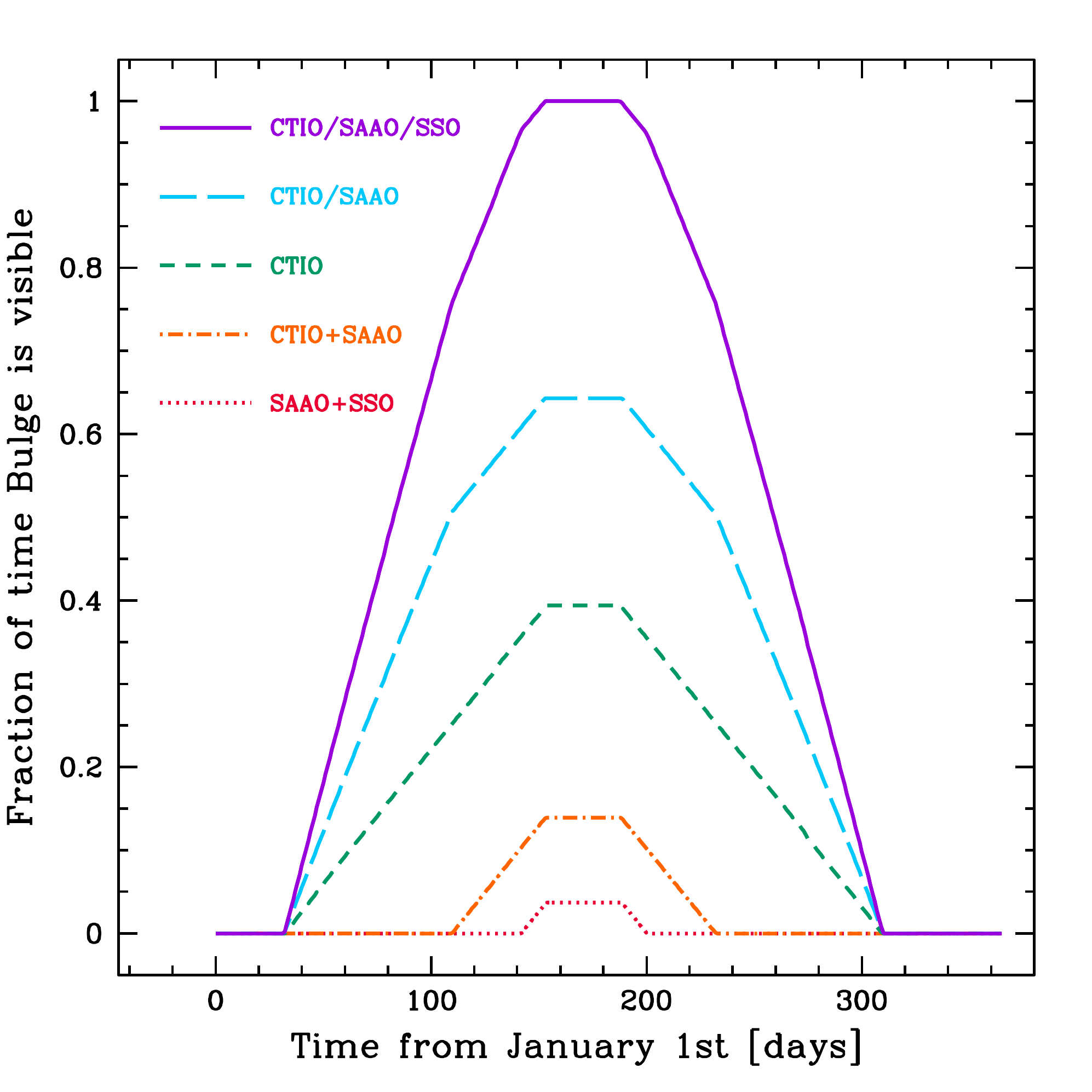}
\includegraphics[width=9cm]{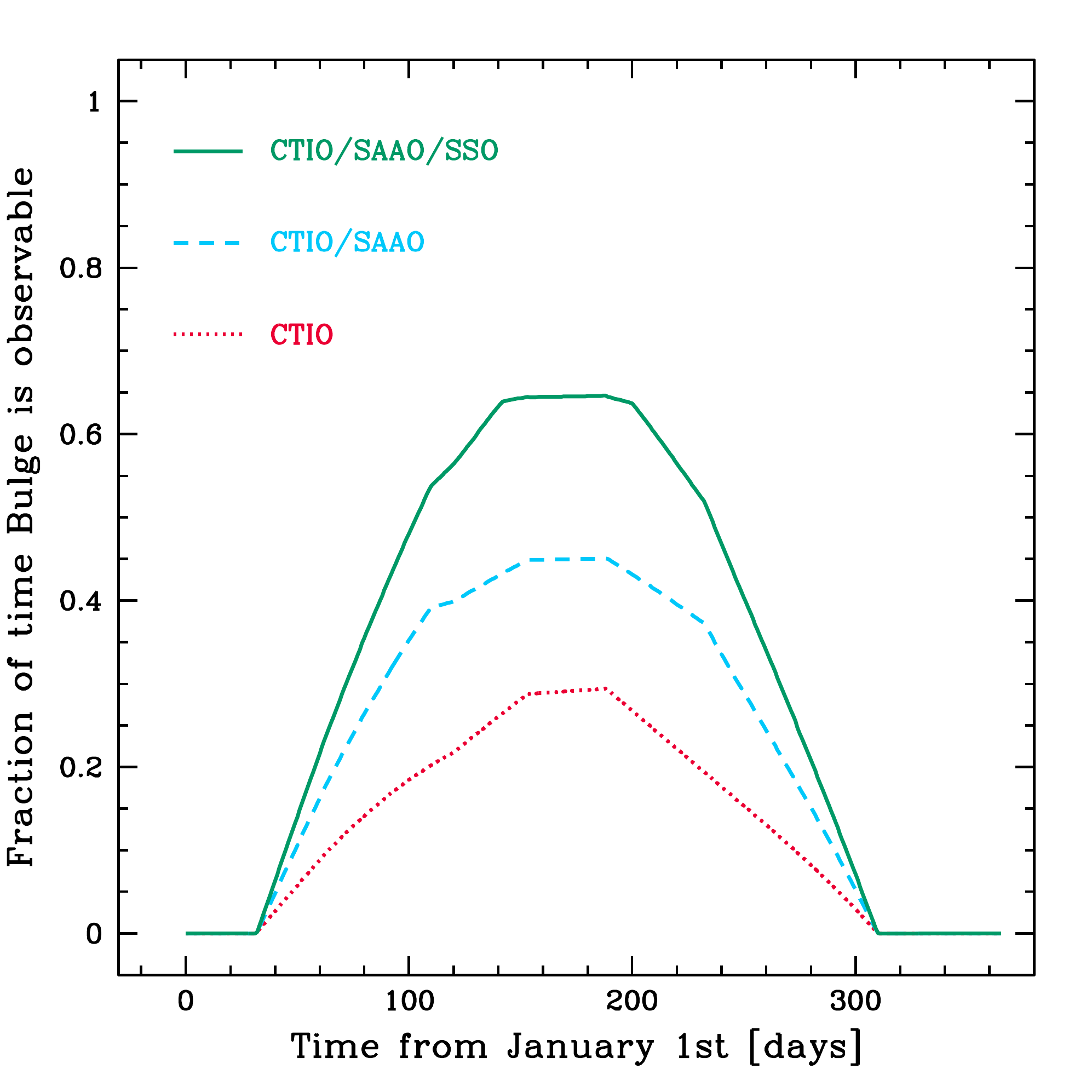}
}
\caption{
\footnotesize{
The fraction of time that the Bulge is observable before (left) and after (right) accounting for weather, for different combinations of observatories for the order in which the KMTNet sites will come online.
We take the center of the Bulge to be $(\alpha,\delta) = (18^{\rm h},~-29^{\circ})$, which corresponds to $(l,b) \approx (1.5^{\circ},~-2.7^{\circ})$.
In the left panel we assume it is visible if it has an airmass $<2.0$ and the Sun is at least
12 degrees below the horizon (nautical twilight).
The upper three curves, in green, blue, and purple, show when the Bulge will be visible from at least one observatory and they follow the order in which each site will come online.
The bottom two curves show when the Bulge will be visible from two observatories simultaneously.
We have not included a line for CTIO+SSO because the coupling of their longitudinal separation with our airmass cut sets an upper limit to their possible observational overlap of $<$0.01$\%$ for our assumed position of the Bulge.
In the right panel we additionally require that the weather is clear at that time, and thus this panel represents the convolution of the visibility of the Bulge, shown in the left panel, with our assumed weather patterns for each observatory site, shown in Figure \ref{fig:clear_v_cloudy_day}.
Even when we include gaps in the data due to weather, we find that the full network will have significant longitudinal coverage for a sizable fraction of the year, with the ability to view the Bulge with at least one telescope for $>$50$\%$ of each day/night for about four and a half months, from the middle of April through the end of August.
}
}
\label{fig:bulge_visobs}
\end{figure*}
Consequently, KMTNet will maintain a constant observational cadence across all target fields.
The first observatory will come online in August 2014 at Cerro Tololo Inter-American Observatory (CTIO) near La Serena, Chile, the second in December 2014 at South Africa Astronomical Observatory (SAAO) at Sutherland, South Africa, and the third in February 2015 at Siding Spring Observatory (SSO) in Coonabarabran, Australia.
Table \ref{tab:site_parms} specifies the location of each observatory.
Figure \ref{fig:earth_kmtnet} shows the location of each of the three KMTNet observatories and also indicates the fraction of the night during which the Bulge will be visible (airmass $<2.0$ and nautical twilight) from each site during the peak of the Bulge observing season.
\begin{deluxetable}{cccc}
\tablecaption{KMTNet Site Parameters}
\tablehead{
\colhead{Site}             &
\colhead{Longitude}        &
\colhead{Latitude}         &
\colhead{Altitude}         \\
\colhead{}                 &
\colhead{[ddd:mm:ss.ss]}   &
\colhead{[dd:mm:ss.ss]}    &
\colhead{[m]}
}
\startdata
CTIO    &    70:42:06      &   -29:00:01.2    &   2400   \\
SAAO   &   339:11:21.5    &   -32:22:46      &   1798    \\
SSO    &   210:56:19.70   &   -31:16:24.10   &   1149
\enddata
\label{tab:site_parms}
\end{deluxetable}

The full network should first be on sky for much of the 2015 Bulge season and will have significant longitudinal coverage for a sizable fraction of the year.
Figure \ref{fig:bulge_visobs} shows the visibility of the Bulge following the observatory chronology both including and excluding the effects of weather.
KMTNet will be able to view the Bulge with at least one telescope for $>$75$\%$ of each day/night for roughly four and a half months from the middle of April through the end of August (excluding the effects of weather).
When the full KMTNet is online, the Bulge will be continuously visible from at least one observatory for $\sim$35 days, from early June through early July.
This represents an unprecedented step forward in the ability of a dedicated microlensing survey to obtain complete longitudinal coverage.
Figure \ref{fig:bulge_visobs} also shows when the Bulge will be visible from two observatories simultaneously.

\section{Simulation Overview} \label{sec:sim_over}

The primary goal of this paper is to simulate a large number of microlensing light curves that resemble, as closely as possible, those that KMTNet will obtain.
From these simulated microlensing events we can then determine the total number of events (per year) that will result in planet detections, given an assumed planet population.
In order to accomplish this we must estimate the contribution of each of our simulated events to the total microlensing event rate, given realistic assumptions about the population of lenses and sources toward the target fields in the Galactic Bulge.

We estimate the event rate toward a given line of sight as follows.
Following \citet{peale98}, we consider a slab of thickness d{\dl} located a distance {\dl} from the observer.
The number of potential lenses d$N_{l}$ in this slab with mass within d{\ml} of {\ml} and within a solid angle d$\Omega$ is
\begin{equation}
   {\rm d}N_{l}=\frac{{\rm d}n_{l}({\ml},{\dl})}{{\rm d}{\ml}}{\rm d}{\ml}{\rm d}V,
\end{equation}
where
\begin{equation}
{\rm d}V={\rm d}\Omega{\rm d}{\dl}D_{l}^{\rm 2}
\end{equation}
is the volume element at a distance {\dl} and $n_{l}$ is the volume number density of compact objects with mass within d{\ml} of {\ml} at a distance {\dl}.

We define a microlensing event to occur if a source at distance {\ds} passes within an angular separation of $u_{0,{\rm max}}\thetae$ of a given lens, where $u_{0,{\rm max}}$ is the impact parameter in units of {\thetae}.
Conventionally, the microlensing optical depth and event rate are defined for $u_{0,{\rm max}} = 1$, which corresponds to a minimum magnification of $\simeq$1.34 for a single lensing mass and a point-like source.
However, microlensing events are detectable for smaller magnifications with sufficiently high cadence and/or sufficiently good photometric precision, so we will therefore allow for an arbitrary value of the impact parameter.
The solid angle covered by a lens per unit time within which it is possible for a microlensing event to happen is given by
2{\thetae}$u_{0,{\rm max}}\mu_{\rm rel}$, where $\mu_{\rm rel}$ is the geocentric relative lens-source proper motion.

Thus, the microlensing event rate per solid angle d$\Omega$, from lenses located within d{\dl} of {\dl} and with mass within d{\ml} of {\ml}, for sources at a distance {\ds}, and with relative lens-source proper motion within d$\mu_{\rm rel}$ of $\mu_{\rm rel}$ is
\begin{equation}
   \frac{{\rm d}\Gamma}{{\rm d}\mu_{\rm rel}{\rm d}{\ml}{\rm d}\Omega{\rm d}{\dl}{\rm d}N_{s}} = \frac{{\rm d}n_{l}({\ml},{\dl})}{{\rm d}{\ml}}D_{l}^{\rm 2}2 u_{\rm 0,max}{\thetae}{\mu_{\rm rel}}.
\end{equation}
Using equation (\ref{eq:re}) and the relation between $\mu_{\rm rel}$ and the relative transverse velocity, {\vrel}, between a given source and an intervening lens,
\begin{equation}
   {\vrel}=\mu_{\rm rel}{\dl},
\end{equation}
we can rewrite this as
\begin{equation} \label{eq:rate_onesource}
   \frac{{\rm d}\Gamma}{{\rm d}v_{\rm rel}{\rm d}{\ml}{\rm d}\Omega{\rm d}{\dl}{\rm d}N_{s}} = \frac{{\rm d}n_{l}({\ml},{\dl})}{{\rm d}{\ml}}2 u_{\rm 0,max}{\re}{\vrel}.
\end{equation}

We next consider a distribution of source magnitudes and distances.
We adopt a luminosity function (LF) $\Phi_{*}$, which gives the number of sources with absolute magnitude within d$M_{I,s}$ of $M_{I,s}$ per unit solid angle d$\Omega$.
As discussed further in \S \ref{sec:lf}, we employ the LF of \citet{holtzman98}, which is an empirical determination of the number of stars per absolute magnitude per solid angle toward Baade's Window (BW).
We call this LF $\Phi_{\rm *,BW}$.
To obtain the LF $\Phi_{*}$ toward an arbitrary line-of-sight (l.o.s.), we use our Galactic density models, which are primarily based on the models of \citet{han95a,han95b,han03} and discussed in \S \ref{sec:bulgedisk}, to compute $\xi$, the ratio of the total integrated mass density along the l.o.s.~toward the given ($l$,$b$) to that toward BW, and assume that the LF scales by that ratio,
\begin{equation}
   \Phi_{*}=\xi\Phi_{\rm *,BW}.
\end{equation}

We assume the areal LF $\Phi_{*}$ applies at all {\ds}, but weight the source distances by the fraction of sources $f_{s}$ at each distance, given a volume density of sources $\rho_{s}({\ds})$ as a function of {\ds}.
The volume element increases as $D_{s}^{2}$d{\ds}, making the fraction of sources within d{\ds} of {\ds}
\begin{equation} \label{eq:source_frac}
   f_{s} = \frac{\rho_{s}({\ds})D_{s}^{2}{\rm d}{\ds}}{\int_{0}^{\infty}\rho_{s}({\ds})D_{s}^{2}{\rm d}{\ds}}.
\end{equation}
Thus, each bin of the LF represents a population of sources with fixed luminosity, and within each bin we allow for the sources to be distributed across the full range of distances being considered.
The differential number density of sources at a distance {\ds} with a given luminosity is then
\begin{equation}
   {\rm d}N_{\rm s} = \Phi_{*}f_{s}{\rm d}M_{I,s}.
\end{equation}

Combining this with equation (\ref{eq:rate_onesource}), the differential event rate for a population of lenses and sources is
\begin{multline} \label{eq:rate_pop}
   \frac{{\rm d}\Gamma}{{\rm d}v_{\rm rel}{\rm d}{\ml}{\rm d}\Omega{\rm d}{\dl}{\rm d}{\ds}{\rm d}M_{I,s}} =
   \\
   \frac{{\rm d}n_{l}({\ml},{\dl})}{{\rm d}{\ml}}2u_{\rm 0,max}{\re}{\vrel}\xi\Phi_{\rm *,BW}\frac{\rho_{s}({\ds})D_{s}^{2}}{\int_{0}^{\infty}\rho_{s}({\ds})D_{s}^{2}{\rm d}{\ds}}.
\end{multline}
The total event rate is then given by
\begin{multline} \label{eq:rate_integral}
   \Gamma = \int{\rm d}\Gamma
   \\
   = \int{\rm d}v_{\rm rel} \int{\rm d}{\ml} \int{\rm d}\Omega \int{\rm d}{\dl} \int{\rm d}{\ds} \int{\rm d}M_{I,s}
   \\
   \frac{{\rm d}\Gamma}{{\rm d}v_{\rm rel}{\rm d}{\ml}{\rm d}\Omega{\rm d}{\dl}{\rm d}{\ds}{\rm d}M_{I,s}}.
\end{multline}
However, we are interested in simulating individual events and thus are interested in the differential contribution of each event to the total event rate.

To estimate the microlensing event rates, we perform a Monte Carlo (MC) simulation of a large number of microlensing events.
In general, there are two possible approaches to creating an ensemble of microlensing events that accounts for the various contributions to the differential event rate in equation (\ref{eq:rate_pop}).
One approach would be to draw the microlensing event parameters according to their contributions to the event rate.
The second is to draw parameters from, e.g., uniform distributions, and then weight each event by equation (\ref{eq:rate_pop}).
We adopt a hybrid approach: we draw some variables from our assumed input distribution functions while others are drawn uniformly and weighted accordingly.

An outline of our MC simulation is as follows.
We assume a population of planetary companions with fixed mass {\mpl} and on a circular orbit with semimajor axis $a$.
Then, we begin by stepping through each absolute magnitude bin $j$ of the LF.
For each bin $j$, we simulate a large number of MC trials $N_{{\rm MC},j}$.
For the $i$-th MC trial we independently draw $D_{l,i}$ and $D_{s,i}$ uniformly, giving
\begin{equation} \label{eq:dl_ds_nmc}
   {\rm d}{\dl}{\rm d}{\ds} = \frac{\Delta{\dl}\Delta{\ds}}{N_{{\rm MC},j}},
\end{equation}
where $\Delta{\dl}$ and $\Delta{\ds}$ represent the full range of {\dl} and {\ds} being considered, respectively.
We draw lens and source velocities from the distributions described in \S \ref{sec:bulgedisk} and then  compute $v_{{\rm rel},i}$ from these velocities and $D_{l,i}$ and $D_{s,i}$.

We assume that the mass function of lenses is independent of location in the Galaxy, and thus separate the volume number density of lenses into two components,
\begin{equation}
   \frac{{\rm d}n_{\rm l}({\ml},{\dl})}{{\rm d}{\ml}}{\rm d}{\ml} = \frac{{\rm d}N}{{\rm d}{\ml}}n_{l}.
\end{equation}
Here ${\rm d}N/{{\rm d}{\ml}}$ is the (normalized) mass function of lenses, i.e., the fraction of lenses with mass within ${\rm d}{\ml}$ of {\ml}, and $n_{l}$ is the number density at a distance $D_{l,i}$.
Our models actually specify the mass volume density $\rho_{l}$, and therefore we substitute $n_{l} = \rho_{l}/{\ml}$ and draw a value of $M_{l,i}$ from ${\ml}({\rm d}N/{\rm d}{\ml})$, where ${\rm d}N/{\rm d}{\ml}$ is the \citet{gould00} mass function as described in \S \ref{sec:lensmass}.
From $M_{l,i}$, $D_{l,i}$, and $D_{s,i}$ we compute $R_{{\rm E},i}$ according to equation (\ref{eq:re}).
We then evaluate $\rho_{l}$ at the value of $D_{l,i}$.
Each event is randomly assigned a pair of Galactic coordinates ($l_{i}$,$b_{i}$) within the FoV of the detector.
With $D_{l,i}$ and ($l_{i}$,$b_{i}$) we use our Galactic models to compute $\rho_{l,i}$, the mass density of lenses at $D_{l,i}$ in the direction of ($l_{i}$,$b_{i}$).

We similarly use our density models to compute $\rho_{s,i}$ as well as the total mass of sources, $M_{s,{\rm tot},i}$, across $\Delta{\ds}$ toward ($l_{i}$,$b_{i}$),
\begin{equation} \label{eq:ms_tot}
   M_{s,{\rm tot},i} \equiv \int_{\Delta{\ds}}\rho_{s}({\ds})D_{s}^{2}{\rm d}{\ds}.
\end{equation}
Combining equations (\ref{eq:dl_ds_nmc}) and (\ref{eq:ms_tot}) with $\rho_{s,i}$ at $D_{s,i}$ allows us to specify equation (\ref{eq:source_frac}) via
\begin{equation}
   f_{s,i} = \frac{\Delta{\ds}\rho_{s,i}D_{s,i}^{2}}{N_{{\rm MC},j}M_{s,{\rm tot},i}}.
\end{equation}

Finally, the contribution to the event rate from the $i$-th MC trial is then
\begin{equation} \label{eq:rate_mc}
   \Delta \Gamma_{i} = \frac{2\Delta{\dl}\Delta{\ds}}{N_{{\rm MC},j}{M_{s,{\rm tot},i}}} \frac{u_{\rm 0,max} R_{{\rm E},i}v_{{\rm rel},i}\rho_{l,i}(D_{l,i})\rho_{s,i}(D_{s,i})D_{s,i}^{2}}{M_{l,i}}.
\end{equation}
Fundamentally, equation (\ref{eq:rate_mc}) gives the weight of a microlensing event that is taken to be representative of all possible events with physical characteristics within the same infinitesimal range of parameter values.
This must be multiplied by $\xi_{i}\Phi_{*,{\rm BW},j}$ to account for the number of sources toward that ($l_{i}$,$b_{i}$) with the same fixed luminosity.
The total planet detection rate is given by summing across all MC trials, all LF bins, and finally all target fields {\nfld}, yielding
\begin{multline} \label{eq:rate_tot_p}
   \Gamma_{\rm tot} = u_{\rm 0,max}{\omegafov} \sum_{k}^{\nfld} \sum_{j}^{N_{\rm LF~bins}}\Phi_{*,{\rm BW},j}
   \\
   \sum_{i}^{\nmc}\xi_{i}\Delta\Gamma_{i} H({\dchisqpri}>{\dchisqprith})\cdot H({\dchisqslf}>{\dchisqslfth}),
\end{multline}
where {\omegafov} represents KMTNet's FoV.
We have also included two Heaviside step functions $H(x)$.
The first requires that the improvement in $\chi^2$ for a microlensing fit relative to a constant fit, $\dchisqpri$, is larger than some minimum threshold $\dchisqprith$ in order to detect the primary event.
The second requires that the improvement in $\chi^2$ for a binary-lens fit relative to a single-lens fit $\dchisqslf$ is larger than some threshold $\dchisqslfth$ in order to subsequently detect the planetary signature.
In practice, if we are interested solely in the overall microlensing event rate, we do not include the second step function.
We also include a few additional cuts on our events that we discuss below, but do not specify explicitly here.

Our MC simulation includes many different ingredients, including
\begin{itemize}
   \item using Galactic models to generate populations of source and lens stars with physical properties such as mass densities, distances, velocities, masses, and apparent magnitudes that match empirical constraints,
   \item populating each lens system with a planetary companion and assigning microlensing parameters in order to compute the magnification of the given binary microlensing event as a function of time, accounting for the effects of a source of finite size when appropriate,
   \item using realistic observing conditions to create ``observed'' light curves for these binary microlensing events by determining the photon rate normalization, including all contributing sources of noise and background such as the Moon, the dark sky, the lens, and unassociated blend stars, and modeling the effects of visibility, gaps due to weather, and seeing at each site, and
   \item implementing a detection algorithm for each light curve to determine first whether the primary microlensing event is detected and if so whether the signal of the planetary perturbation is subsequently robustly detected.
\end{itemize}

We thus divide the discussion of our simulation into these four primary components.
The first generates a population of lens and source stars drawn from a Galactic model that matches empirical constraints.
In the second we compute parameters for binary microlensing events.
For each of these microlensing events we calculate the magnification as a function of time and then simulate realistic observing conditions and effects, including all relevant sources of uncertainty in the flux measurements, and compute the observed light curve.
Finally, we subject these simulated light curves to a series of detection criteria.
We describe the details of each of these components in the subsequent subsections.

\subsection{Galactic Model} \label{sec:galmod}

The first step in our MC simulation is to generate microlensing events with parameter distributions that are consistent with those expected based on empirical constraints on Galactic structure.
As described in \S \ref{sec:sim_over}, we do this by drawing and weighting individual event parameters by their contribution to the total microlensing event rate.
This requires the following ingredients: a source LF, density distribution models for the Galactic Disk and Bulge, models for the kinematics of Disk and Bulge stars, and a mass function of lenses.
Furthermore, in order to predict the lens and source fluxes and the flux of blended light, we must adopt an extinction map as well as a mass-luminosity relation for the lenses.
Finally, we must include a radius-luminosity relation to obtain the physical and angular size of the sources.

\subsubsection{Luminosity Function} \label{sec:lf}

\begin{figure}
\centerline{
\includegraphics[width=9cm]{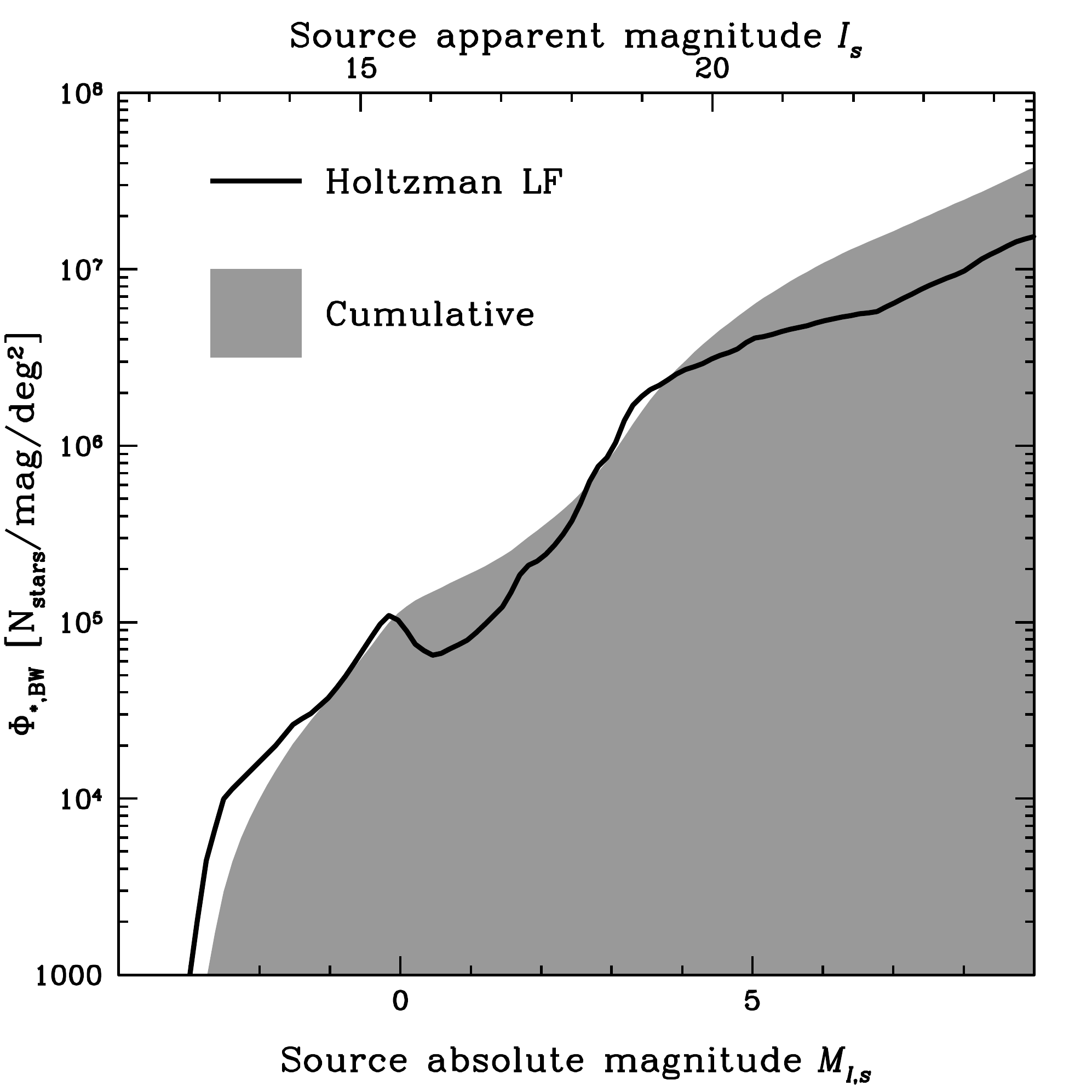}
}
\caption{
\footnotesize{
The differential and cumulative luminosity function of \citet{holtzman98} toward BW.
The apparent magnitude assumes a distance of 8.2 kpc \citep{nataf13} and an extinction of ${\ai} = 1$.
Here we see the bump due to red clump giants at $M_{I,s} \approx 0$, which leads into the subgiant branch and ultimately the main sequence, at $M_{I,s} \approx 3$.
For $M_{I,s} \gtrsim 4$ ($I_{s} \gtrsim 19$) the source population is dominated by main sequence stars.
}
}
\label{fig:lf}
\end{figure}

We use the LF of \citet{holtzman98}, who use \textit{Hubble Space Telescope} data to obtain $\Phi_{*,{\rm BW},j}(M_{I,s})$, the number density of stars for different bins, $j$, of absolute $I$-band magnitude, $M_{I,s}$, toward BW near the Galactic Bulge.
Figure \ref{fig:lf} shows both the cumulative and differential LF.
For the $j$-th bin we generate $N_{{\rm MC},j}$ microlensing events, where $N_{{\rm MC},j} = C\Phi_{*,{\rm BW},j}(M_{I,s}) M_{p}^{-1/2}$.
We scale $N_{{\rm MC},j}$ with $\Phi_{*,{\rm BW},j}$ to ensure that the number of simulated events is proportional to the number of sources with a given absolute magnitude $M_{I,s}$ and thus proportional to the event rate, thereby producing a fixed fractional accuracy in the event rate per bin of $M_{I,s}$ of the LF.
The scaling with $M_{p}^{-1/2}$ arises from the fact that the duration of a planetary perturbation is given by
\begin{equation} \label{eq:tep}
   \Delta t_{p} \approx q^{1/2}{\te},
\end{equation}
where $q$ is the mass ratio of the lens system, given by
\begin{equation} \label{eq:q}
   q = \frac{\mpl}{\ml}.
\end{equation}
We consequently expect the observational coverage and thus the planet detection rate to roughly scale as $M_{p}^{\nu}$, where $\nu \approx 1/2$.
In reality, as we will discuss further in \S \ref{sec:sim_fid}, the planet detection rate scaling is closer to $\nu \approx 3/4$.
Regardless, for simplicity we scale our number of sampled events according to our naive expectation that $\nu = 1/2$.

\subsubsection{Bulge and Disk Models} \label{sec:bulgedisk}

We base our Galactic Bulge model on that of \citet{han95a}, derived from the ``boxy'' Gaussian triaxial G2 model of \citet{dwek95}, which has the functional form (see equations (3) and (4) of \citealt{dwek95})
\begin{equation}
  \rho_{B}=\rho_{0,B}~\mathrm{exp}(-0.5r_{s}^{\rm 2}),
\end{equation}
where
\begin{equation}
   r_{s}=\left\{\left[\left(\frac{x'}{x_{\rm 0}}\right)^{\rm 2}+\left(\frac{y'}{y_{\rm 0}}\right)^{\rm 2}\right]^{\rm 2}+\left(\frac{z'}{z_{\rm 0}}\right)^{\rm 4}\right\}^{\rm 1/4}.
\end{equation}
Here the origin is at the Galactic Center (GC) and the three axes $x'$, $y'$, and $z'$ point along the three axes of the triaxial Bulge.
The values for the scale lengths of the three different axes, $x_{\rm 0}$, $y_{\rm 0}$, and $z_{\rm 0}$, as well as the normalization for the Bulge stellar mass density, $\rho_{0,B}$, are given in Table \ref{tab:density_model_parms}.
We adopt the values of the scale lengths from the 2.2 $\mu$m fit of \citet{dwek95} (see their Table 1) but renormalize them to a Galactocentric distance $R_{\rm GC}$ of 8.2 kpc \citep{nataf13}.
We take the position angle of the major axis of the triaxial Bulge to be 25$^{\circ}$ \citep{nataf13} and normalize the stellar mass density of the Bulge, which includes main sequence stars (MSSs), brown dwarfs (BDs), and remnants---white dwarfs (WDs), neutron stars (NSs), and black holes (BHs)---such that we obtain a column density of stars, BDs, and remnants in the Bulge toward BW equal to the value of 2086 ${\rm M}_{\sun}~{\rm pc}^{\rm -2}$ obtained by \citet{han03}.
\begin{deluxetable*}{cccccccccccc}
\tablecaption{Density Model Parameters}
\tablewidth{0pt}
\tablehead{
\multicolumn{6}{c|}{Bulge}   &
\multicolumn{6}{c}{Disk}    \\
\colhead{$x_{\rm 0}$}   &
\colhead{$y_{\rm 0}$}   &
\colhead{$z_{\rm 0}$}   &
\colhead{$\rho_{0,B}$}   &
\colhead{$D_{l,{\rm min}}$}   &
\multicolumn{1}{c|}{$D_{l,{\rm max}}$}   &
\colhead{$R_{\rm GC}$}   &
\colhead{$R_{0}$}   &
\colhead{$z_{0,D}$}   &
\colhead{$\rho_{0,D}$}   &
\colhead{$D_{l,{\rm min}}$}   &
\colhead{$D_{l,{\rm max}}$}   \\
\colhead{[pc]}   &
\colhead{[pc]}   &
\colhead{[pc]}   &
\colhead{[${\rm M}_{\odot}$ pc$^{\rm -3}$]}   &
\colhead{[pc]}   &
\multicolumn{1}{c|}{[pc]}   &
\colhead{[pc]}   &
\colhead{[pc]}   &
\colhead{[pc]}   &
\colhead{[${\rm M}_{\odot}$ pc$^{\rm -3}$]}   &
\colhead{[pc]}   &
\colhead{[pc]}
}
\startdata
1580   &   620   &   430   &   1.25   &   4200   &   12200   &   8200   &   3500   &   325   &   0.06   &   0   &   12200
\enddata
\label{tab:density_model_parms}
\end{deluxetable*}

For our Galactic Disk model we follow the prescription of \citet{han95b} and adopt the \citet{bahcall86} model, which has the form
\begin{equation}
   \rho_{D}=\rho_{0,D}~\mathrm{exp}\left[-\left(\frac{R-R_{\rm GC}}{R_{0}}+\frac{z}{z_{0,D}}\right)\right],
\end{equation}
where $R=(x^{2}+y^{2})^{\rm 1/2}$, $R_{\rm 0}$ is the radial scale length of the disk, $z_{0,D}$ is the vertical scale height of the disk, and $\rho_{0,D}$ is the mass density in the Solar neighborhood, all of which are specified in Table \ref{tab:density_model_parms}.
The values for $R_{0}$, $R_{\rm GC}$, $z_{0,D}$, and $\rho_{0,D}$ come from \citet{han95a,han95b}.
This coordinate system has its origin at the GC, and the $x$-axis points toward Earth, the $y$-axis toward increasing Galactic longitude, and the $z$-axis toward the North Galactic Pole.

We simulate microlensing events for two populations of lens systems and assume the source is in the Bulge for both.
In the first case we assume the lens to also be located in the Bulge and refer to these events as Bulge-Bulge (BB) events.
The second case consists of lens systems in the Disk and are called Disk-Bulge (DB) events.
In each case, we randomly draw the Galactic coordinates ($l$,$b$) of the event from within the FoV.
We randomly draw the distance from the observer to the lens, {\dl}, from the range
\begin{equation} \label{eq:dist_lens}
   D_{l,{\rm min}} \leq {\dl} \leq D_{l,{\rm max}},
\end{equation}
where $D_{l,{\rm min}}$ and $D_{l,{\rm max}}$ are different for BB and DB events.
Their values are specified in Table \ref{tab:density_model_parms}.
We assume that all sources are in the Bulge and so, for both BB and DB events, draw the distance from the observer to the source, $D_{s}$, from the same range as $D_{l}$ for BB events.
Events for which ${\dl} \geq {\ds}$ are discarded.
We compute the total mass of sources across the full range of $D_{s}$ toward the given ($l$,$b$), $M_{s,{\rm tot}}$, according to equation (\ref{eq:ms_tot}).
Using the appropriate models for BB and DB events, we then calculate the mass density of sources, $\rho_{s}({\ds})$, and lenses, $\rho_{l}({\dl})$, along the l.o.s.~to and at the distance of the source and lens, respectively.

With {\ds} and {\dl} in hand we calculate {\vrel} for each population of lenses, BB and DB.
We assume a Gaussian velocity distribution for both the $y$- and $z$-direction of motion with mean and dispersion adopted from \citet{han95b} and listed in Table \ref{tab:vrel_parms}.
Finally, we add the two components in quadrature, obtaining the relative velocity of the lens-source system in the plane of the sky, {\vrel}.
\begin{deluxetable}{ccccc}
\tablecaption{Velocity Distribution Parameters}
\tablewidth{0pt}
\tablehead{
\colhead{Location}                           &
\colhead{$\mu_{\rm v_{\rm y,rel}}$}          &
\colhead{$\sigma_{\rm v_{\rm y,rel}}^{2}$}   &
\colhead{$\mu_{\rm v_{\rm z,rel}}$}          &
\colhead{$\sigma_{\rm v_{\rm z,rel}}^{2}$}
}
\startdata
Bulge   &   $-220(1 - \eta)$   &   $82.5^{2}\cdot(1+\eta^{2})$   &   0   &   $66.3^{2}\cdot(1+\eta^{2})$   \\
Disk   &   $200\eta$   &   $30^{2}+(82.5\eta)^{2}$   &   0   &   $20^{2}+(66.3\eta)^{2}$
\enddata
\tablecomments{
All values are in km s$^{-1}$ and $\eta \equiv \frac{\dl}{\ds}$.
}
\label{tab:vrel_parms}
\end{deluxetable}

\subsubsection{Lens Mass} \label{sec:lensmass}

\begin{figure}
\centerline{
\includegraphics[width=9cm]{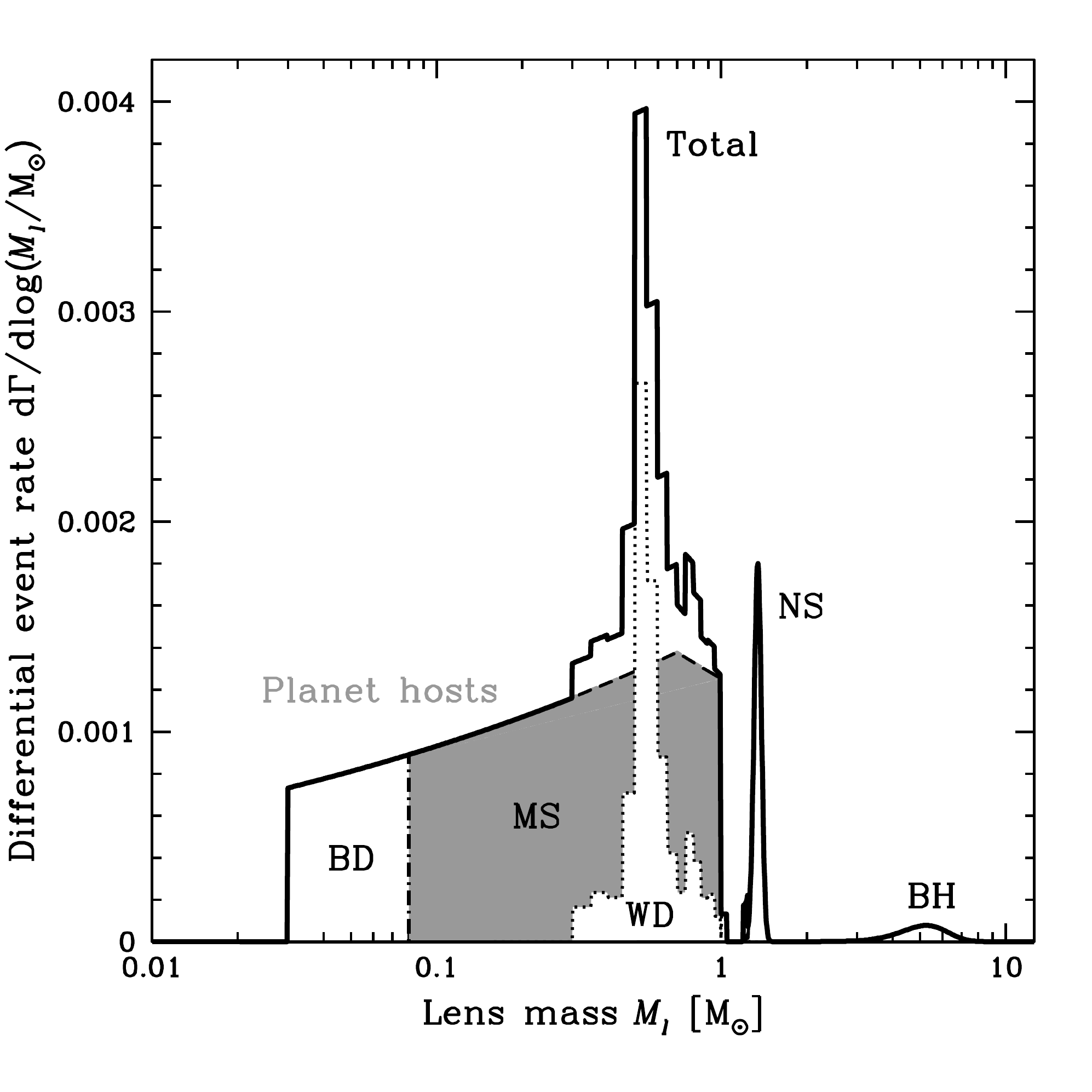}
}
\caption{
\footnotesize{
Our input event rate distribution as a function of the lens mass {\ml}, which goes as $M_{l}^{1/2}\frac{{\rm d}N}{{\rm d}M_{l}}$, adapted from \citet{gould00}.
We draw $M_{l}$ from the mass range $0.03 < {\ml}/{\rm M}_{\odot} \leq 10$.
We exclude BDs and remnants as planetary hosts but include them in the total microlensing event rate.
}
}
\label{fig:lensmass}
\end{figure}

We draw {\ml} from the mass function of \citet{gould00}.
Specifically, we adopt a power-law mass function of the following form
\begin{equation}
   \frac{{\rm d}N}{{\rm d}{\ml}} \propto \left(\frac{\ml}{M_{\rm brk}}\right)^{\epsilon},~~M_{\rm brk}=0.7 {\rm M}_{\sun},
\end{equation}
where
\begin{subequations}
  \begin{align}
   \epsilon = -1.3~~(0.03 < \frac{\ml}{{\rm M}_{\sun}} < M_{\rm brk}),     &   \\
   \epsilon = -2.0~~(M_{\rm brk} < \frac{\ml}{{\rm M}_{\rm \sun}} \lesssim 100.0). &
  \end{align}
\end{subequations}
As in \citet{gould00}, we assume that all MSSs in the range $1 < {\ml}/{\rm M}_{\sun} < 8$ have become WDs, in the range $8 < {\ml}/{\rm M}_{\sun} < 40$ have become NSs, and in the range $40 < {\ml}/{\rm M}_{\sun} < 100$ have become BHs, and adopt the same distributions for each class of remnants, which are shown in Figure \ref{fig:lensmass}.
We include all classes of objects---BDs, MSSs, and stellar remnants---in the calculation of the total microlensing event rate.
However, we only consider MSSs as planet hosts and thus exclude objects with mass outside of the range $0.08 < {\ml}/{\rm M}_{\odot} < 1$ as well as WDs with masses in this range.
We obtain $M_{I,l}$, the absolute magnitude of the lens, from {\ml} using a 1 Gyr isochrone of \citet{baraffe98, baraffe02}.

We have hitherto described the models and corresponding parameters we use to determine $\rho_{s}$, $\rho_{l}$, {\dl}, {\ds}, $M_{s,{\rm tot}}$, {\vrel}, and {\ml}.
These are the physical characteristics of the lens and source necessary to compute the weight of an individual microlensing event via equation (\ref{eq:rate_mc}), which describes the rate at which events with parameters in the same infinitesimal range occur.

\subsubsection{Extinction Map} \label{sec:extinctionmap}

\begin{figure}
\centerline{
\includegraphics[width=9cm]{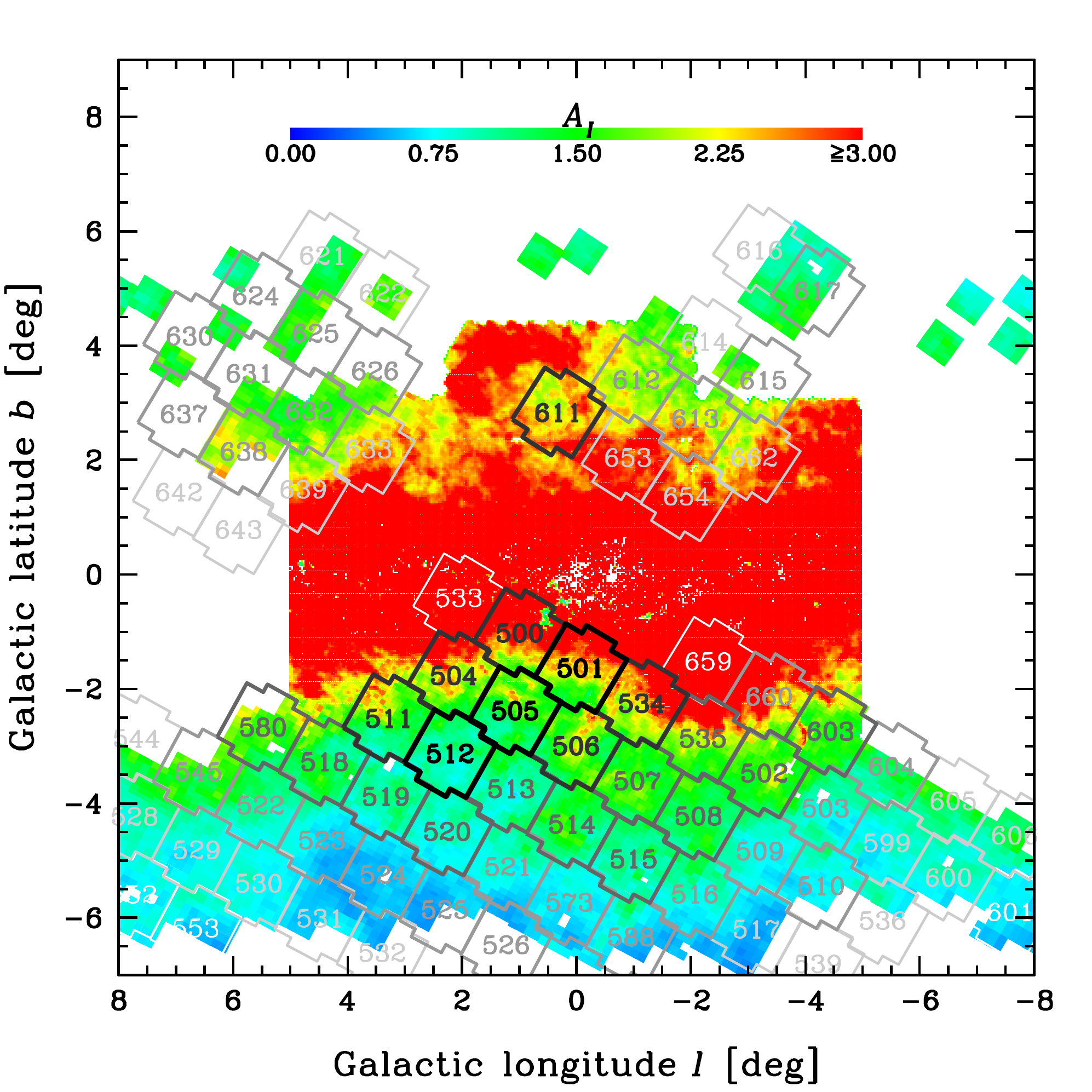}
}
\caption{
\footnotesize{
Our extinction map derived from red clump giants, which covers a significant fraction of the inner Galactic Bulge.
Extinction data come from the $I$-band map of \citet{nataf13} and the mid-IR and NIR map of \citet{majewski11,nidever12}.
Note the deleteriously high extinction for $|b| \lesssim$ 2.
We have overlaid the OGLE-IV target fields, which are grey-scaled according to cadence, with white representing occasional observations and black representing 10--30 observations per night.
}
}
\label{fig:fulldustmap_ogleivfields}
\end{figure}

The dust map we employ combines two different methods of using red clump giants (RCGs) to determine the Galactic extinction in the $I$-band, {\ai}.
The first is the Bulge RCG-derived map of \citet{nataf13} that uses optical and near-IR (NIR) photometry to derive {\ai} for the inner Milky Way.
However, this map is incomplete in the region $|b/{\rm deg}| \lesssim 2$ due to high values of {\ai}.
We thus complement it with an extinction map that uses mid-IR and NIR data \citep{majewski11,nidever12} to determine {\ak} for $|l/{\rm deg}|\leq5$ based on the Rayleigh-Jeans Color Excess (RJCE) method.
To convert {\ak} to {\ai} we sample the overlap region of these two maps, 4595 points at the resolution of the optical map, and find a best-fit slope of ${\ai}/{\ak} = 4.78$ (with an error in the mean of $\pm$0.03) from the median RCG population of the RJCE map, which we apply to the RJCE map.
Our final dust map thus covers a significant fraction of the inner Bulge and is shown in Figure \ref{fig:fulldustmap_ogleivfields}.

It should be noted that our resulting extinction map does not contain any information about the distribution of the dust along the l.o.s., forcing us to estimate {\ai} specifically at {\dl} and {\ds} for every ($l$,$b$) that we might sample.
The optical map of \citet{nataf13} explicitly and exclusively targets RCGs in the Bulge, and we have utilized the RCG stellar population of the RJCE map.
While the median distance to the RCG sample of the IR map lies $\sim$3 kpc in front of the Bulge (see Figure 5 of \citet{nidever12}), we assume that the bulk of the dust lies much nearer to the local Solar neighborhood, making such an offset negligible.
We can therefore estimate the value of {\ai} at a given distance $D$ by adopting a model for the dust distribution and normalizing to the total extinction at the GC.
Though for the bulk of microlensing events the source will lie behind all of the extinction and the lens will be faint, we include this treatment to account for those events with sources on the near side of the Bulge and/or unusually bright lenses.

To do so, we assume that our extinction map provides $A_{I}$ at the distance of the Galactic Center, $D_{\rm GC} = 8.2$kpc, and that the dust responsible for the total extinction toward a given l.o.s.\ is distributed along that l.o.s.\ such that the component of this total extinction at a given distance $D<D_{\rm GC}$ is given by the total column density of dust to $D$.
We assume the dust is distributed exponentially with vertical distance from the plane and integrate the dust volume density along the l.o.s.\ to $D$ in order to obtain the column density.
Then, in order to estimate {\ai} along a given l.o.s.~at $D$, we take the extinction to be distributed as
\begin{equation}
   {\ai}=-2.5\mathrm{log}[e^{\tau(D) \cdot \kappa}],
\end{equation}
where $\tau$ is the optical depth to dust at $D$, which we take as proportional to the dust column density, and $\kappa$ is a normalization factor.
For a given ($l$,$b$) we first integrate along the l.o.s.~to $D_{\rm GC}$ to obtain $\tau$ at the distance of the Bulge.
We use a dust distribution model that is distributed vertically with an exponential profile but that is constant with radius at a given height above the plane.
We assume a vertical scale height of 125pc \citep{marshall06}.
Given that we have assumed our dust map to give {\ai} at $D_{\rm GC}$, we interpolate across the map to determine {\ai} for the given ($l$,$b$) and compute $\kappa$, the value of which is required to reproduce {\ai} toward the given l.o.s.~for the optical depth $\tau(D_{\rm GC})$.
For a source at {\ds}, we subsequently integrate our dust model along the l.o.s.~once more, this time to {\ds}, to obtain $\tau$ at {\ds}, and finally apply $\kappa$ to get {\ai} at {\ds}.
Given {\ds}, $M_{I,s}$, and $A_{I,s}$, we calculate the apparent magnitude of the source, $I_{s}$.
We similarly compute the apparent magnitude of the lens, $I_{l}$.

\subsection{Microlensing Parameters} \label{sec:ulensparm}

After the physical parameters of each lensing event and its event rate contribution have been determined, we assign the microlensing parameters, which we ultimately use to compute the magnification of the source as a function of time.
First we calculate the basic single lens parameters.
Then we add a planetary companion to the lens star and determine the static binary lens parameters.
For all cases we do not consider higher-order dynamical effects such as parallax, xallarap, or lens orbital motion, in our simulations.

\subsubsection{Primary Event} \label{sec:ulensprim}

We refer to the magnification structure that arises from a microlensing event that is due to a single lensing mass as the primary event.
There are four parameters that specify such a single-lens primary event and allow for the derivation of the magnification as a function of time.
They are {\tnot}, the time of closest approach of the source to the lens, {\unot}, the angular distance of the closest approach of the source to the lens, normalized by {\thetae}, the Einstein crossing time {\te}, and $\rho$, the angular size of the source star normalized to {\thetae}.
In our simulations we compute the annual planet detection rate, so we compute time in the reference frame of a generic year and randomly draw {\tnot} from the range
\begin{equation}
   0.0 \leq \frac{\tnot}{\rm days} \leq 365.25.
\end{equation}
We draw {\unot} randomly from the range
\begin{equation}
   0.0 \leq {\unot} \leq u_{\rm 0,max},
\end{equation}
adopting a maximum impact parameter of $u_{\rm 0,max} = 3$.
The Einstein crossing time {\te} is calculated as
\begin{equation}
   {\te} \equiv \frac{\thetae}{\murel}.
\end{equation}
The set of these three parameters, {\tnot}, {\unot}, and {\te}, is sufficient for microlensing events in which the source is point-like.
For the case of a single-lens event, the lens-source separation as a function of time, $u(t)$, is given by
\begin{equation}
   u^{2}(t)=u_{0}^{2}+\left(\frac{t-{\tnot}}{\te}\right)^{2}.
\end{equation}

The magnification for a point-source is then \citep{paczynski86}
\begin{equation} \label{eq:mag_pssl}
   A[u(t)] = \frac{u^{2}+2}{u\sqrt{u^{2}+4}}.
\end{equation}
If the source passes sufficiently near the lens mass such that there is a significant second derivative of the magnification across its surface and the size of the source can be resolved, the additional parameter $\rho$ must be specified.
We use the 10 Gyr isochrone of \citet{girardi00}, assuming solar metallicity, to obtain a relation between $M_{I,s}$ and $R_{*}$.
As mentioned in \citet{gaudi00}, reasonable variations in age and metallicity do not have appreciable effects on the conversion between $M_{I,s}$ and $R_{*}$.
We use $R_{*}$ to determine the physical size of the source star, given its absolute magnitude from the LF.
The angular size of the source star normalized to {\thetae} is then
\begin{equation}
   \rho = \frac{\theta_{*}}{\thetae},
\end{equation}
where
\begin{equation}
   \theta_{*} = \frac{R_{*}}{D_{s}}.
\end{equation}

\subsubsection{Binary Lens} \label{sec:ulensbin}

The next step is to populate the lens system with a planet and compute the three additional parameters that determine a static binary lens.
These are $q$, the mass ratio of the lens system, $s_{0}$, the instantaneous projected separation of the lens components in units of {\thetae} at the time of the event, and $\alpha_{0}$, the angle of the source trajectory with respect to the binary axis at the time of the event.
The binary axis points from the primary, the lens star, to the secondary, the planet.
The mass ratio of the lens system is given by
\begin{equation}
   q=\frac{\mpl}{\ml}.
\end{equation}
We assume a circular orbit for the planetary companion and compute $s_{0}$ as
\begin{equation}
   s_{0}=\frac{a}{\re}\sqrt{1 - {\rm cos}^{2}\zeta},
\end{equation}
where $\zeta$ is the angle between the semimajor axis $a$ and the plane of the sky.
For randomly oriented orbits, ${\rm cos}\zeta$ is uniformly distributed.
We therefore draw ${\rm cos}\zeta$ from a uniform random deviate in the range [0$-$1].
The trajectory angle $\alpha_{0}$, which specifies the direction of the lens-source relative motion, is measured counter-clockwise from the binary lens axis.
We draw $\alpha_{0}$ randomly from the range
\begin{equation}
   0.0 \leq \alpha_{0} \leq 2\pi.
\end{equation}

\subsubsection{Magnification Calculation} \label{sec:magcalc}

We then calculate the magnification of the source due to the static binary lens system as a function of time.
We first check whether it is appropriate to make use of either of two approximations that use a series of point-source calculations to approximate a source of finite size.
In each case it is necessary to solve a complex fifth-order polynomial, whose coefficients are given by \citet{witt95}, in order to obtain the magnification of a point-like source due to a binary lens (PSBL).
To expedite this procedure we employ the root-solving method of \citet{skowron12}, which is of order a few times faster than the root-solving subroutine ZROOTS contained within {\it Numerical Recipes}.
This process allows us to circumvent using a computationally expensive algorithm to calculate the full finite-source binary-lens (FSBL) magnification for the vast majority of data points without loss of precision, in turn boosting the number of light curves we can simulate per unit time and improving our derived primary event and planet detection rate statistics.

We employ a tiered magnification algorithm that balances computational efficiency with robustness in order to efficaciously model the large number and wide variety of binary lens systems our simulations generate.
At a given time $t$ we first compute the PSBL magnification $A_{\rm psbl}$.
We also estimate the finite-source magnification using the quadrupole approximation \citep{pejcha09}, which uses five point-source magnification calculations---one at the center of the source and four at equally spaced points along the perimeter of the source---to approximate the magnification of a source of extended size.
If the fractional difference between the point-source and quadrupole approximations for the magnification due to a binary lens is below the tolerance $\delta A \equiv \lvert\frac{A_{\rm quad}-A_{\rm psbl}}{A_{\rm quad}}\rvert \leq A_{\rm tol}$, where $A_{\rm tol}=10^{-5}$ is our arbitrary but conservative choice, we adopt the quadrupole magnification for that data point.
Otherwise, we compute the magnification using the hexadecapole approximation \citep{pejcha09,gould08}, which approximates an extended source using thirteen point-source magnification calculations---eight equally spaced along the perimeter of the source, four equally spaced along the perimeter defined by $\rho/2$, and one at the center of the source.
If the fractional difference between the quadrupole and hexadecapole magnifications is below $A_{\rm tol}$ we use the hexadecapole magnification for that point.

If both the quadrupole and subsequently the hexadecapole approximations fail the fractional tolerance criterion, the magnification at that time $t$ requires the use of a full FSBL magnification computation.
We utilize an inverse ray-shooting algorithm that ``shoots'' rays from the images of the magnified source on the image plane and computes the FSBL magnification by using the binary lens equation to determine how many of these rays can be traced back to the interior of the unmagnified source star on the source plane.
In order to further expedite our inverse ray-shooting algorithm, we use the hexadecapole approximation at each time to help determine the appropriate geometry of our coordinate system.
If $A_{\rm hex}$ satisfies $A_{\rm hex} \leq A_{\rm thresh}$, we presume the resulting FSBL magnification of the source will be sufficiently low that it is optimal to create a grid in a rectangular coordinate system.
If $A_{\rm hex} > A_{\rm thresh}$, the FSBL magnification is taken to be sufficiently high that a grid in a polar coordinate system centered on the primary lens mass is more appropriate, as the majority of the magnification arises from two images that form extended arcs centered on the more massive lens component at the distance $\sim${\thetae}.
In this high-magnification regime we utilize a variable axis ratio that decreases the grid resolution in the angular direction relative to that in the radial direction to more accurately capture the image morphology and increase computational efficiency without loss of precision, following \citet{bennett10b}.
We adopt a threshold of $A_{\rm thresh} = 100$ based on the fact that \citet{bennett10b} find that the precision of a polar-based algorithm increases with the axis ratio for magnifications higher than this.
We set the resolution of each grid such that even the largest numerical errors due to finite sampling from either inverse ray-shooting algorithm are, fractionally, $\leq$10$^{-4}$, more than an order-of-magnitude below the fractional photometric precision expected from KMTNet.

\subsubsection{Finite Source Effects in the Single Lens Model} \label{sec:finitesource}

As described in \S\ref{sec:planpert}, in order to determine whether a given planetary perturbation is detectable, we fit our simulated binary-lens lightcurve to a single lens model.
Before doing so, however, we must first determine whether or not we need to consider finite-source effects in the comparison model.
In most cases, a point-source single-lens (PSSL) model provides a sufficiently good approximation.
However, if the source passes very near to or over the primary lens, and the central caustic due to the planet is sufficiently small, the resulting light curve will closely resemble that due to a single lens with finite-source effects (FSSL) and have no significant deviations from the planet (at least during the peak of the event).
Thus, if one were to fit such a light curve to a PSSL model, one would find large deviations, resulting in a spurious planet detection.

For each event, we first determine whether or not ${\unot} \geq 25\rho$.
If so, then we assume that finite-source effects for a single lens are completely negligible and therefore adopt a PSSL  model as the best-fit comparison, leaving {\tnot}, {\unot}, and {\te} as free parameters and computing the magnification according to equation (\ref{eq:mag_pssl}).
In particular, we do not include $\rho$ as a free parameter in this case.
It is straightforward to demonstrate that, for a single lens, the fractional deviation in magnification between a point-like source and a finite source at the closest point of approach $u = {\unot}$ is $\la 2 \times 10^{-4}$ for ${\unot} \geq 25\rho$, assuming $\rho \ll 1$.
This is well below the photometric precision achievable by KMTNet.

If ${\unot} < 25 \rho$, we determine whether or not the fractional difference between the magnification for a point-like source, computed according to equation (\ref{eq:mag_pssl}), and that for a finite source, computed numerically via the elliptic integrals given in \citet{witt94}, is greater than our tolerance $A_{\rm tol}$ for at least one data point.
If so, and if the primary event passes the initial detection criteria, we use a FSSL model to fit to the lightcurve, including $\rho$ as a free parameter.
If not, then we assume finite-source effects are negligible, again adopt a PSSL model, and do not include $\rho$ as a free parameter.

\subsection{Light Curve Creation} \label{sec:lccreate}

With all of the physical and microlensing parameters in hand, the next task is to generate the light curve for each event.
This requires turning the magnification of the source as a function of time into a measured flux by determining the flux contributions of the lens, the source, and all sources of blended and background light, accounting for the effects of the Moon and weather, and accurately modeling the flux measurement uncertainties.

\subsubsection{Observational Parameters} \label{sec:obsparm}

For each observatory we divide a generic year into the total number of possible data points, assuming a constant cadence (the choice of which is discussed in \S \ref{sec:obsparm_opt}).
For each target field we determine whether that field is observable for each of the total possible data points from each observatory.
The criteria are that the field center be at or above an airmass of 2.0 and that the Sun is at least 12 degrees below the horizon (nautical twilight).

\begin{figure}
\centerline{
\includegraphics[width=9cm]{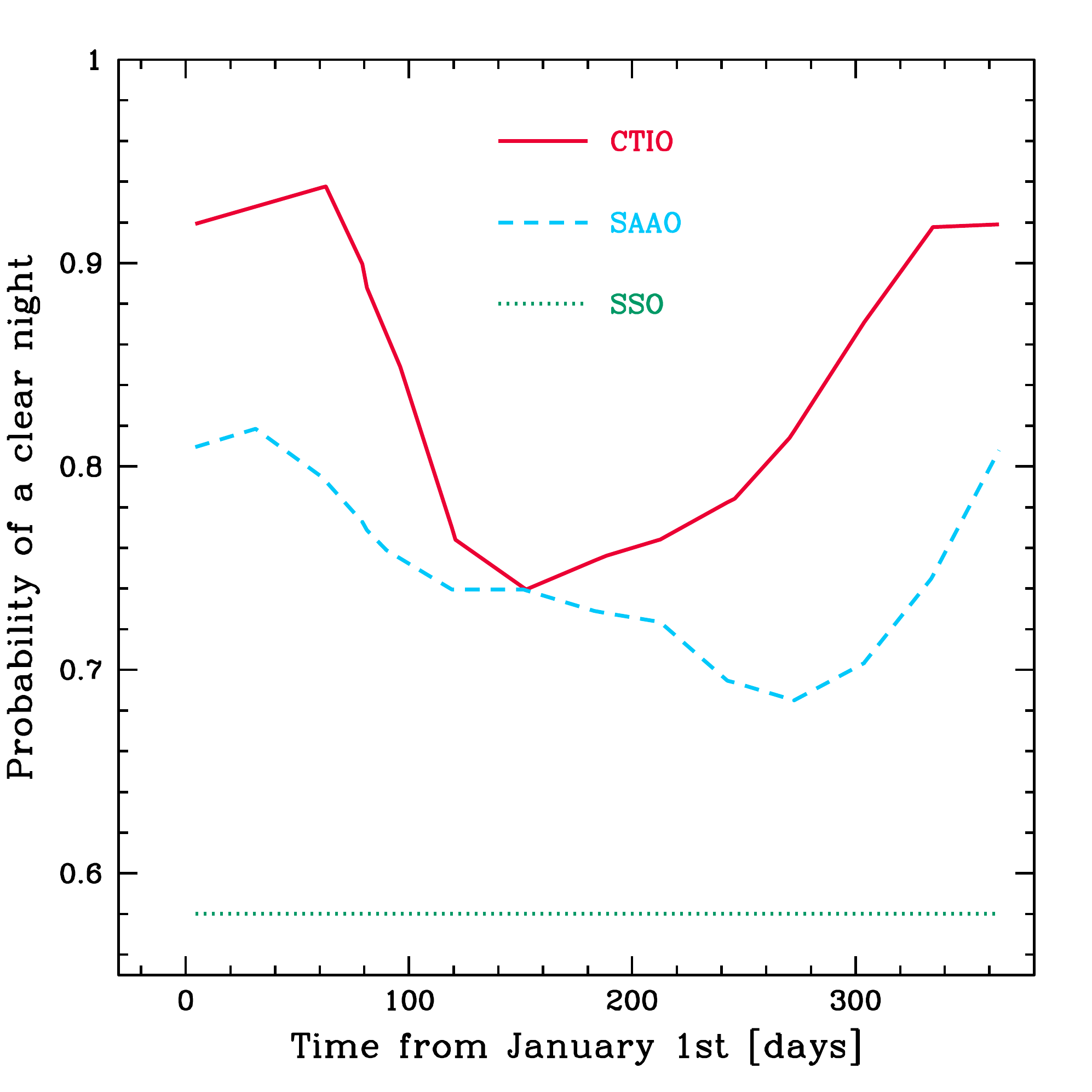}
}
\caption{
\footnotesize{
The fraction of clear nights for each of the three KMTNet observatory sites as a function of time.
These data were taken from \citet{peale97} and we assume the weather at La Silla is a good approximation for the weather at the KMTNet observatory at CTIO.
}
}
\label{fig:clear_v_cloudy_day}
\end{figure}

We model the length of weather patterns as a Poisson process and compute the cumulative distribution function (CDF) of their duration as $e^{-\lambda}\sum_{i{\rm = 0}}^{k}\lambda^{i}/i!$.
We adopt a mean weather pattern coherence length of $\lambda=4$ days and compute the CDF to $k=20$.
Beginning with the first data point for each observatory, we randomly draw from the CDF to obtain the duration of a given weather pattern at the given observing site.
Then, using the fraction of clear and cloudy nights for each site, taken from \citet{peale97} and shown in Figure \ref{fig:clear_v_cloudy_day}, we randomly draw to determine whether it will be cloudy or clear for the weather pattern.
Here we have assumed that the fraction of clear nights at La Silla is a good approximation of the same fraction for CTIO.
If cloudy, we skip all data points that occur during the pattern and repeat this process beginning with the next data point after the weather pattern.
Otherwise, the weather for those data points is considered clear.
Figure \ref{fig:bulge_visobs} shows the observability of the Bulge for the chronology of the KMTNet sites, which convolves the visibility of the Bulge with the distribution of clear nights at each site that is shown in Figure \ref{fig:clear_v_cloudy_day}.

For each observable data point we draw the seeing from a Gaussian with site-dependent minimum, mean, and sigma seeing values, listed in Table \ref{tab:see_fid}, and modify the seeing as (airmass)$^{0.6}$ \citep{woolf82}, where airmass = sec($z$) and $z$ is the zenith angle.
We calculate the total background sky brightness in $I$-magnitudes per square arcsecond, including the contributions from the mean dark sky at zenith, which we assume is $\mu_{\rm sky} = 19.9~{\rm mag}/\square \arcsec$ for each site, and the phase and distance of the Moon to the field center, according to the prescription of \citet{krisciunas91}.
\begin{deluxetable}{cccc}
\tablecaption{Fiducial Site-dependent Seeing Distribution Parameters}
\tablewidth{0pt}
\tablehead{
\colhead{Site}    &
\colhead{min.}    &
\colhead{$\mu$}   &
\colhead{$\sigma$}
}
\startdata
CTIO   &   0.8   &   1.4   &   0.26   \\
SAAO   &   0.9   &   1.6   &   0.30   \\
SSO    &   1.3   &   2.0   &   0.40
\enddata
\tablecomments{All values are in arcseconds.}
\label{tab:see_fid}
\end{deluxetable}

\subsubsection{Flux Determination} \label{sec:fluxdeter}

The final photometric reduction pipeline that KMTNet will implement is based on difference image analysis (DIA) photometry \citep{alard98, alard00}.
Aperture photometry breaks down in crowded stellar fields, and even the approach of point-spread function (PSF) photometry becomes quite difficult in the regime of extreme blending that is typical of the Galactic Bulge.
To circumvent this issue, DIA constructs a reference template frame by combining the subset of images with the best seeing and then measures the PSF solely on this reference image.
For each observation, it then determines a convolution kernel that transforms the PSF of the reference image into that of the given frame, ``matching'' the two, and subtracts the given image from the convolved reference image.
Each resulting difference image thus yields the difference in flux between the given observation and the template image, which causes only photometrically variable objects to have a non-zero difference flux.
Performing PSF photometry on the reference image then provides a flux zero-point (see \S3.1 of \citealt{hartman04} for a more complete discussion of DIA).

In the case of a microlensing event, the object flux that is measured on the reference image includes the light from the source, the lens, and any other interloping stars that are unassociated with the event but still unresolved and thus blended with the source.
For typical Bulge fields with high stellar density, there will also be a quasi-smooth background flux produced by faint unresolved stars scattered across the entire frame, even for a reference image with excellent seeing.
DIA treats this ``sea'' of unresolved stars in the same way it does the Moon and the dark sky, fitting and subtracting these smooth backgrounds from the reference image prior to the measurement of the object flux.
These sources of background flux---the Moon, the dark sky, and the ``sea'' of faint unresolved stars---will thus not contribute to the flux measured on the reference image.
They will, however, still contribute to the measured flux uncertainty for each individual frame in a way that varies with the seeing of the image.
Often there are interloping stars blended with the event that are brighter than the limiting magnitude that defines this ``sea'' of faint unresolved stars, so their brightness is not (entirely) subtracted off with the smooth stellar background and so does not vary with seeing from image to image.

We approximate these populations of unassociated and unresolved stars as a dichotomy between faint stars that contribute solely to the noise of an individual flux measurement and bright stars that contribute to the object flux as well as the noise.
Under this dichotomy we assume each microlensing event to be blended with, on average, one bright interloping star as well as a smooth surface brightness of faint stars.
The flux from the interloping star will contribute to the object flux measured on the reference frame as well as its uncertainty, while the flux from the smooth stellar background will only contribute to the flux measurement uncertainty in a manner that depends on the seeing.
To determine their respective flux contributions, we estimate the apparent brightness above which there is, on average, one unassociated and unresolved star per seeing disc.
The first step in this process is to simulate the construction of a reference image.

As shown in Table \ref{tab:see_fid}, CTIO will have the best seeing of the three KMTNet sites, so we assume that the template will be comprised entirely of images taken at CTIO.
From 12 and 10 random light curves from 2011 and 2012 OGLE-IV data, respectively, we find that the 1st percentile value of seeing is $0.93\arcsec$, which we take as the seeing of the reference image, $\sigma_{\rm ref}$.
We make the approximation that all non-lens stellar blend flux, resolved or unresolved, is due to stars at the distance of the Bulge and interpolate across our dust map to get {\ai} at the location of the event assuming a distance of 8.2kpc.
We then modify our LF to give the areal density of stars toward a given l.o.s.~as a function of apparent magnitude.
To do so, first we apply the 8.2 kpc distance to the Bulge and the computed extinction.
Then we use our Bulge model, described in \S \ref{sec:bulgedisk}, to calculate the ratio of the stellar surface density for a given l.o.s.~to that of BW, $\xi_{*}$, the region for which \citet{holtzman98} determined the LF.

Next we use a Moffat function to model the star's light profile \citep{moffat69}.
A Gaussian falls off more steeply at larger radii and so is insufficient for capturing the full extension of the wings of a realistic PSF.
We find that adopting a Moffat profile is crucial for regions of such high stellar density in the Bulge.
The intensity $I$ as a function of radius $r$ from the center of the profile is
\begin{equation}
   I(r)=I_{0}\left[1+\left(\frac{r}{\alpha}\right)^{2}\right]^{-\beta},
\end{equation}
where $I_{0}$ is the intensity at the central peak, $\alpha$ is the width parameter, related to the full width at half maximum (FWHM), i.e., the seeing, of the light profile via
\begin{equation}
   \alpha=\frac{FWHM}{2\cdot\sqrt{2^{1/\beta}-1}},
\end{equation}
and $\beta$ is the atmospheric scattering parameter.
From examining bright and isolated stars across a series of OGLE-III Disk reference images, we find $\beta=3.0$ to provide a good empirical fit.
Following the prescription of \citet{king83}, we find the effective area over which a star contributes noise to the background to be
\begin{equation} \label{eq:moffat_noise_area}
   {\omegaeff}=\frac{5\pi\alpha^{2}}{4}=\frac{5\pi\cdot FWHM^{2}}{16\cdot(2^{1/3}-1)}
\end{equation}
for $\beta=3.0$.
This corresponds to an increase in {\omegaeff} by a factor of $\sim$1.7 when compared to a Gaussian with the same $FWHM$.

\begin{figure}
\centerline{
\includegraphics[width=9cm]{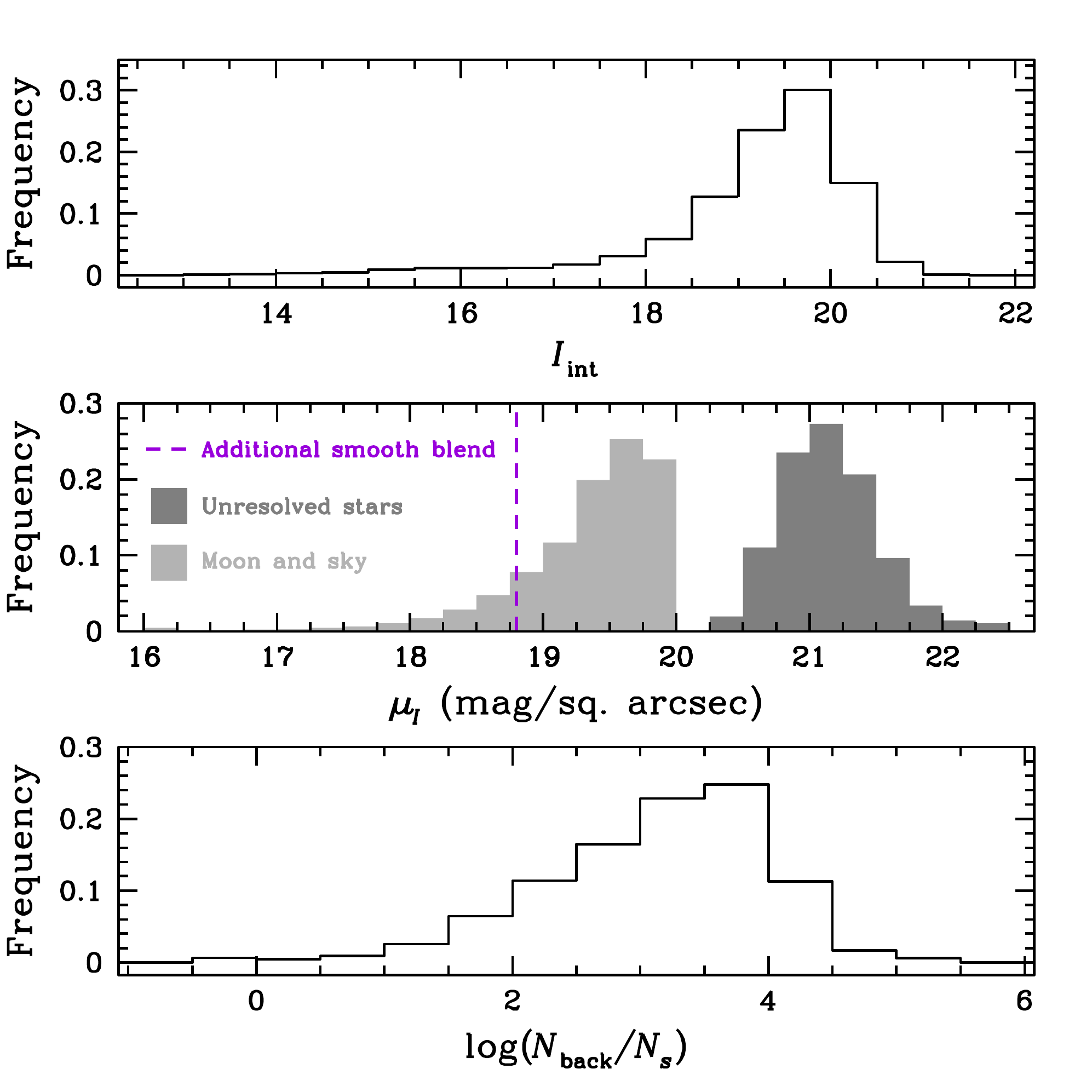}
}
\caption{
\footnotesize{
The distributions of the different sources of blend flux across the final set of target fields for all three observatories.
The top panel shows the distribution of the apparent magnitude of the bright interloping blend star, $I_{\rm int}$.
We see that, under our assumptions about and treatment of unassociated but blended stars, there is a floor for the brightness of objects that KMTNet will detect toward the Bulge at $I_{\rm obj} \approx 20$.
The middle panel panel gives the contributions to the total surface brightness from the Moon, the dark sky at zenith, faint unresolved stars, and the additional smooth blend we include to match the OGLE-III photometric uncertainties.
These will contribute to the noise of each data point, but not to the flux measurement itself.
The contribution from the Moon and the sky overwhelms that due to unresolved stars, although the additional smooth blend is the dominant contributing source of noise from the smooth backgrounds.
In the bottom panel we sum the photon rate from all sources of background---the lens, the interloping blend star, the Moon and sky, the faint sea of unresolved stars, and also the additional smooth blend---and compare it to the photon rate of the source.
In doing so we see that KMTNet will be background-dominated.
}
}
\label{fig:blend_components}
\end{figure}

Using this and taking $FWHM = \sigma_{\rm ref}$ to be the diameter of a seeing disc, we start with the brightest bin of our LF and sum across bins of decreasing brightness to obtain the cumulative number of stars that fall in a seeing disc, according to
\begin{equation}
   N_{\rm *,disc}=\sum_{j {\rm = 1}}10^{\Phi_{*,{\rm BW},j}}~{\omegaeff}~\xi_{*}~{\rm d}m.
\end{equation}
The apparent magnitude at which $N_{\rm *,disc} = 1$ defines the cutoff magnitude $I_{\rm cut}$.
We take there to be one interloping star brighter than $I_{\rm cut}$ that is unresolved and unassociated but blended with the event.
We draw from the cumulative distribution $N_{*,{\rm disc}}$ to obtain $I_{\rm int}$, the brightness of the interloping blend star.
Stars below $I_{\rm cut}$ contribute to the blend flux with a smooth surface brightness equal to the cumulative flux below $I_{\rm cut}$ divided by {\omegaeff}.
It should be noted that this approach will underestimate the number of blended interloping stars in some cases but will overestimate the total blend flux.
Figure \ref{fig:blend_components} shows our resulting distributions of $I_{\rm int}$ and the background surface brightness contributions from the Moon, the dark sky at zenith, and faint unresolved stars for our final set of target fields, discussed in \S \ref{sec:obsparm_opt}.

For each data point, the total object flux is calculated as the combination of the flux of the source, $F_{s}$, the lens, $F_{l}$, and the interloping blend star, $F_{\rm int}$,
\begin{equation}
   F_{\rm obj}(t) = {\fls}A(t) + F_{l} + F_{\rm int},
\end{equation}
where $A$ is the magnification at time $t$.
The photometric uncertainty is the combination of the Poisson photon uncertainty of the number of all object photons collected in an exposure,
\begin{equation}
   N_{\rm obj} = \dot{\gamma}_{\rm obj}{\texp},
\end{equation}
where $\dot{\gamma}_{\rm obj}$ is the combined photon rate of the magnified source, the lens, and the interloping blend star and {\texp} is the exposure time, and the total smooth background
\begin{equation}
   N_{\rm back} = \dot{\omega}_{\rm back}{\omegaeff}{\texp},
\end{equation}
where $\dot{\omega}_{\rm back}$ is the total photon rate per steradian of the Moon, the dark sky, and the smooth stellar background.
Here we use the seeing of the individual data point as modified by the airmass, described in \S \ref{sec:obsparm}, as the $FWHM$ for equation (\ref{eq:moffat_noise_area}).
The fractional Poisson photometric uncertainty $\sigma_{\rm Poi}$ is then
\begin{equation}
   \sigma_{\rm Poi} = \frac{\sqrt{N_{\rm obj} + N_{\rm back}}}{N_{\rm obj}}.
\end{equation}
We also include a fractional systematic uncertainty, {\sigsys}, to account for the limit of precision with which it is possible to measure even the brightest stars due to uncertainties from scintillation, flat-fielding, the determination of the PSF on the reference image, and other factors.
The final fractional uncertainty $\sigma_{\rm obs}$ on a given flux measurement is given by
\begin{equation}
   \sigma_{\rm obs} = \sqrt{\sigma_{\rm Poi}^{2} + \sigma_{\rm sys}^{2}}.
\end{equation}

\begin{figure}
\centerline{
\includegraphics[width=9cm]{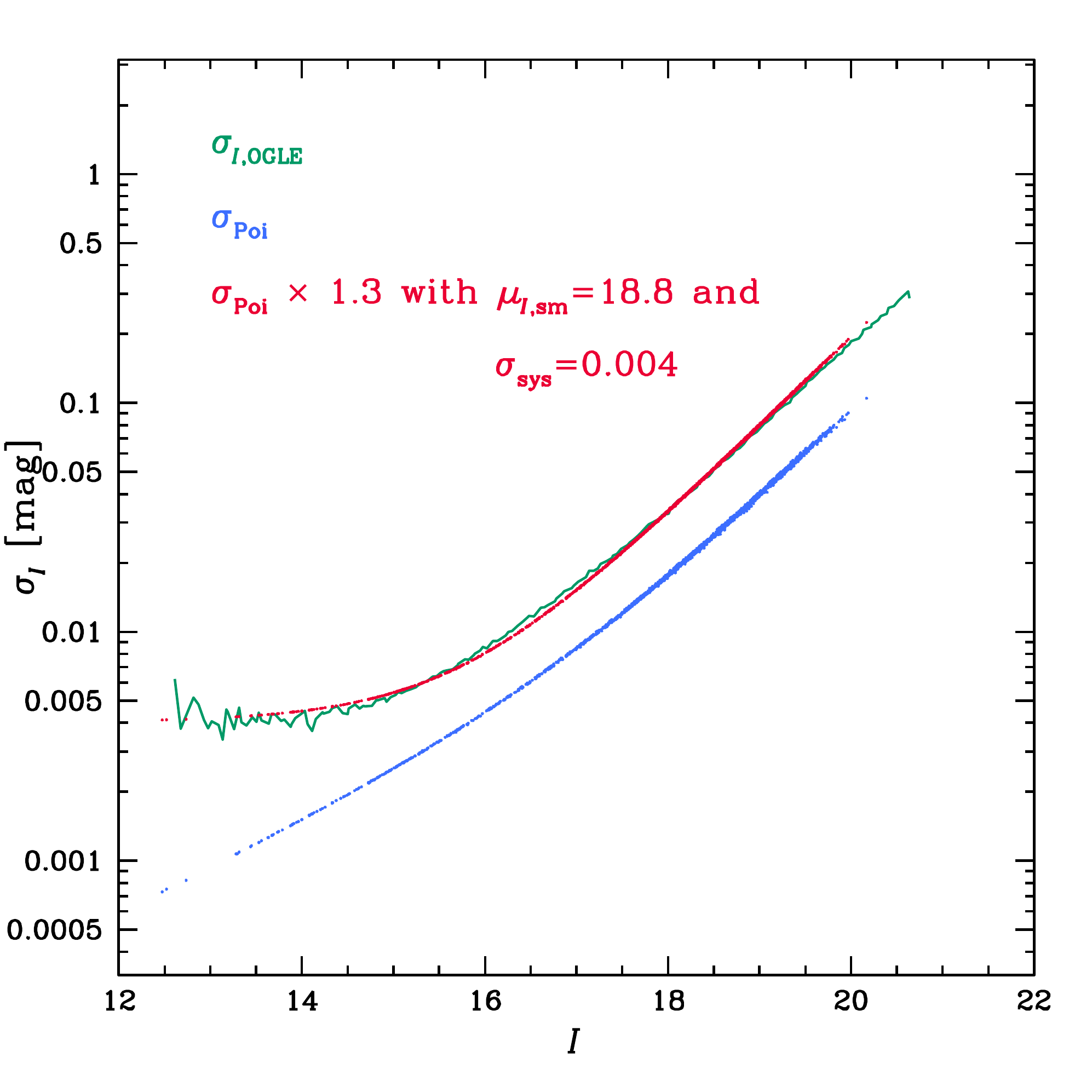}
}
\caption{
\footnotesize{
The result of matching the expected KMTNet photon rate normalization to OGLE-III photometric data.
The green line represents the 5th percentile of the RMS seen in OGLE-III light curves and the blue points represent our simulated KMTNet photometric data assuming the same photon rate normalization (2.11 ph s$^{-1}$ at $I=22.0$), aperture size (1.3m), and exposure time (${\texp} = 120$s) as OGLE-III.
We find that in order to match their photometric uncertainties we have to including a fractional systematic error floor of $\sigma_{\rm sys} = 0.004$ mag, scale the flux measurement Poisson uncertainties up by a factor of 1.3, and include an additional smooth background blend component of $\mu_{I,{\rm sm}} = 18.8$ mag/$\square\arcsec$.
The OGLE-III data shown here are for chip 5 from OGLE-III field 190, centered on (l,b)=(1.2136,~-3.8694), near BW.
}
}
\label{fig:matchphot}
\end{figure}

To calibrate the expected KMTNet flux in a realistic fashion we use OGLE-III photometric data.
The 1.3m aperture of OGLE-III obtains a photon rate of $2.11\dot{\gamma}$ for $I=22$.
We find that this photon rate normalization alone does not account for the photometric uncertainties seen in OGLE-III photometry.
We find that we are able to match the reported OGLE uncertainties only by introducing an additional smooth blend component of $\mu_{I,{\rm sm}} = 18.8~{\rm mag}/\square \arcsec$, scaling the resulting Poisson uncertainty up by a factor of 1.3, and including a systematic error floor of ${\sigsys} = 0.004$ magnitudes.
Figure \ref{fig:matchphot} shows an example using data from chip 5 of OGLE-III field 190, centered on $(l,b) = (1.2136,-3.8694)$, near BW.
Beyond exceptionally high amounts of scattered light, we are unable to account for the origin of the additional smooth background component that contributes so significantly to the noise.
Nevertheless, we include $\mu_{I,{\rm sm}}$ in our fiducial simulations to account for realistic observational hurdles.
We also use ${\sigsys} = 0.004$ mag as a pessimistic assumption of what the precision limit of KMTNet might be and scale the Poisson photometric uncertainties by the same factor of 1.3 as a conservative estimate.
We run another set of simulations, discussed in \S \ref{sec:rates_opt}, in which we remove this extra source of background, lower the value of {\sigsys}, and make other optimistic assumptions in an attempt to bracket our expectations of the detection yields that KMTNet will obtain.

We base the final photon rate of KMTNet on OGLE-IV, which obtains $3.24\dot{\gamma}$ for $I=22$, higher than for OGLE-III due to newer-generation CCD chips and an improved photometry pipeline.
Scaling this by the ratio of the KMTNet to OGLE apertures, $(1.6/1.3)^{2}$, yields a final photon rate normalization of
\begin{equation}
   \dot{\gamma}=4.91~{\rm ph~s^{-1}}\cdot10^{-0.4(I-22.0)}.
\end{equation}

For both our fiducial and more optimistic simulation results we do not include any scatter in the photometric measurements and instead take them to be exactly equal to what is expected for the binary-lens model.
If we were to include scatter, finding the best-fit FSBL model for each light curve would be prohibitively time-consuming.
Furthermore, adding such noise to the measured fluxes would cause different subsets of light curves to be scattered into and out of our sample of planet detections for each realization generated by the simulations.
We are interested in a more precise global estimate of the planet detection rate that KMTNet will obtain, one that is not dependent on such fluctuations, and so do not include any Gaussian (or otherwise) random noise in our simulated photometry.
This is equivalent to assuming noise in the photometry that is uncorrelated and scattered symmetrically about the binary lens model and averaging over a large number of realizations of our simulations.
In such a case, for each realization of the KMTNet detection rates, some detections would be scattered into, and others out of, the sample due to the inclusion of symmetric photometric noise.
On average, these competing effects would cancel out.
Thus, by not including any noise in our photometry we are both expediting the fitting process---we know the best-fit FSBL model exactly---and implicitly assuming our predictions for the detection rates to be an average over a large number of realizations of our simulations that include uncorrelated symmetric scatter, Gaussian or otherwise, in the photometry.

\subsection{Detection Algorithm} \label{sec:detalg}

A microlensing event will ultimately result in a planet detection only if the initial increase in brightness due to the primary event is detected and if the light curve subsequently displays sufficiently significant deviations from a single-lens microlensing event.
There are several criteria that must be satisfied for this to be true.
The most important are that the microlensing event itself must be detected and the perturbation due to the planetary companion must distinguish itself from a best-fit single-lens event, both according to predetermined {\dchisq} thresholds.
If a given microlensing event passes these cuts, it is considered robustly detected and its event rate is added to the total event rate of detected planets.

\subsubsection{Detection of the Primary Event} \label{sec:primev}

Before a microlensing event can be probed for planetary signatures, the initial increase in brightness due to the primary microlensing event must be reliably detected.
We establish three criteria to determine this.
We first compute the error-weighted mean flux of the light curve and subsequently the difference in $\chi^{2}$ between this constant model and the best-fit binary-lens model, {\dchisqpri}.
Since we do not include any scatter in our photometry, as discussed in \S \ref{sec:fluxdeter}, the best-fit binary-lens model is simply the light curve itself, and this curve has ${\dchisq} \equiv 0$.
Thus, {\dchisqpri} is just the {\dchisq} of the constant (error-weighted mean) model.
For the microlensing event to be initially detected, {\dchisqpri} must satisfy
\begin{equation}
   {\dchisqpri} \geq {\dchisqprith},
\end{equation}
where we choose a threshold of ${\dchisqprith} = 500$.
It should be noted that there will be a subset of events that will pass this threshold solely due to the planetary perturbation.
For certain low-magnification events the primary event will not be sufficiently distinct from a constant model, given our choice of {\dchisqprith}.
However, a significant planetary signature can itself increase {\dchisqpri} to cause the microlensing event to be initially detected, despite having a weakly magnified primary event.
FFPs represent the extreme limit of this case, when the signal due to the planet is the sole source of magnification over the course of the event.
Our second criterion is that there must be more than 100 points in the light curve, which we establish as a crude proxy for the precise determination of the lensing parameters.
Thirdly, {\tnot} must fall within the time coverage of the light curve, which also improves upon the precision of the parameters measured from the light curve.
This final criterion is not automatically satisfied due to the limited visibility of the Bulge at the beginning and end of each year.
If the event satisfies all of these criteria, its event rate is added to the total microlensing event rate, which is given by equation (\ref{eq:rate_tot_p}) without the second Heaviside step function.
This is separate from our final planet detection rate, as it considers solely the identification of primary microlensing events, independent of whether their planetary signature is ultimately detected.

\subsubsection{Detection of the Planetary Perturbation} \label{sec:planpert}

The next step is to search the primary event for the signature of a planetary companion.
First we use the input values for {\tnot}, {\unot}, and {\te} from the event and compute a comparison PSSL light curve.
We then compute the {\dchisq} of the FSBL light curve from this initial PSSL model, {\dchisqsli}.
We discard events with 
\begin{equation}
   {\dchisqsli} \leq {\dchisqslith},
\end{equation}
where we choose a threshold of ${\dchisqslith} = 100$.
We make this cut because we do not consider any events below {\dchisqslith} to be robustly detected, and finding a best-fit model will only decrease from the value of {\dchisqsli} that was found using the values from the binary lens as input for the parameters {\tnot}, {\unot}, and {\te}.
Moreover, finding a best-fit light curve is generally the second-most computationally expensive operation, after the FSBL magnification calculation, and extraneous fitting should be avoided.
Otherwise, if ${\dchisqsli} > {\dchisqslith}$, we determine a best-fit PSSL or FSSL model, depending on the algorithm described in \S \ref{sec:finitesource}.

In the case of a PSSL model, we must find the PSSL light curve whose observed flux best matches that of the FSBL light curve.
The model flux is given by
\begin{equation}
F(t) = {\fls} \cdot A(t) + {\flb}
\end{equation}
where {\fls} is the base flux of the un-magnified source, $A(t)$ is the magnification, and {\flb} is the total blend flux.
For a single-lens microlensing event, $F(t)$ is uniquely determined by the five parameters {\tnot}, {\unot}, {\te}, {\fls}, and {\flb}.
The observed flux is a linear function of {\fls} and {\flb} but is non-linear in {\tnot}, {\unot}, and {\te}, which together specify $A(t)$.
We use a downhill-simplex method \citep{press92} to explore the parameter space for combinations of {\tnot}, {\unot}, and {\te}.
At each point of the simplex, which represents a unique set of ({\tnot}, {\unot}, {\te}), we use matrix inversion to solve for $F_{s}$ and $F_{b}$ at each observatory \citep{gould03}.
We determine the PSSL light curve that yields the lowest {\dchisqslf} between itself and the simulated light curve and take that to be the best-fit PSSL model.
The procedure is identical when finding the FSSL model except that $\rho$ is included as a fourth non-linear free parameter during the downhill-simplex process.
If the {\dchisqslf} between this best-fit model and the FSBL light curve is greater than {\dchisqslfth}, we count the planetary event as being detected and its event rate is added to the total event rate of detected planets, according to equation (\ref{eq:rate_tot_p}).
We adopt {\dchisqslfth} = 160 as our canonical value \citep{bennett02} and note that while this value is appropriate for detections that arise from perturbations due to the source passing over a planetary caustic, it is likely too low for detections resulting from high-magnification events (\citealt{gould10,yee13}).
However, we aver this to be a valid threshold since the vast majority of planet detections will result from events with a peak magnification of $A \lesssim 100$, as discussed in \S \ref{sec:parm_dist}.

\begin{figure}
\centerline{
\includegraphics[width=9cm]{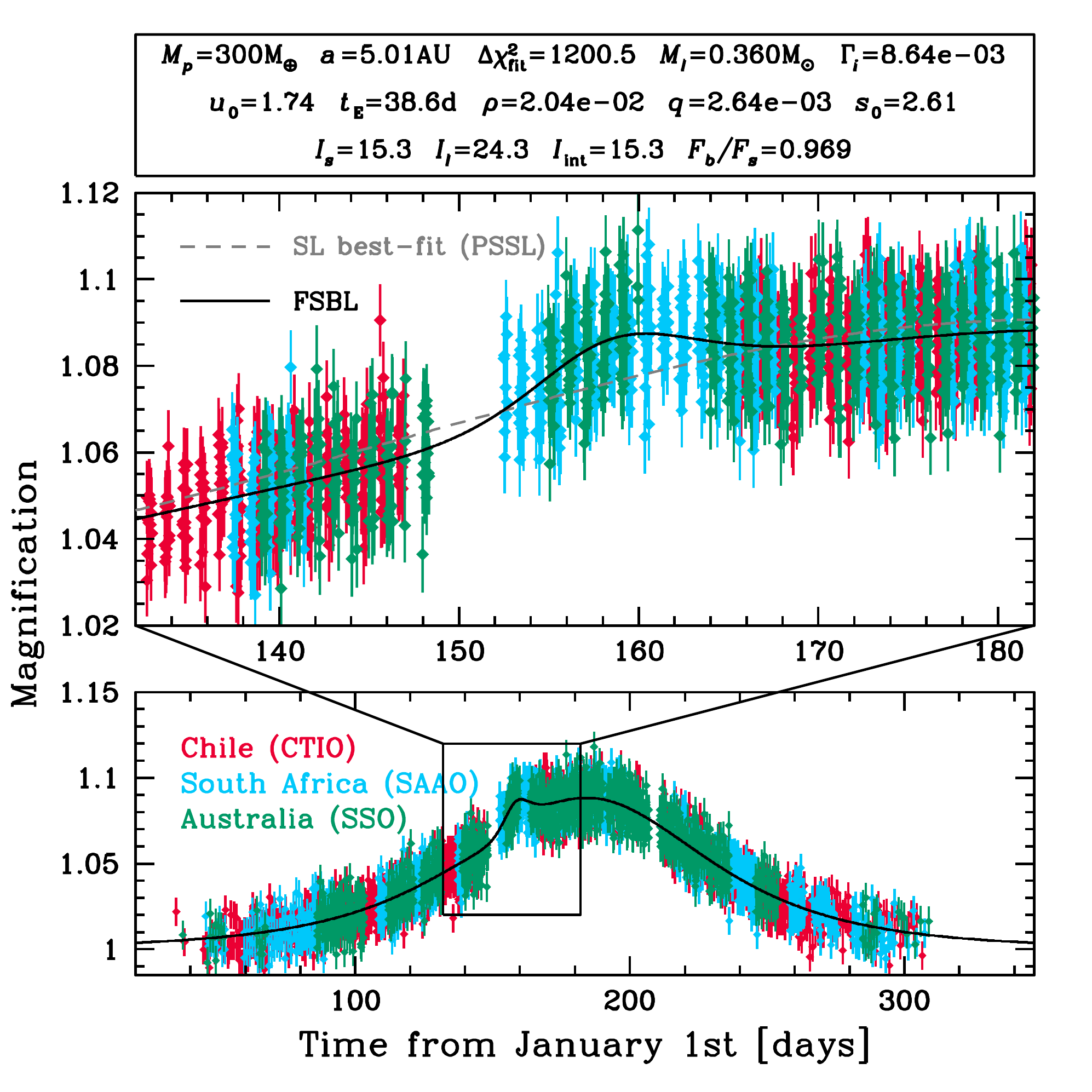}
}
\caption{
\footnotesize{
An example light curve for an analog of Jupiter.
The top panel shows the physical, lensing, and observational parameters for this event.
We have included Gaussian scatter in the photometry solely for the purposes of visualizing the quality of data that KMTNet will actually obtain.
This type of perturbation, with a smaller amplitude and a longer time scale, is common for larger planet masses.
Even a perturbation with such a low amplitude is robustly distinguished from the best-fit single-lens model due to the combination of KMTNet's photometric precision and cadence, allowing for precise and dense coverage of such deviations.
}
}
\label{fig:lc_jupiter}
\end{figure}

\begin{figure}
\centerline{
\includegraphics[width=9cm]{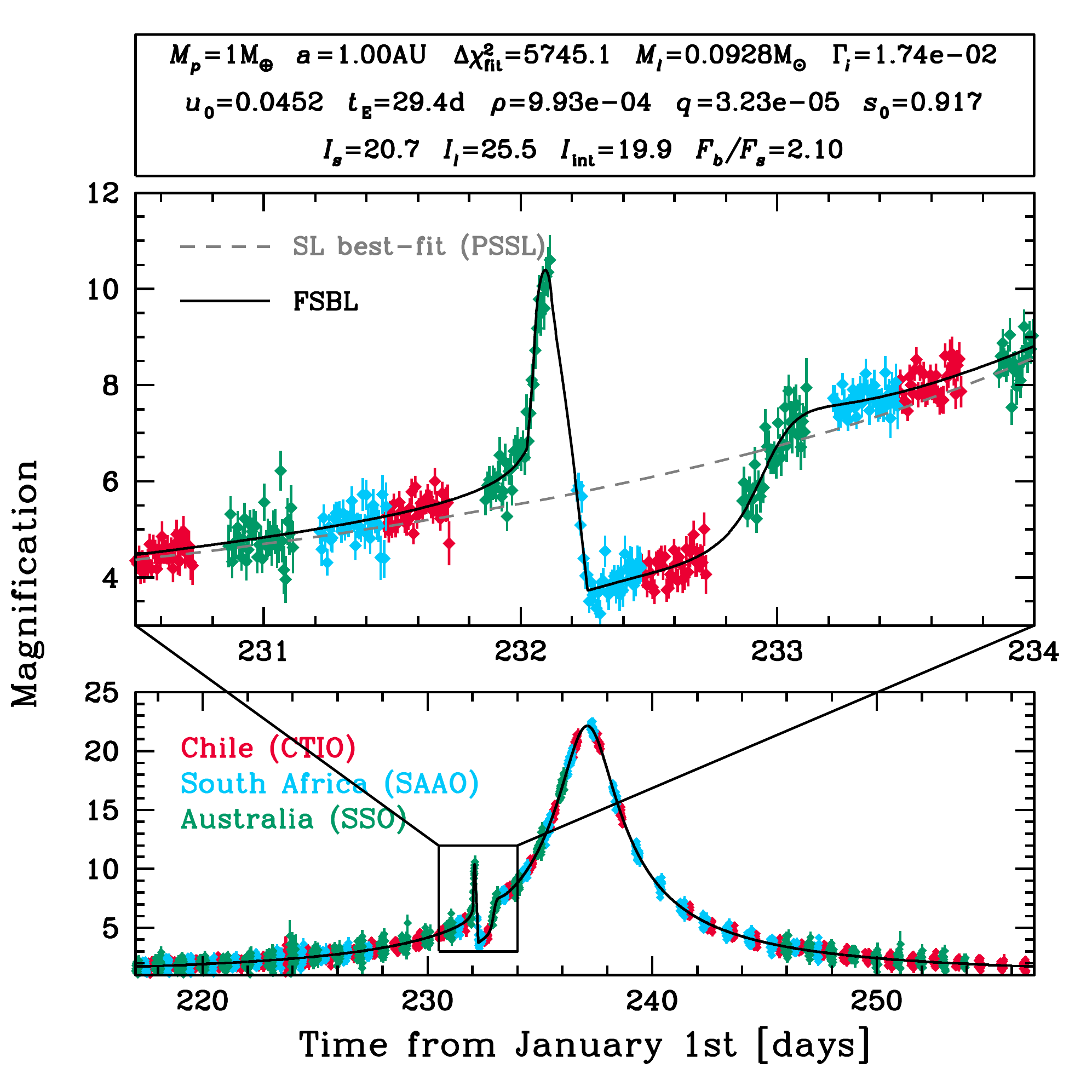}
}
\caption{
\footnotesize{
An example light curve for an analog of Earth.
The top panel shows the physical, lensing, and observational parameters for this event.
We have included Gaussian scatter in the photometry solely for the purposes of visualizing the quality of data that KMTNet will actually obtain.
This type of perturbation, with a larger amplitude and a shorter time scale, is common for smaller planet masses.
Even a planet for an event with such a faint source ($I_{s} \approx 21$) and short perturbation ($\Delta t_{p} \approx 1$d) can be robustly detected with KMTNet's photometric precision and cadence.
}
}
\label{fig:lc_earth}
\end{figure}

\begin{figure}
\centerline{
\includegraphics[width=9cm]{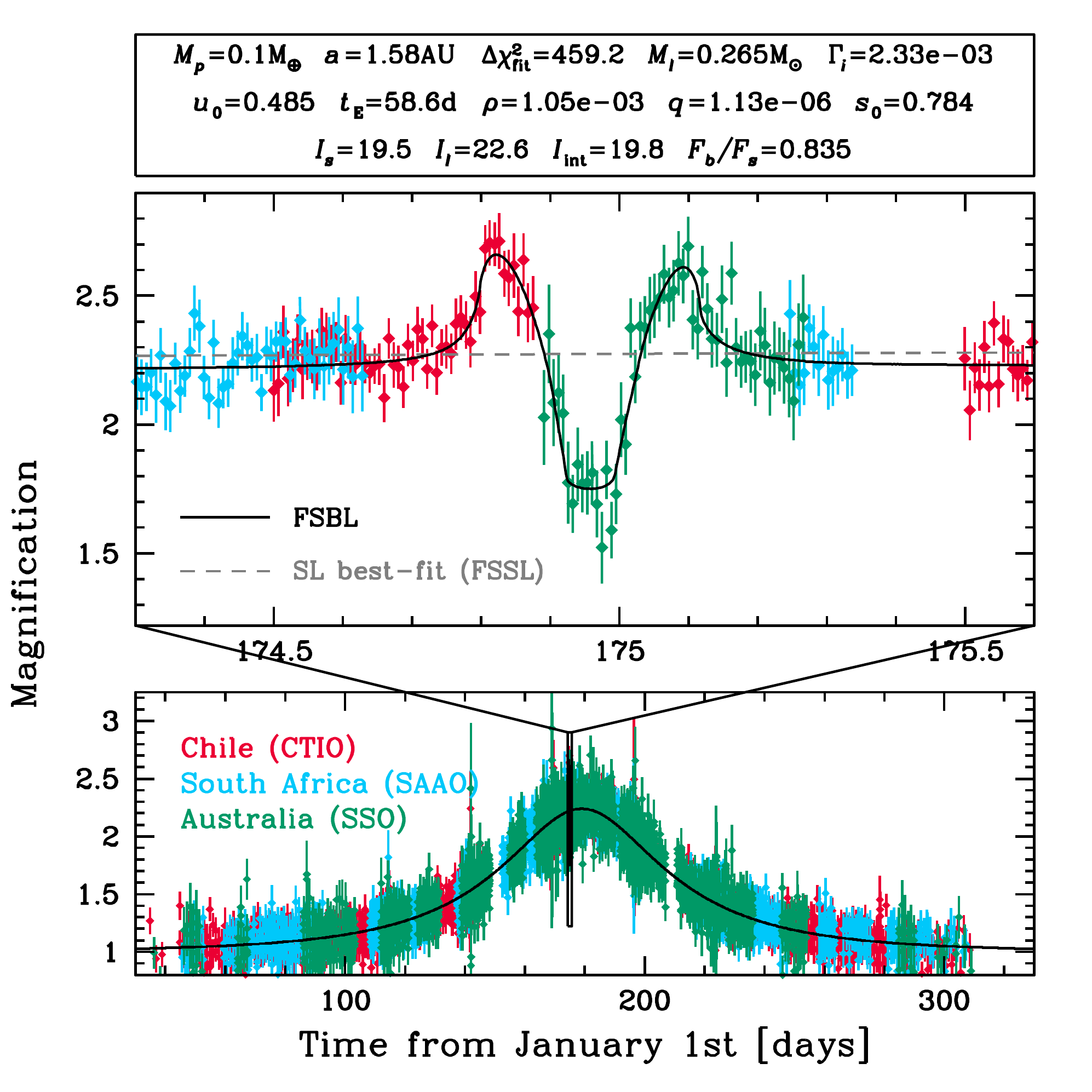}
}
\caption{
\footnotesize{
An example light curve for an analog of Mars.
The top panel shows the physical, lensing, and observational parameters for this event.
We have included Gaussian scatter in the photometry solely for the purposes of visualizing the quality of data that KMTNet will actually obtain.
Here the best-fit single-lens model is for a source of finite size.
Even such a low-mass planet can be robustly detected given the photometric precision and cadence of KMTNet.
}
}
\label{fig:lc_mars}
\end{figure}

Figures \ref{fig:lc_jupiter}--\ref{fig:lc_mars} show a representative sample of light curves for planet detections that are analogs of Solar System planets (in terms of planet mass and semi major axis).
In all cases we have included Gaussian scatter in the photometry and have done so solely for the purposes of visualizing the quality of data that KMTNet will actually obtain.
Figure \ref{fig:lc_jupiter} shows a light curve for a planet of mass ${\mpl} \approx 300{\rm M}_{\Earth}$ and $a \approx 5$AU, an analog of Jupiter.
This illustrates the types of perturbations that are common for planets of larger mass and are characterized by longer time scales and lower amplitudes.
For such bright sources, likely giants, the cadence and precision of KMTNet will facilitate robust detections even for perturbations with such low amplitudes.
Figure \ref{fig:lc_earth} shows a light curve for an Earth analog.
The planetary perturbation, while short, is of sufficiently high amplitude that the detection is clear, even given that the source is quite faint, with $I_{s} \approx 21$.
Finally, Figure \ref{fig:lc_mars} highlights a modest detection of a planet that is an analog to Mars.
In this case the combination of $u_{\rm 0}$, $\rho$, and $q$ indicates that finite-source effects are important, and in this case the best-fit is an FSSL model.

\section{Observational Parameter Optimization} \label{sec:obsparm_opt}

We use the simulations as described above to first determine the optimal observing strategy for KMTNet.
The parameters over which we optimize are the locations of the target fields, the number of target fields, {\nfld}, and the exposure time, {\texp}.
KMTNet is designed to conduct a uniform survey to explore exoplanet demographics in the regimes of parameter space to which microlensing is most sensitive.
Given its homogeneous approach, we perform our optimizations assuming each field to have the same value of {\texp}, which we ultimately optimize.
First we determine the importance of field placement, then we investigate trade-offs between different numbers of fields as well as observational cadence.

\subsection{Field Locations} \label{sec:field_location}

There is an arbitrary number of possible tilings for a given number of target fields.
Fortunately, as we will show, the precise locations of the field centers do not significantly affect the planet detection rate, provided that we restrict attention to fields with high event rate and low extinction.
For definiteness, we consider five tilings with a maximum of 13 fields each.
In choosing the field centers we avoid regions where ${\ai} \gtrsim 3.0$ but we otherwise maximize event rates by choosing regions with high stellar density and event rates.
All fields are chosen by-eye and generally constrained to be located within our dust map, which encapsulates all high-cadence OGLE-IV fields and the regions of highest measured microlensing optical depth.
Finally, we only consider tilings with non-overlapping fields.

The locations of the field centers in the different tilings generally differ by shifts that are of order half the detector size, or $\sim$1$^{\circ}$.
Determining the maximum planet yield across a large grid of planet mass and separation pairings for each of the five tilings would be prohibitively time-consuming.
Therefore, we instead use the total primary microlensing event rate down to $I_{s} = 22$ as a proxy for the planet detection rate.
For each field in each of these tilings we run 10$^{5}$ MC trials, drawing sources and lenses from within the field and computing their parameters and contributions to the event rate as described in \S \ref{sec:sim_over}.
The total event rate for all sources brighter than $I_{s} = 22$ per field is thus the product of the total event rate for all events, the solid angle of the FoV, and the number density of source stars down to $I_{s} = 22$ for the ($l$,$b$) of each event,
\begin{equation}
   \Gamma_{{\rm tot} ,I_{s} \leq 22}={\omegafov}\sum_{i = 1}^{\nmc}\Delta \Gamma_{i} \xi_{i} \Phi_{*,{\rm BW},I_{s} \leq 22,i},
\end{equation}
where, in $\Delta \Gamma_i$ (see equation (\ref{eq:rate_mc})), we have taken $u_{\rm 0,max} = 1$.

\begin{figure}
\centerline{
\includegraphics[width=9cm]{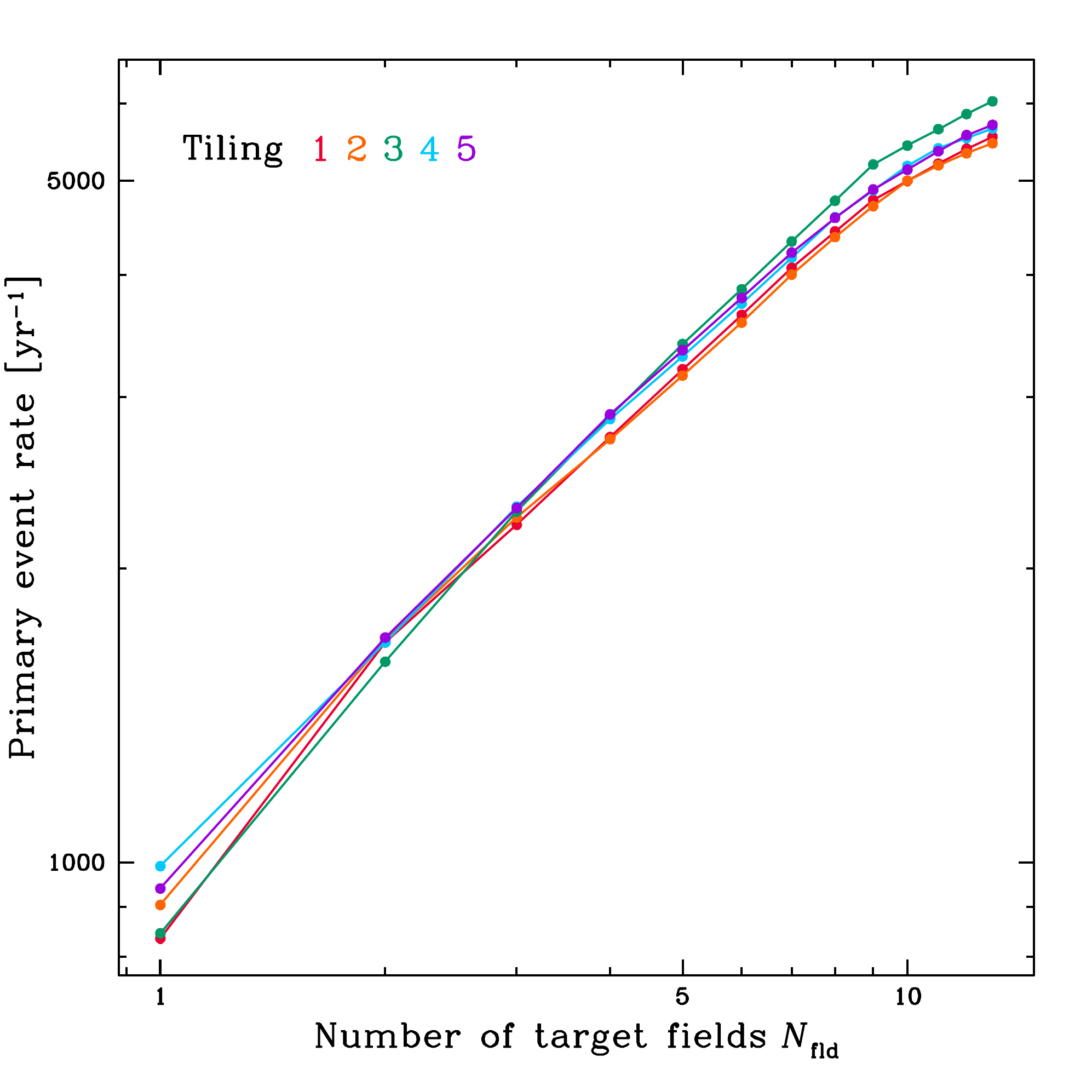}
}
\caption{
\footnotesize{
The cumulative microlensing event rate for sources with $I_{s} \leq 22$ as a function of number of fields {\nfld} for our five different tilings of 13 fields each.
While there is a slight preference for different tilings at certain fixed numbers of fields, the differences between all tilings for a given value of {\nfld} are nearly all within three sigma of the Poisson error on the rates.
Given this, we consider the event rates across all tilings to be essentially equivalent.
}
}
\label{fig:tiling_comparison}
\end{figure}

We then rank all fields within a given tiling by this event rate.
Figure \ref{fig:tiling_comparison} shows the cumulative event rate for each of the five different tilings.
While there is a slight preference for different tilings at certain fixed numbers of fields, the differences between all tilings for a given value of {\nfld} are nearly all within three sigma of the Poisson error on the rates.
Given this, we consider that the event rates across all tilings are essentially equivalent.
In other words, from our model and set of assumptions there does not appear to be an optimal tiling of target fields, provided they are chosen to lie within regions of sufficiently low extinction and sufficiently high stellar density.
This is a consequence of KMTNet's FoV, which is generally larger than the features in the morphology of the optical depth and stellar density in the regions of the Bulge we are considering.
To converge on an optimal tiling we turn to the results of \citet{sumi13}, who use two years of MOA-II data to measure the microlensing event rate and optical depth toward the Bulge.
They identify and model a peak in the event rate per square degree per year that is located at $(l,b) \approx (1.0, -2.5)$ (see their Figures 3 and 12).
The most highly ranked field for tiling 4 is roughly coincident with this peak, so we thus select tiling 4 as our fiducial tiling and utilize it to conduct the further optimizations.

\begin{figure}
\centerline{
\includegraphics[width=9cm]{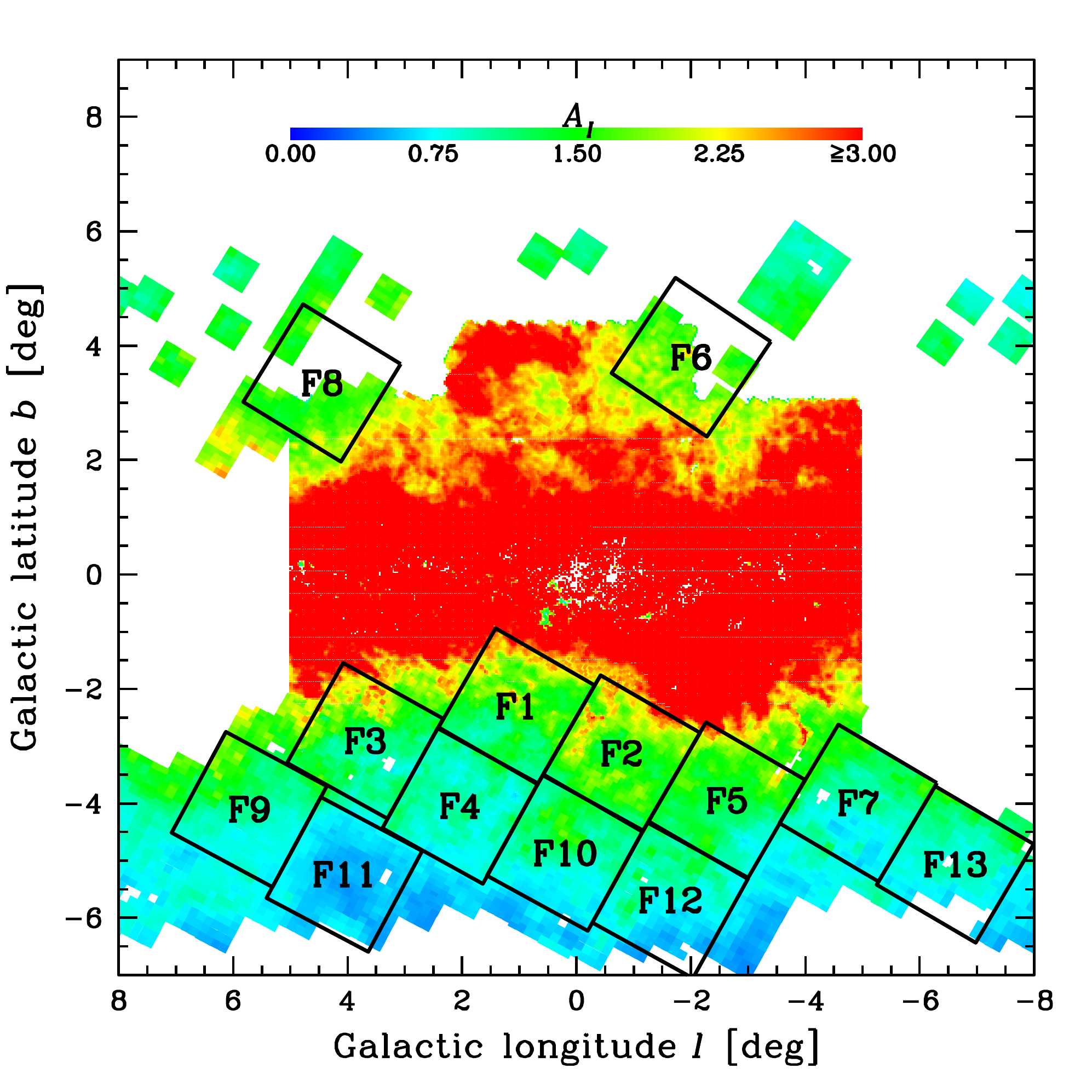}
}
\caption{
\footnotesize{
Our fiducial tiling of 13 KMTNet fields, ranked by their primary event rate, overlaid on our extinction map.
}
}
\label{fig:tiling_fid}
\end{figure}

In order to more robustly determine the ranking of the fields within this tiling, we run our full simulation across a 3x3 grid of planet mass and planet-star separation, with each variable equally separated in log-space.
This grid specifically includes planet masses of ${\rm log}({\mplme}) = 0.00, 1.00,~{\rm and}~2.00$ and planet-host star separations of log($a$/AU) = 0.15, 0.40, and 0.65 for each mass.
In linear units this corresponds to planet masses of {\mpl} = 1, 10, and 100${\rm M}_{\Earth}$ and planet-host star separations of $a \approx$ 1.41, 2.51, and 4.47AU.
We assume observations are only taken in one field, and then for each field in turn, we create and fit the light curves and implement the {\dchisq} cuts discussed in \S \ref{sec:detalg}.
We then rank the fields according to their primary event rate as well as their planet detection rate.
The rankings using the two different metrics are nearly identical, and for those field rankings that are not, the differences in the rates between the two different fields with the same rank are within one sigma of the Poisson uncertainty in the rates.
We ultimately select the primary event rate as our indicator for field ranking and order the fields accordingly.
These final rankings and the corresponding coordinates are given in Table \ref{tab:tiling_fid}.
Figure \ref{fig:tiling_fid} shows the field locations of this tiling, overlaid on our extinction map, with field rankings labeled.

\begin{deluxetable}{ccccc}
\tablecaption{Fiducial Tiling of KMTNet Fields}
\tablewidth{0pt}
\tablehead{
\colhead{Field Rank}   &
\colhead{$l$}          &
\colhead{$b$}          &
\colhead{$\alpha$}     &
\colhead{$\delta$}     \\
\colhead{}             &
\colhead{[deg]}        &
\colhead{[deg]}        &
\colhead{[deg]}        &
\colhead{[deg]}
}
\startdata
1    &    1.0500   &   -2.3000   &   269.28090   &   -29.20837   \\
2    &   -0.7846   &   -3.1263   &   269.04706   &   -31.20837   \\
3    &    3.6969   &   -2.9066   &   271.36107   &   -27.20837   \\
4    &    2.0361   &   -4.0408   &   271.57224   &   -29.20837   \\
5    &   -2.6225   &   -3.9495   &   268.80282   &   -33.20837   \\
6    &   -2.0000   &    3.8000   &   261.52480   &   -28.57250   \\
7    &   -4.9212   &   -3.9919   &   267.44505   &   -35.20837   \\
8    &    4.4500   &    3.3500   &   265.81065   &   -23.39383   \\
9    &    5.7235   &   -4.1018   &   273.62035   &   -26.00837   \\
10   &    0.2021   &   -4.8673   &   271.38545   &   -31.20837   \\
11   &    4.0587   &   -5.2401   &   273.86816   &   -28.00837   \\
12   &   -1.6355   &   -5.6909   &   271.19321   &   -33.20837   \\
13   &   -6.6162   &   -5.0639   &   267.52308   &   -37.20837
\enddata
\tablecomments{The fields are ranked by primary event rate.}
\label{tab:tiling_fid}
\end{deluxetable}

\subsection{Number of Fields} \label{sec:num_fields}

Now that a single tiling of target fields has been selected, we determine the optimal number of fields, {\nfld}, within this tiling.
We run our full simulation over the same 3x3 grid of {\mpl} and $a$ as in \S \ref{sec:field_location}, again creating and fitting the light curves and implementing the {\dchisq} cuts.
We vary the total number of fields from 1--13 and add fields according to the ranks determined in \S \ref{sec:field_location}, as shown in Table \ref{tab:tiling_fid}.
We include the overhead time for KMTNet of $t_{\rm over} = 30$s (Table \ref{tab:kmt_cam_parms}).
In addition, we assume a fixed exposure time of ${\texp} = 120$s.
As a result, the cadence  $t_{\rm cad}$ with which the light curves in a given field are sampled increases with the number of fields $\nfld$ as $t_{\rm cad} = 150{\rm s}~{\nfld}$.

\begin{figure}
\centerline{
\includegraphics[width=9cm]{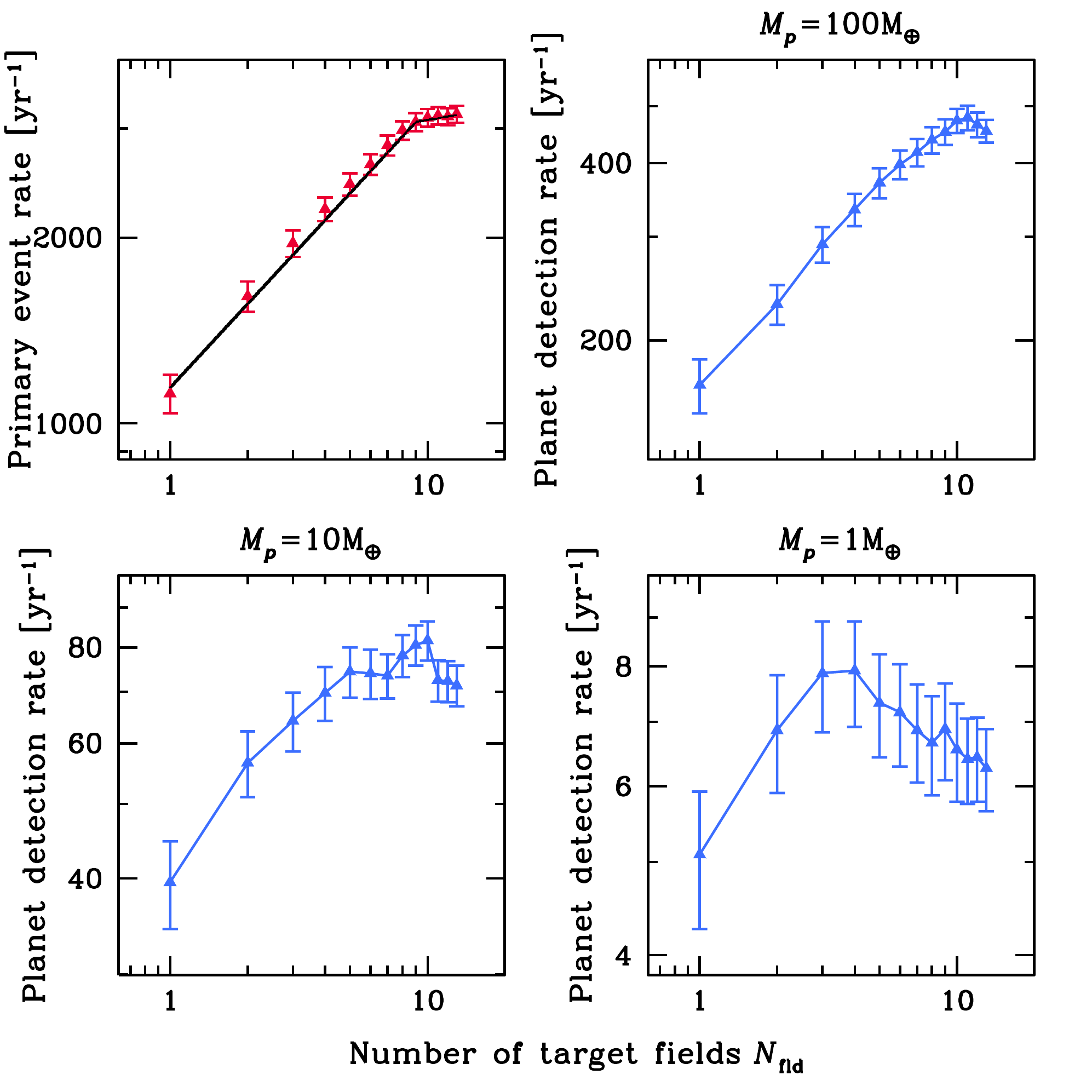}
}
\caption{
\footnotesize{
The upper left panel shows the cumulative primary event rate as a function of the number of fields {\nfld} for our fiducial tiling.
Here the black line shows the broken power-law fit, with a slope of $\alpha \approx 0.45$ for $1 \leq {\nfld} \leq 9$ and $\alpha \approx 0.069$ for $9 \leq {\nfld} \leq 13$.
The remaining three panels show the planet detection rate for three different planet masses, summed across semimajor axis $a$, as a function of {\nfld}.
Here we have assumed a constant exposure time of $120$s and an overhead of $30$s, and thus the cadence varies with the number of fields as 150s {\nfld}.
The cumulative primary event rate increases with {\nfld} as a broken power-law with the shift in slope occurring at ${\nfld} \approx 9$.
On the other hand, the behavior of the planet detection rate as a function of {\nfld} shows a dependence on planet mass, reaching maxima at lower numbers of target fields for planets of lower mass.
}
}
\label{fig:numfields_rates}
\end{figure}

Figure \ref{fig:numfields_rates} shows the cumulative primary event rate and planet detection rate as a function of the number of target fields, the latter for three different planet masses.
The cumulative primary event rate increases monotonically with {\nfld} as a power-law ${\nev} \propto N_{\rm fld}^{\alpha}$ with $\alpha \approx 0.45$ for ${\nfld} \lesssim 9$, at which point the slope becomes more gradual, $\alpha \approx 0.069$.
We find that the detailed behavior of the variation in the planet detection rate with {\nfld} depends on the planet mass.
For ${\mpl} = 100{\rm M}_{\Earth}$, the planet detection rate increases almost monotonically with additional target fields up until ${\nfld} \approx 11$, at which point the slope becomes negative and the number of detected planets decreases with additional target fields.
For ${\mpl} = 10{\rm M}_{\Earth}$, the detection rate has approximately leveled off by ${\nfld} \approx 5$.
For ${\mpl} = 1{\rm M}_{\Earth}$, the planet detection rate peaks at ${\nfld} \approx$ 3--4 and decreases for ${\nfld} > 4$.

The behavior of the planet yield with the number of fields can be understood by examining the two competing factors that affect the planet detection rate.
On one hand, the cumulative primary event rate increases with {\nfld} for a fixed value of {\texp}, as shown in Figure \ref{fig:numfields_rates}.
In other words, the total number of primary microlensing events monitored and detected increases monotonically as the number of fields increases, at least up to $\nfld = 13$, and therefore there is a larger number of primary events in which to detect perturbations from planetary companions.
On the other hand, each of these primary light curves will be more poorly sampled, resulting in a decrease in the signal-to-noise ratio (i.e., {\dchisqslf}) of the planetary perturbations.
According to equation (\ref{eq:tep}), the planetary Einstein ring crossing time $\Delta t_{p}$ for an Earth-mass planet orbiting a 0.3${\rm M}_{\odot}$ host star ($q \sim 10^{-5}$) for an event with a time scale of ${\te} = 20$d is $\sim$90 minutes, and the durations of the planetary perturbations are a factor of a few times larger than this.
Therefore, given the scaling of $t_{\rm cad}$ with {\nfld}, this would lead to many dozens of observations on a typical perturbation for a single target field, but fewer than 10 observations for 13 target fields.

To investigate these two contributions more quantitatively, we first examine the dependence of the primary event rate on {\nfld}.
Revisiting the broken power-law behavior of Figure \ref{fig:numfields_rates}, we find that the cumulative primary event rate goes as ${\nev} = k_{F} N_{\rm fld}^{\alpha}$, with a change in $\alpha$ at ${\nfld} \approx 9$.
For $1 \leq {\nfld} \leq 9$, $k_{F} \approx 1140$ and $\alpha \approx 0.45$.
Over the range $9 \leq {\nfld} \leq 13$ the slope flattens to $\alpha \approx 0.069$, and the coefficient increases to $k_{F} \approx 2640$.
We see that $\alpha$ is always positive, indicating that including additional target fields acts to increase the microlensing event rate monotonically up to at least ${\nfld} \approx 13$.
However, the break at ${\nfld} \approx 9$ indicates a transition to a regime in which adding additional fields becomes significantly less advantageous with regard to the total microlensing event rate, because one must monitor fields with lower stellar density and lower event rate.
Over this entire range of {\nfld}, $\alpha < 1$ because we have ordered the fields by decreasing event rate.

\begin{figure}
\centerline{
\includegraphics[width=9cm]{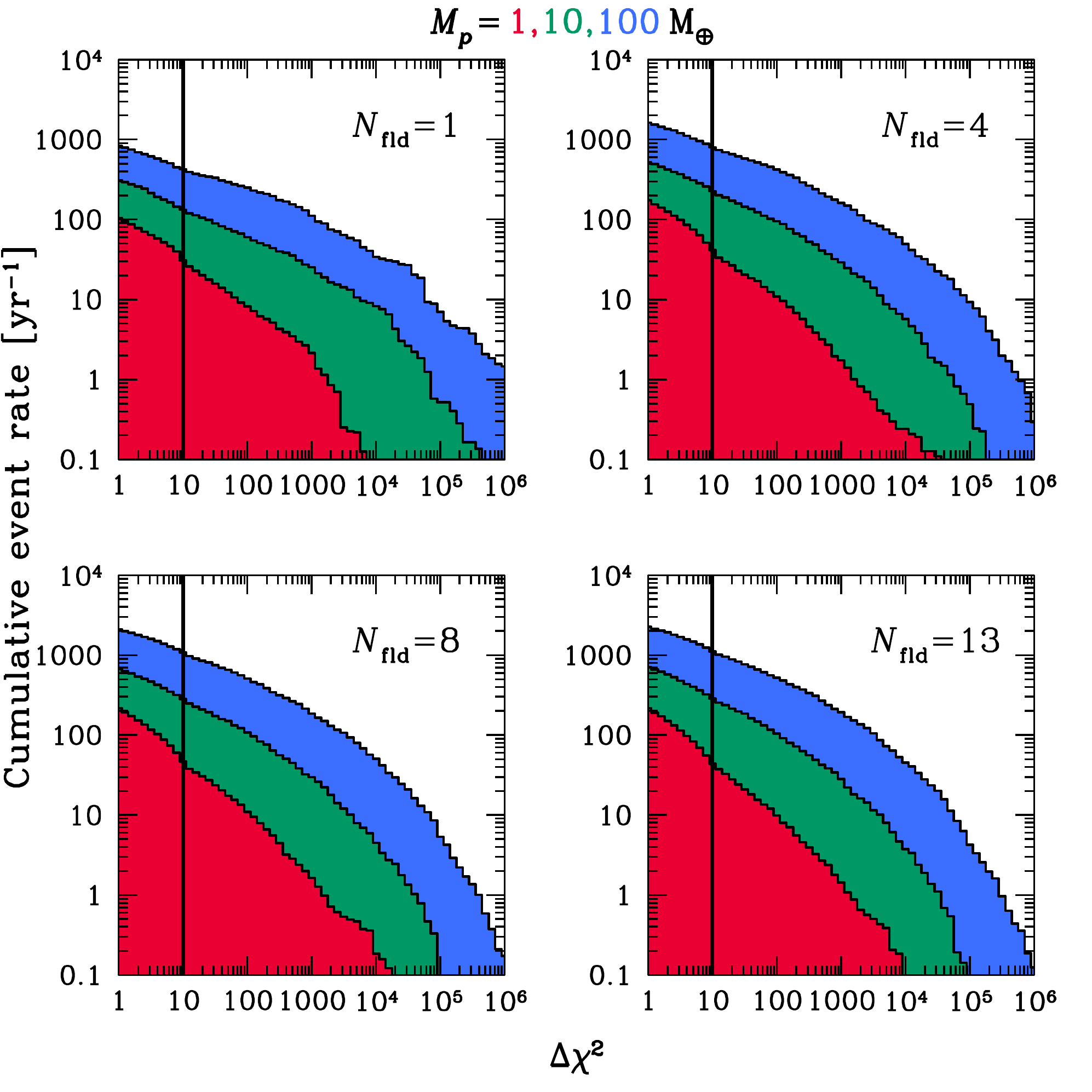}
}
\caption{
\footnotesize{
Primary event rate as a function of the final {\dchisq} for each event, summed across semimajor axis $a$, for four different numbers of fields using a value of ${\dchisqprith}=10$, denoted by the solid black line.
Above {\dchisqprith} and up until some {\dchisqslf} cutoff, the event rate obeys a power law whose slope increases in magnitude with decreasing planet mass.
We find that the slope of this power law increases with {\nfld} and is steeper for smaller planet masses, indicating the increasing importance of finite-source effects.
}
}
\label{fig:numfields_chi2}
\end{figure}

To better understand the underlying {\dchisqslf} distribution, we run a set of higher-fidelity simulations across the same grid of {\mpl} and $a$ for each number of fields.
We increase {\nmc} and lower the primary event detection threshold to ${\dchisqprith} = 10$ in order to increase the total number of simulated events and decrease the statistical noise.
Figure \ref{fig:numfields_chi2} shows the cumulative microlensing event rate as a function of {\dchisq} for four values of {\nfld} for each value of {\mpl}, summed across $a$.
We see that above {\dchisqprith} the {\dchisq} distribution is well-characterized by a power-law, such that the fraction of events with $\dchisqslf$ above a minimum value $\Delta\chi^2_{\rm min}$ scales as $f_{\rm det}\propto \left({\Delta\chi^2_{\rm min}}\right)^{-\beta}$ with $\beta>0$ up to some cutoff value, {\dchisqslfcut}, with the distribution dropping off precipitously above this value due to finite-source effects.
This cutoff occurs at lower values of {\dchisqslf} for lower planet masses because as $q$ decreases, the presence of finite-source effects increasingly hampers our ability to detect the planetary signals.

Even conservative choices of {\dchisqslfcut} encapsulate $\geq$90$\%$ of the primary event rate for each number of fields, so we take the power-law regime to be a robust proxy for the total microlensing event rate.
Because we assume a fixed {\texp}, $t_{\rm cad} \propto {\nfld}$, which in turn implies that the number of data points per unit of time, and thus ${\dchisqslf}$, is inversely proportional to the number of fields ${\dchisqslf}\propto N_{\rm fld}^{-1}$.
As a result, for our approximate power law form for the cumulative distribution of $\dchisqslf$, the fraction of events with $\dchisqslf > \Delta\chi^2_{\rm min}$ scales as $f_{\rm det}\propto N_{\rm fld}^{-\beta}$.

Finally, we can combine these two results to estimate the total planet detection rate,
\begin{equation}
   N_{\rm det} = {\nev}f_{\rm det} \propto N_{\rm fld}^{\alpha-\beta}.
\end{equation}
Both $\alpha$ and $\beta$ are positive for all values of {\dchisq}, {\nfld}, and $M_{p}$ we examined.
Thus, when $\alpha > \beta$, including additional target fields will increase the planet detection rate, because the loss in the number of detections due to the poorer sampling of the planetary perturbations is more than compensated for by the increase in the number of primary events.
On the other hand, if $\alpha <\beta$, then adding more fields will result in a net decrease in the number of detected planets since the additional fields have an optical depth that is sufficiently low that the increase in the number of primary events they contribute is overwhelmed by the decrease in planet detections due to the poorer sampling.
If $\alpha \simeq \beta$, the total number of planet detections is essentially independent of the number of fields.
Pictorially, we can understand the result of these two competing effects by examining the morphology of Figure \ref{fig:numfields_chi2}.
As {\nfld} increases, the overall {\dchisqslf} distribution gets shifted up to higher event rates due to the increase in the number of detected primary events and to the left to lower values of {\dchisqslf} due to the increase in $t_{\rm cad}$ and thus decrease in the number of data points and resulting {\dchisqslf} per light curve.

The value of $\beta$ gradually increases as {\nfld} increases from {\nfld} = 1 to {\nfld} = 13, steepening from $\beta \approx 0.3$ to $\beta \approx 0.4$ for ${\mpl} = 100{\rm M}_{\Earth}$ over the range of {\nfld}, $\beta \approx 0.4$ to $\beta \approx 0.5$ for ${\mpl} = 10{\rm M}_{\Earth}$, and $\beta \approx 0.56$ to $\beta \approx 0.75$ for ${\mpl} = 1{\rm M}_{\Earth}$.
This indicates that the drop-off of planet detection rate with {\dchisqslf} becomes more severe as {\nfld} increases and as $M_{p}$ decreases.
Recalling that $\alpha \approx 0.45$ for $1 \leq {\nfld} \leq 9$ and $\alpha \approx 0.069$ for $9 \leq {\nfld} \leq 13$, we see that $\alpha < \beta$ for for the entire range of {\nfld} for ${\mpl} = 1{\rm M}_{\Earth}$ and that $\alpha < \beta$ by ${\nfld} \approx 5$ for ${\mpl} = 10{\rm M}_{\Earth}$, causing the planet detection rate to level off.
For ${\mpl} = 100{\rm M}_{\Earth}$, the balance is not reached over our range of {\nfld}.
The expectations based on these approximate analytic scalings are roughly in accord with the detailed results shown in Figure \ref{fig:numfields_rates}. 

Thus we conclude that there is no unique value for the optimum number of fields, because the choice that maximizes the planet detection rate depends on the planet mass.
Furthermore, there are additional considerations.
In all cases, even though the planet detection rate may remain roughly constant or increase slightly as one increases the number of fields beyond 2--4, it is clear that these perturbations will be less well sampled, and therefore less well characterized.
Therefore, we choose four target fields as our fiducial optimal value because although the detection rates for massive planets may increase slightly for a larger number of fields, the increase is modest, the sampling of the perturbations is worse, and these detections will be less well characterized.
Furthermore, we are more interested in low-mass planets, whose detection rates and characterizations will be improved for fewer fields.

\subsection{Exposure Time} \label{sec:exp_time}

In the absence of overheads (e.g., detector readout time and telescope slew and settle time), pixel saturation, and systematic precision limits, conservation of information dictates that the exposure time should not affect the total microlensing rate or the planet detection rate, provided that {\texp} is sufficiently shorter than the duration of typical magnification features.
However, in reality, the value of {\texp} can significantly affect both the number and type of microlensing events detected, given the existence of overheads, saturation, precision limits, and planetary signatures of varying durations.
The choice of {\texp} is, then, a balance between number of data points per light curve and the resulting uncertainties in the flux measurement, both of which contribute to the {\dchisq} of the event and ultimately the uncertainties in the derived parameters.
We now investigate the effect of varying {\texp} for a fixed number of fields.
The combined slew, settle, and readout time for KMTNet will be $t_{\rm over} = 30$s, so we investigate the effect that choices of {\texp} within a factor of several of $t_{\rm over}$ will have on the event and detection rates.

We run our simulation across the same 3x3 grid of {\mpl} and $a$ for a range of cadences.
In \S \ref{sec:num_fields} we found that a smaller number of fields is preferable for planets with ${\mpl} \lesssim 10{\rm M}_{\Earth}$.
Higher-mass planets have only a slight increase in detection rates with more target fields, and that boost would be mitigated to some extent by larger parameter uncertainties.
We thus explore cadences of 300, 600, 900, 1200, and 1500s for 1--6 fields.

\begin{figure}
\centerline{
\includegraphics[width=9cm]{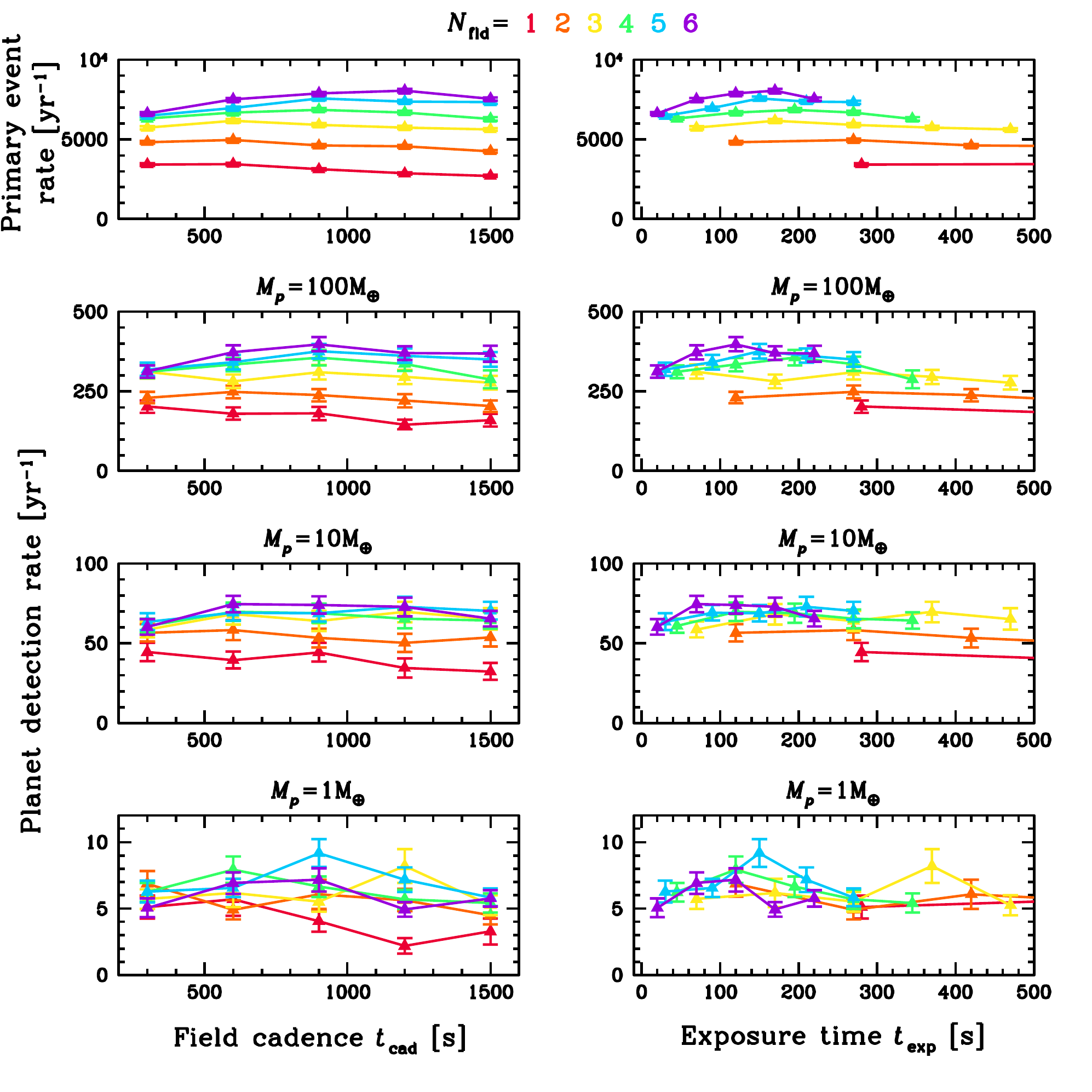}
}
\caption{
\footnotesize{
The upper two panels show the primary event rate as a function of the cadence $t_{\rm cad}$ (left) and the exposure time {\texp} (right) for five values of $t_{\rm cad}$.
The lower three rows each represent the planet detection rate for a different planet mass.
Each line represents a different value of {\nfld} and is color-coded according to the legend at the top.
For all planet masses and cadences there is a preference for $100 \lesssim {\texp}/{\rm s} \lesssim 250$, where there is a broad maximum in event rate.
In the panels on the right the lines for smaller values of {\nfld} continue to gradually decay for higher values of {\texp}.
}
}
\label{fig:tcad_texp}
\end{figure}

Figure \ref{fig:tcad_texp} shows how the primary event rate and planet detection rate for each planet mass {\mpl}, summed across semimajor axis $a$, changes as a function of cadence and exposure time for different numbers of fields.
For {\mpl} = 10 and 100${\rm M}_{\Earth}$ for ${\nfld} \gtrsim 4$, the event rate increases with exposure time up to ${\texp} \approx 200$s.
For ${\texp} \gtrsim 200$s, the event rate begins to decrease.
Smaller numbers of fields, specifically ${\nfld} \lesssim 3$, show a steady decline of event rate with {\texp}.
For ${\mpl} = 1{\rm M}_{\Earth}$, the same structure is present for ${\nfld} \gtrsim 4$, but the decline with {\texp} for ${\nfld} \lesssim 3$ is less pronounced.

The drop in event rate as {\texp} decreases for ${\nfld} \gtrsim 4$ arises from the fact that as {\texp} approaches $t_{\rm over}$, the total time spent on a given observation becomes dominated by overhead time, so fewer photons are collected, driving the measured flux uncertainties up and consequently lowering the {\dchisq}.
The drop in event rate as ${\texp}/t_{\rm over}$ increases is slightly more subtle.
Though a larger {\texp} decreases the number of data points per light curve, information should be conserved, as the extant data points would have reduced noise.
However, this gain in precision is bounded by both the saturation depth of an individual pixel, which renders individual data points with sufficiently high flux unusable, as well as the systematic error floor, which establishes the lower precision limit of the photometry.

We note that since our simulation does not strongly prefer a specific choice of {\texp}, any value in the range $\sim$100--250s appears to produce equivalent microlensing event and planet detection rates.
Given that we do not find evidence for a sharply optimal exposure time, we choose 120s ($4 \times {t_{\rm over}}$, corresponding to a cadence of 10 minutes for 4 fields) as the fiducial exposure time to minimize noise and maximize number of data points.

\section{Fiducial Simulation Run} \label{sec:sim_fid}

Having converged on the field locations, {\nfld}, and the exposure time {\texp}, we run a set of simulations across a larger 9x17 grid of planet mass and planet-star separation, with each variable again equally separated in log-space.
This grid spans the range {\mprange} and {\arange}, with 0.50 and 0.10 dex spacing, respectively.
In linear units, to which we will refer for the remainder of the text, this corresponds to planet masses of $0.1 \leq {\mplme} \leq 1000$ and planet-star separations of $0.398 \lesssim a/{\rm AU} \lesssim 15.8$.
For the sake of brevity we will limit each quoted linear value of {\mpl} and $a$ for all grid points to a maximum of three significant figures.
This grid of {\mpl} and $a$ constitutes our fiducial results, which we take to be a conservative estimate of the KMTNet event and planet detection rates.
For this final run we implement the methodology described in \citet{penny13a} to expedite our simulations.
Their Caustic Region Of INfluence (CROIN) parametrization of binary microlensing events identifies an area centered on the planetary caustics outside of which there will be no (detectable) deviation due to the presence of a planetary companion.
We use their CROIN parametrization to avoid modeling events whose source does not pass through the CROIN but we include high-magnification events, which have ${\unot} \ll 1$.
We also run a more optimistic simulation on the same grid, discussed in \S \ref{sec:rates_opt}, and anticipate that these two will bracket the microlensing event rates and planet detection rates that KMTNet will obtain.

After running our full grid of simulations, we find a handful of cases for which our magnification computation algorithm, described in \S \ref{sec:magcalc}, fails, yielding data points with incorrect FSBL magnification according to the known underlying FSBL model.
A systematic search reveals that this affects $\lesssim$10 points for 28 total detections across the grid points for the three lowest masses, $M_{p} = 0.1,~0.316,~{\rm and}~1 {\rm M}_{\Earth}$.
The sum of the detection rate for the affected events, from equation (\ref{eq:rate_tot_p}), is $\sim$0.016 for $M_{p} = 0.1{\rm M}_{\Earth}$, $\sim$0.20 for $M_{p} = 0.316{\rm M}_{\Earth}$, and $\sim$0.32 for $M_{p} = 1{\rm M}_{\Earth}$, which is only $\sim$2\%, $\sim$4\%, and $\sim$2\% of the total planet detection rate for the respective planet masses.
Furthermore, after recomputing the magnification and refitting the events we find that in all cases the fractional change in the value of {\dchisqslf} is $\lesssim$20\%, and in all but three cases it is $\lesssim$3\%, with none of the events having new values of {\dchisqslf} that are below our threshold of 160.

\subsection{Planet Detection Rates} \label{sec:rates_fid}

\begin{figure}
\centerline{
\includegraphics[width=9cm]{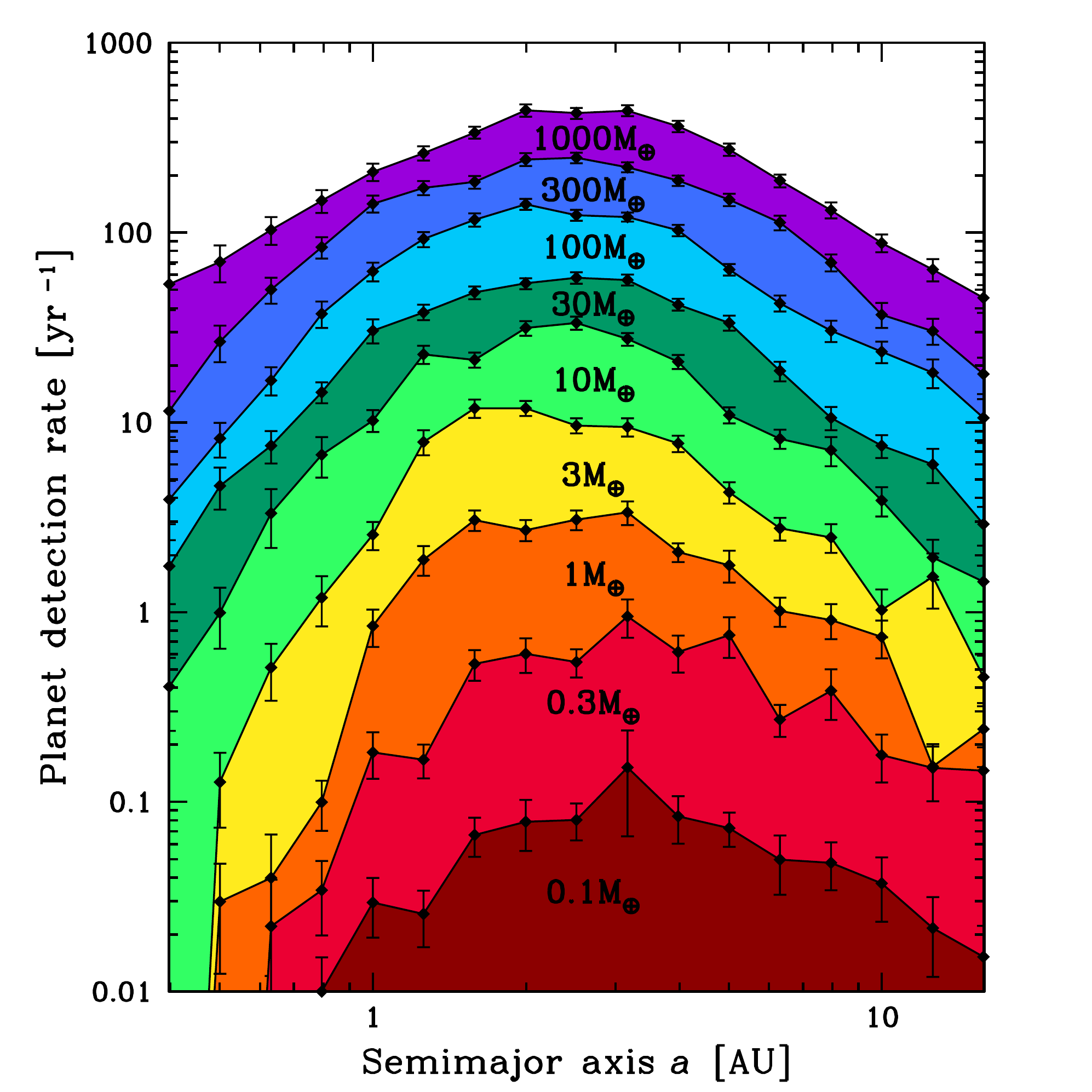}
}
\caption{
\footnotesize{
Planet detection rate as a function of semimajor axis $a$ for different planet masses {\mpl} for the full grid of our fiducial simulations.
Here we have assumed that each lens star hosts exactly one planet, of the specified mass, at the specified separation.
The detection rate increases with planet mass and peaks at a semimajor axis of $a \approx 2.5$AU for all planet masses, falling off at higher and lower values of $a$.
}
}
\label{fig:ndet}
\end{figure}

Figure \ref{fig:ndet} shows the annual planet detection rates as a function of semimajor axis for all planet masses.
Here we assume that each lens star hosts exactly one planet, of the specified mass, at the specified separation.
The detection rate for each [{\mpl}, $a$] grid point is calculated via equation (\ref{eq:rate_tot_p}) and is listed in Table \ref{tab:rates_fid}.
We compute the uncertainty in the detection rate as the Poisson fluctuation of each individual event, weighted by its rate.
Thus, the total detection rate for a given combination of {\mpl} and $a$ is given by equation (\ref{eq:rate_tot_p}) and its corresponding uncertainty by
\begin{equation} \label{eq:rate_err}
   \sigma_{\Gamma_{\rm tot}}=\sqrt{\sum_{i=1}^{N_{\rm det}}\Gamma_{i}^{2}}.
\end{equation}
The peak in sensitivity occurs at $a \approx 2.5$AU, with statistical fluctuations for lower planet masses due to low detection rates.
Under the assumption of one planet of the specified mass, at the specified separation, per star, approximately $76\%$ of the total planet detections would have ${\mpl} \gtrsim {\rm M}_{\rm Jup}$, while planets with ${\mpl} \lesssim {\rm M}_{\Earth}$ would constitute 4$\%$ of the detection rate.
\begin{deluxetable*}{cccccccccc}
\tablecaption{Fiducial Planet Detection Rates}
\tablewidth{0pt}
\tablehead{
\colhead{$a$/AU}   &
\colhead{\mplme}   &
\colhead{\mplme}   &
\colhead{\mplme}   &
\colhead{\mplme}   &
\colhead{\mplme}   &
\colhead{\mplme}   &
\colhead{\mplme}   &
\colhead{\mplme}   &
\colhead{\mplme}
}
\startdata
        &   0.100     &   0.316    &   1.00     &   3.16    &   10.0     &   31.6    &   100.   &   316    &   1000   \\
0.398   &   0.00      &   0.00     &   0.00     &   0.00    &   0.405    &   1.75    &   3.92   &   11.5   &   53.7   \\
0.501   &   0.00      &   0.00     &   0.0298   &   0.127   &   0.994    &   4.62    &   8.25   &   26.6   &   70.3   \\
0.631   &   0.00413   &   0.0221   &   0.0397   &   0.511   &   3.32     &   7.55    &   16.7   &   50.1   &   104    \\
0.794   &   0.0100    &   0.0343   &   0.0996   &   1.19    &   6.75     &   14.4    &   37.4   &   83.9   &   147    \\
1.00    &   0.0295    &   0.182    &   0.845    &   2.56    &   10.2     &   30.4    &   62.5   &   142    &   209    \\
1.26    &   0.0256    &   0.167    &   1.89     &   7.89    &   22.8     &   38.1    &   92.4   &   172    &   263    \\
1.58    &   0.0671    &   0.533    &   3.06     &   11.9    &   21.4     &   48.4    &   117    &   185    &   338    \\
2.00    &   0.0788    &   0.604    &   2.71     &   11.9    &   31.4     &   54.1    &   141    &   243    &   441    \\
2.51    &   0.0804    &   0.546    &   3.07     &   9.62    &   33.4     &   58.0    &   124    &   248    &   426    \\
3.16    &   0.152     &   0.950    &   3.36     &   9.47    &   27.5     &   56.4    &   121    &   221    &   440    \\
3.98    &   0.0838    &   0.616    &   2.07     &   7.75    &   20.9     &   41.8    &   103    &   188    &   364    \\
5.01    &   0.0728    &   0.757    &   1.77     &   4.29    &   10.9     &   33.5    &   64.0   &   149    &   274    \\
6.31    &   0.0495    &   0.272    &   1.01     &   2.76    &   8.19     &   18.7    &   42.5   &   113    &   188    \\
7.94    &   0.0478    &   0.385    &   0.911    &   2.48    &   7.13     &   10.6    &   30.4   &   69.7   &   132    \\
10.0    &   0.0373    &   0.176    &   0.738    &   1.03    &   3.87     &   7.52    &   23.6   &   37.0   &   88.3   \\
12.6    &   0.0217    &   0.151    &   0.154    &   1.54    &   1.94     &   6.02    &   18.3   &   30.3   &   64.1   \\
15.8    &   0.0153    &   0.146    &   0.242    &   0.455   &   1.44     &   2.91    &   10.5   &   18.0   &   45.4
\enddata
\tablecomments{Here we assume one planet per star at each grid point.}
\label{tab:rates_fid}
\end{deluxetable*}

\begin{figure}
\centerline{
\includegraphics[width=9cm]{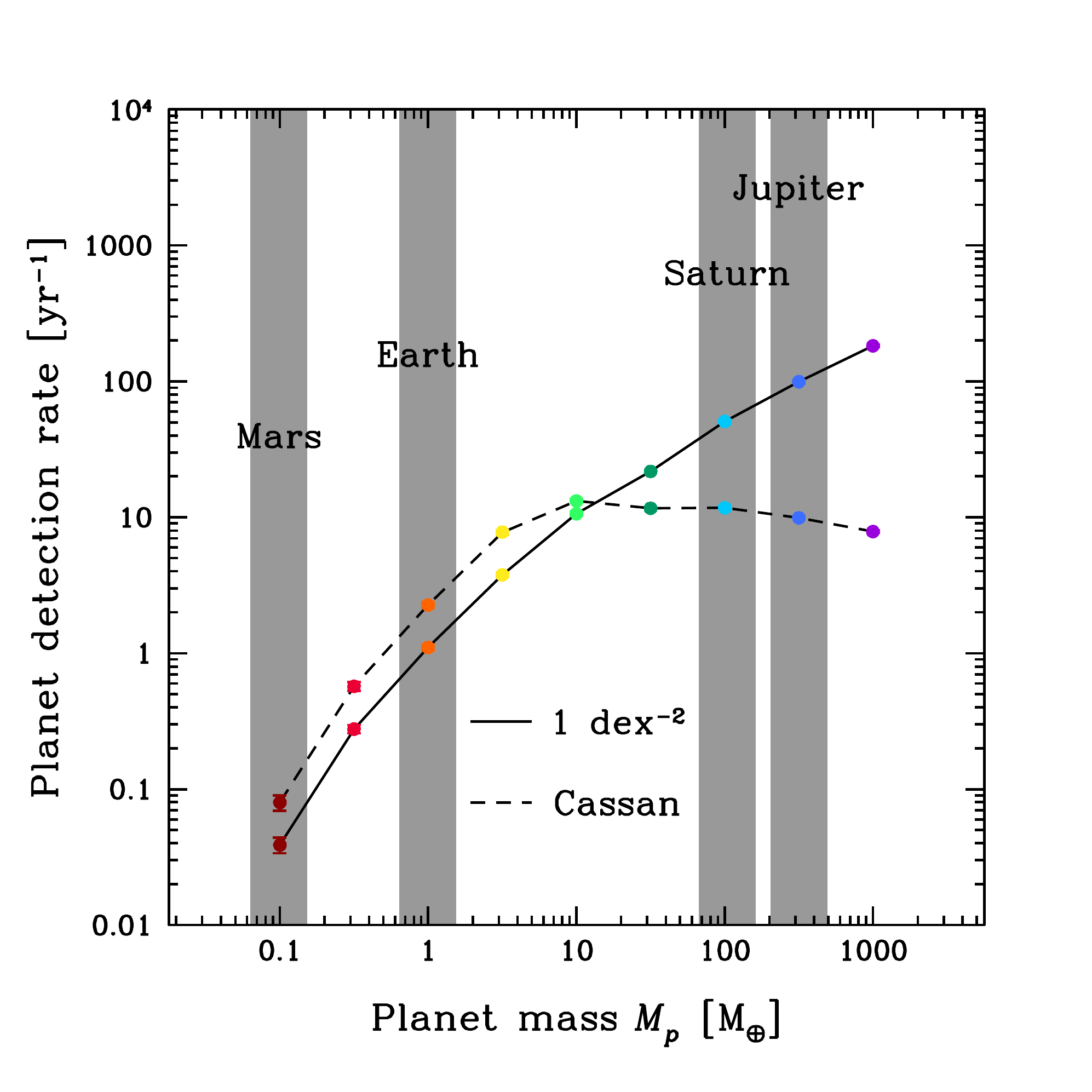}
}
\caption{
\footnotesize{
Planet detection rate as a function of planet mass {\mpl}, summed across semimajor axis $a$, for the full grid of our fiducial simulations.
The solid black line represents an assumed planet frequency of one planet per dex$^{2}$ per star.
The dashed black line represents our predictions for the detection rates after normalizing to the cool-planet mass function of \citet{cassan12}, where we have saturated it at ${\mpl} = 5{\rm M}_{\Earth}$ (corresponding to $\sim$2 planets per dex$^{2}$).
We see that the steep dependence that the detection rates have as a function of mass assuming one planet per dex$^{2}$ is nearly exactly canceled out by the increasing frequency of planets with decreasing planet mass according to the cool-planet mass function for ${\mpl} \gtrsim 5{\rm M}_{\Earth}$.
We find that KMTNet will be approximately uniformly sensitive to planets with mass in the range $5 \leq {\mplme} \leq 1000$ and will detect $\sim$20 planets per year per dex in mass across that range.
For lower-mass planets with mass in the range $0.1 \leq {\mplme} < 5$ we predict that KMTNet will detect $\sim$10 planets per year, the rate being dominated by planets with mass near the upper end of this range.
}
}
\label{fig:ndet_suma_pes}
\end{figure}

We then normalize our planet detection rates assuming an underlying distribution of one planet per dex$^{2}$ per star in log({\mpl}) and log($a$).
The detection rate as a function of log({\mpl}) is approximately a power-law with a slope of $\sim$0.7.
The slope becomes steeper for ${\mpl} \lesssim {\rm M}_{\Earth}$, likely due to finite-source effects.
We also rescale our planet detection rates according to \citet{cassan12}, who combine the three planet detections found in PLANET data from 2002--2007 with previous estimates of the slope of the mass-ratio function \citep{sumi10} and its normalization \citep{gould10} to derive a cool-planet mass function:
\begin{multline} \label{eq:cool_planet}
   f[\mathrm{log}~(a),~\mathrm{log}~({\mpl})] \equiv \frac{\mathrm{d}N_{\rm det}}{\mathrm{d log}~(a)~\mathrm{d log}~(M_{p})}
   \\
   = 10^{-0.62\pm0.22}\left(\frac{M_{p}}{95{\rm M}_{\Earth}}\right)^{-0.73\pm0.17}.
\end{multline}
We saturate the cool-planet mass function in equation (\ref{eq:cool_planet}) at $5{\rm M}_{\Earth}$, corresponding to $\sim$2 planets per dex$^{2}$, as a conservative approximation that is congruent with the fact that \citet{cassan12} have no measurements below that mass.
Figure \ref{fig:ndet_suma_pes} shows our planet detection rates as a function of {\mpl} assuming both an underlying planet population of one planet per dex$^{2}$ per star as well as our modified version of the cool-planet mass function.
Interestingly, the steep dependence that the detection rates have as a function of mass assuming one planet per dex$^{2}$ is nearly exactly canceled out by the increasing frequency of planets with decreasing planet mass that is found by \citet{cassan12}.
This causes the detection rates to be relatively independent of mass for ${\mpl} \gtrsim 5{\rm M}_{\Earth}$.
Adopting equation (\ref{eq:cool_planet}) and saturating it at {\mpl} = 5${\rm M}_{\Earth}$, we find that KMTNet will be approximately uniformly sensitive to planets with mass in the range $5 \leq {\mplme} \leq 1000$ and will detect $\sim$20 planets per year per dex in mass across that range.
For lower-mass planets with mass in the range $0.1 \leq {\mplme} < 5$ we predict that KMTNet will detect $\sim$10 planets per year, the rate being dominated by planets with mass near the upper end of this range.
The detection rates as a function of {\mpl}, assuming both one planet per dex$^{2}$ per star and also applying our modified version of the cool-planet mass function, are given in Table \ref{tab:rates_dex2cassan}.
\begin{deluxetable}{ccc}
\tablecaption{Normalized Planet Detection Rates}
\tablewidth{0pt}
\tablehead{
\colhead{\mplme}                 &
\colhead{$N_{\rm det}$ $^{a}$}   &
\colhead{$N_{\rm det}$ $^{b}$}
}
\startdata
0.100   &   0.038 $\pm$ 0.005  &   0.08 $\pm$ 0.01   \\
0.316   &   0.28 $\pm$ 0.02    &   0.57 $\pm$ 0.04   \\
1.00    &   1.10 $\pm$ 0.05    &   2.3 $\pm$ 0.1     \\
3.16    &   3.8 $\pm$ 0.1      &   7.8 $\pm$ 0.3     \\
10.0    &   10.6 $\pm$ 0.3     &   13.2 $\pm$ 0.4    \\
31.6    &   21.7 $\pm$ 0.6     &   11.6 $\pm$ 0.3    \\
100.    &   50.8 $\pm$ 1       &   11.7 $\pm$ 0.3    \\
316     &   99.5 $\pm$ 2       &   9.9 $\pm$ 0.2     \\
1000    &   182 $\pm$ 4        &   7.8 $\pm$ 0.2
\enddata
\tablenotetext{a}{Here we assume an underlying planet population of one planet per dex$^{2}$ per star and the rates are per year.}
\tablenotetext{b}{Here we apply the cool-planet mass function of \citet{cassan12}, where we have saturated it at ${\mpl} = 5{\rm M}_{\Earth}$ (corresponding to $\sim$2 planets per dex$^{2}$), and the rates are per year.}
\label{tab:rates_dex2cassan}
\end{deluxetable}

\subsection{Parameter Distributions} \label{sec:parm_dist}

\begin{figure}
\centerline{
\includegraphics[width=9cm]{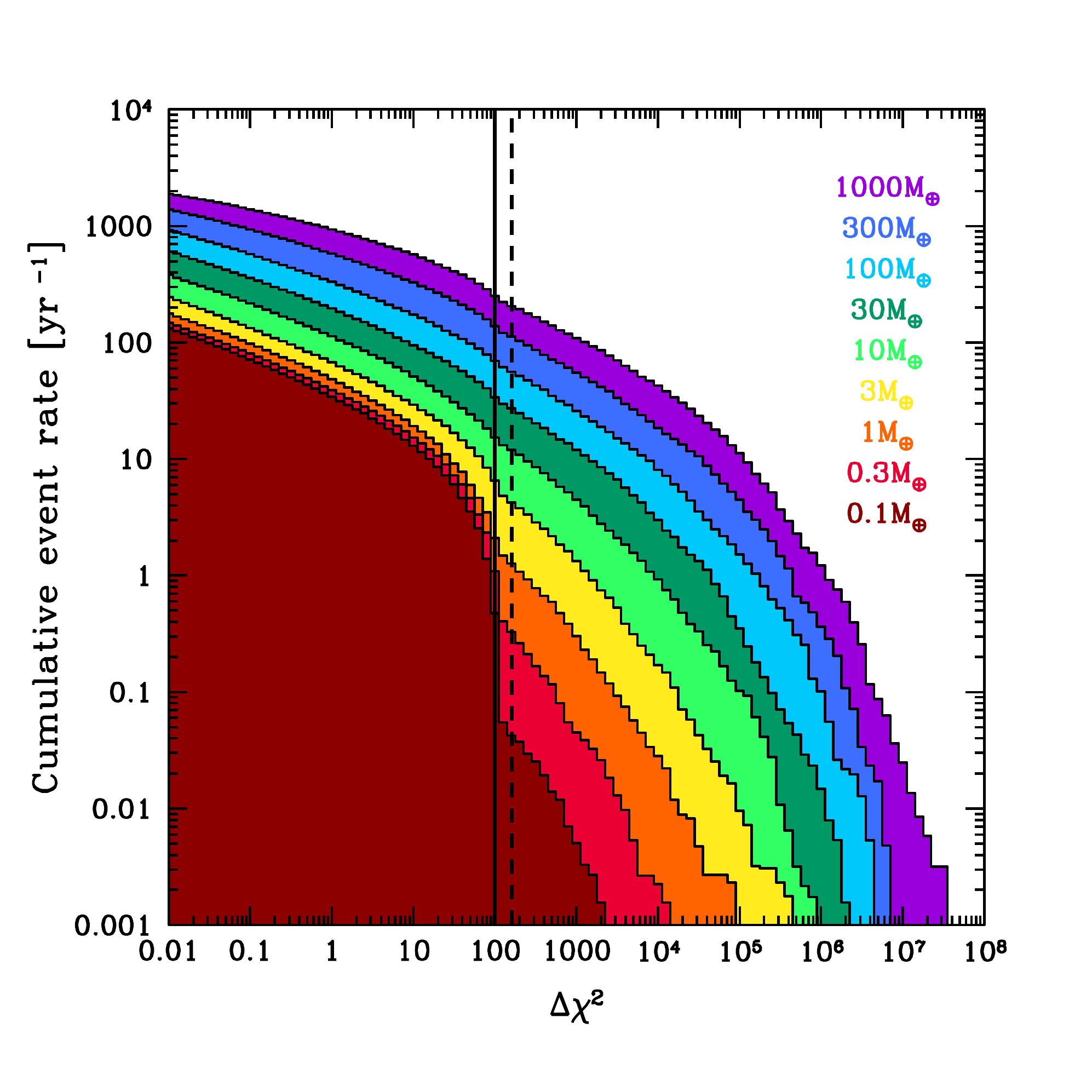}
}
\caption{
\footnotesize{
Microlensing event rate as a function of $\Delta\chi^{2}$ for different planet masses for our fiducial simulations, assuming one planet per star per dex$^{2}$.
The solid black line shows our initial single-lens fitting threshold ${\dchisqslith} = 100$, above which the rates drops off sharply.
The dashed black line denotes our final best-fit detection threshold ${\dchisqslfth} = 160$.
}
}
\label{fig:chi2_apars}
\end{figure}

Figure \ref{fig:chi2_apars} shows the microlensing event rate as a function of {\dchisq} for all planet masses.
For all planet masses there is a pileup in the rates just below {\dchisqslfth}, as expected.
Above {\dchisqslfth} the distribution follows a power-law across several dex of {\dchisq} before falling off steeply due to the prominence of finite-source effects.

\begin{figure}
\centerline{
\includegraphics[width=9cm]{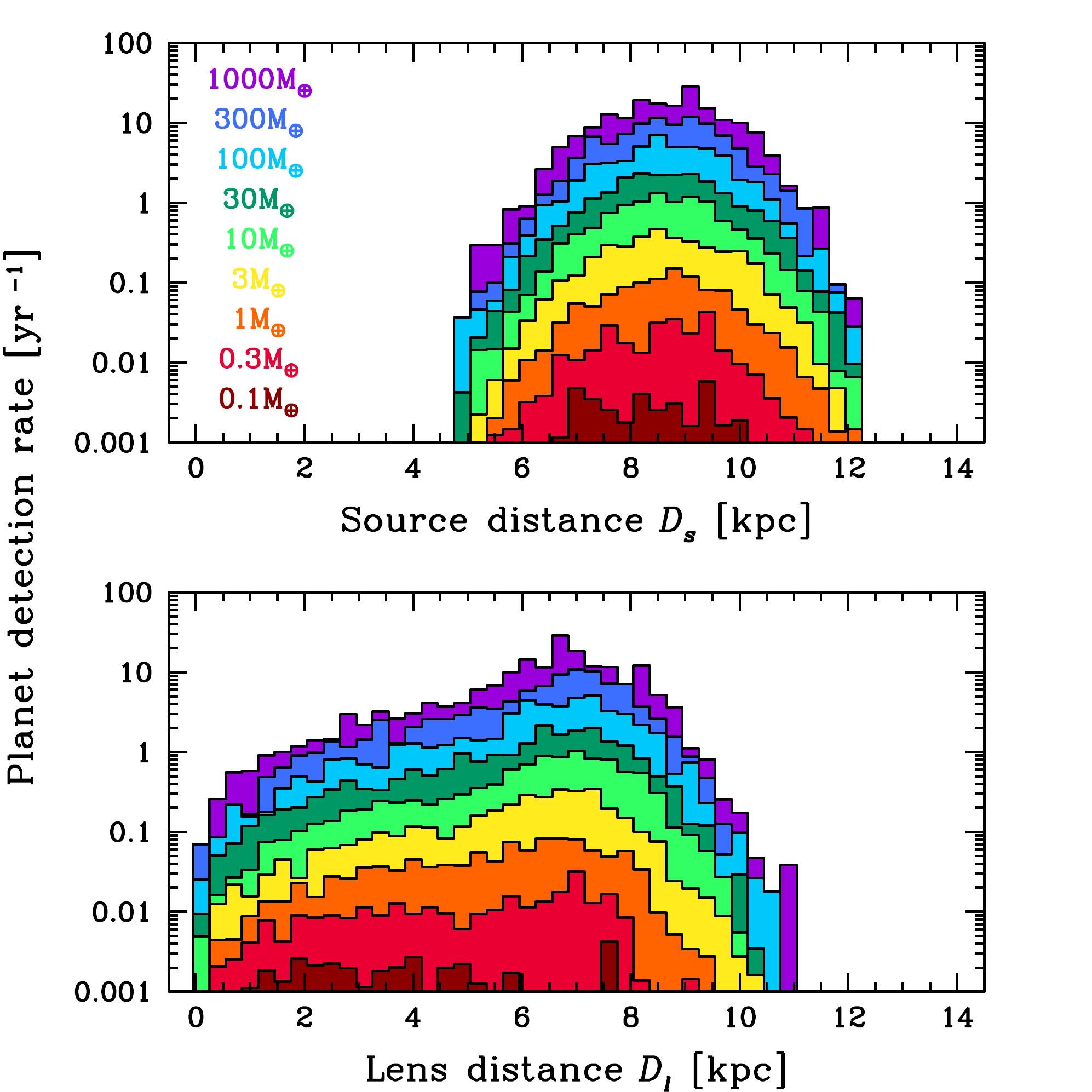}
}
\caption{
\footnotesize{
Planet detection rate as a function of source distance $D_{s}$ and lens distance $D_{l}$ for different planet masses for the full grid of our fiducial simulations, assuming one planet per dex$^{2}$ per star.
The mean value of {\ds} for all planet detections is 8.7kpc, somewhat larger than our adopted value of the Galactocentric distance, $D_{\rm GC} = 8.2$kpc.
We also find that the distribution of {\dl} varies as a function of planet mass.
Lower-mass planets are found around preferentially closer lenses, with the average lens distance decreasing from ${\dl} \approx 6.3$kpc for ${\mpl} = 1000{\rm M}_{\Earth}$ to ${\dl} \approx  4.0$kpc for ${\mpl} = 0.1{\rm M}_{\Earth}$.
For a fixed source distance {\ds} and lens mass {\ml}, smaller values of {\dl} increase the size of the Einstein ring {\thetae}, from equation (\ref{eq:thetae}).
The larger the size of {\thetae}, the less important finite-source effects will be, accentuating the deviations due to the presence of the planetary companion that are typically more subtle for lower-mass planets (i.e., lens systems will smaller mass ratios, $q$).
Consequently, as planet mass decreases, or more appropriately, as $q$ decreases, detections will occur for preferentially closer lens systems.
The percentage of Bulge versus Disk lenses correspondingly also depends on planet mass.
The fractions of Bulge and Disk lenses are 52$\%$ and 48$\%$, respectively, for ${\mpl} = 1000{\rm M}_{\Earth}$.
But as {\mpl} decreases, the lens population becomes increasingly dominated by Disk lenses, with the percentage reaching 68$\%$ for ${\mpl} = 0.1{\rm M}_{\Earth}$.
}
}
\label{fig:dist_ndet}
\end{figure}

Figure \ref{fig:dist_ndet} shows the planet detection rate as a function of {\ds} and {\dl} for all planet masses.
The mean value of {\ds} for all planet detections is 8.7kpc, somewhat larger than our adopted value of the Galactocentric distance, $D_{\rm GC} = 8.2$kpc.
We also find that the distribution of {\dl} varies as a function of planet mass.
Lower-mass planets are found around preferentially closer lenses, with the average lens distance decreasing from ${\dl} \approx 6.3$kpc for ${\mpl} = 1000{\rm M}_{\Earth}$ to ${\dl} \approx  4.0$kpc for ${\mpl} = 0.1{\rm M}_{\Earth}$.
For a fixed source distance {\ds} and lens mass {\ml}, smaller values of {\dl} increase the size of the Einstein ring {\thetae}, from equation (\ref{eq:thetae}).
The larger the size of {\thetae}, the less important finite-source effects will be, accentuating the deviations due to the presence of the planetary companion that are typically more subtle for lower-mass planets (i.e., lens systems will smaller mass ratios, $q$).
Consequently, as planet mass decreases, or more appropriately, as $q$ decreases, detections will occur for closer lens systems.
The percentage of Bulge versus Disk lenses correspondingly also depends on planet mass.
The fractions of Bulge and Disk lenses are 52$\%$ and 48$\%$, respectively, for ${\mpl} = 1000{\rm M}_{\Earth}$.
But as {\mpl} decreases, the lens population becomes increasingly dominated by Disk lenses, with the percentage reaching 68$\%$ for ${\mpl} = 0.1{\rm M}_{\Earth}$.

\begin{figure}
\centerline{
\includegraphics[width=9cm]{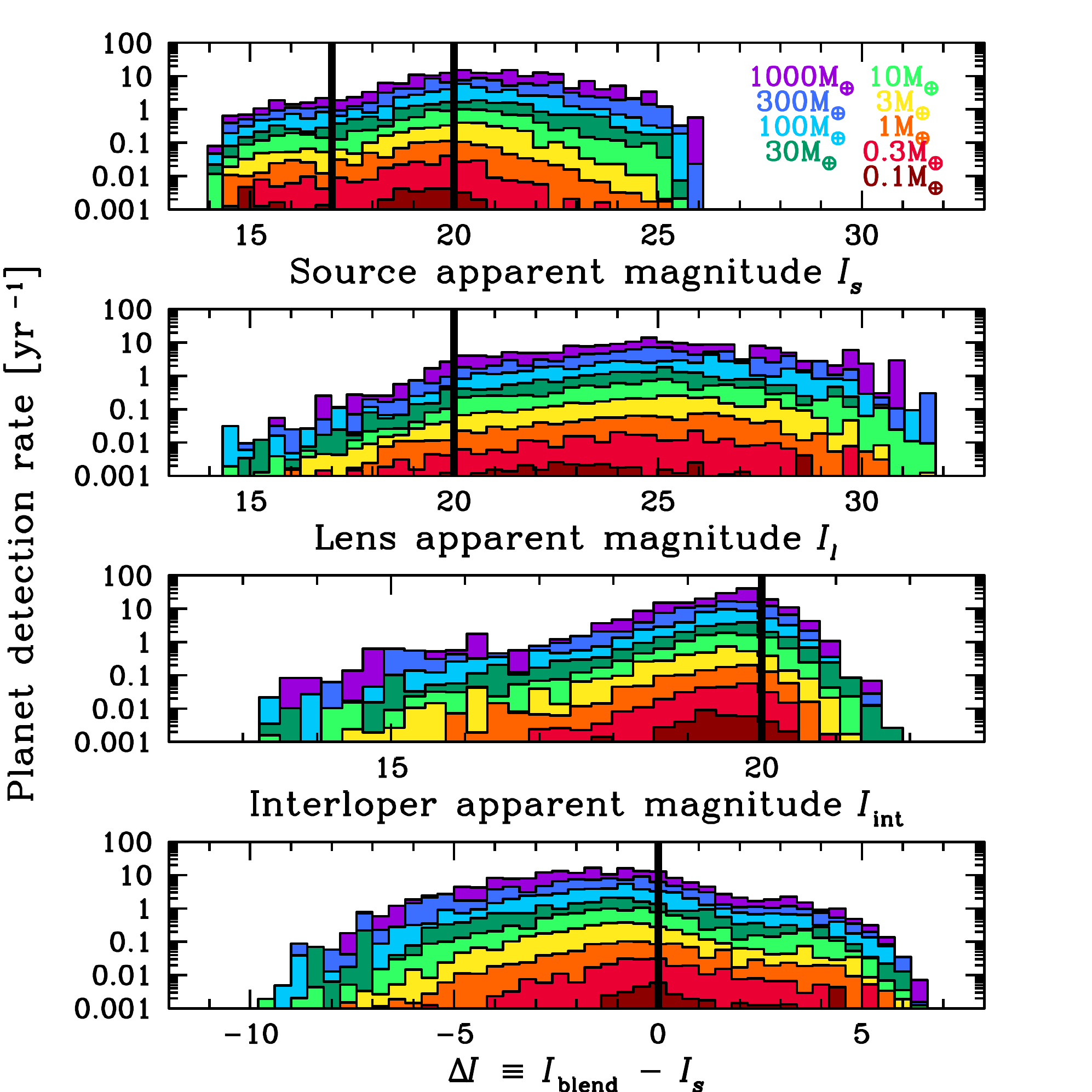}
}
\caption{
\footnotesize{
Planet detection rate as a function of apparent source magnitude $I_{s}$, apparent lens magnitude $I_{l}$, and their difference $\Delta I \equiv I_{l} - I_{s}$ for different planet masses for the full grid of our fiducial simulations, assuming one planet per dex$^{2}$ per star.
The morphology of the source brightness distribution roughly follows that of the LF, shown in Figure \ref{fig:lf}.
Only 6$\%$ of the detection rate arises from sources with $I_{s} \leq 17$, which we take as a crude cut-off for giant stars.
The majority of the detections will come from fainter sources, with 64$\%$ having $I_{s} \geq 20$.
The bulk of the detection rate for lower-mass planets comes from brighter sources, for which the photometric precision is higher, on average, making it easier to robustly detect lower amplitude perturbations.
However, the vast majority of detections will suffer from severe blending, with 76$\%$ of detections coming from events with a source that is fainter than the combination of the lens plus the interloping blend star.
Furthermore, a substantial fraction of the lenses will be faint, with 3$\%$ of detections havine $I_{l} \geq 20$, making it more difficult to follow-up the lens systems and obtain direct flux measurements.
}
}
\label{fig:imag_diff_ndet}
\end{figure}

Figure \ref{fig:imag_diff_ndet} shows the planet detection rate as a function of $I_{l}$, $I_{s}$, and $\Delta I \equiv I_{l} - I_{s}$ for all planet masses.
The morphology of the source brightness distribution roughly follows that of the LF, shown in Figure \ref{fig:lf}.
Only 6$\%$ of the detection rate arises from sources with $I_{s} < 17$, which we take as a crude cut-off for giant stars.
The majority of the detections will come from fainter sources, with 64$\%$ having $I_{s} > 20$.
The bulk of the detection rate for lower-mass planets comes from brighter sources, for which the photometric precision is higher, on average, making it easier to robustly detect lower amplitude perturbations.
However, the vast majority of detections will suffer from severe blending, with 76$\%$ of detections coming from events with a source that is fainter than the combination of the lens plus the interloping blend star.
Furthermore, a substantial fraction of the lenses will be faint, with 3$\%$ of detections havine $I_{l} \geq 20$, making it more difficult to follow-up the lens systems and obtain direct flux measurements.

\begin{figure}
\centerline{\includegraphics[width=9cm]{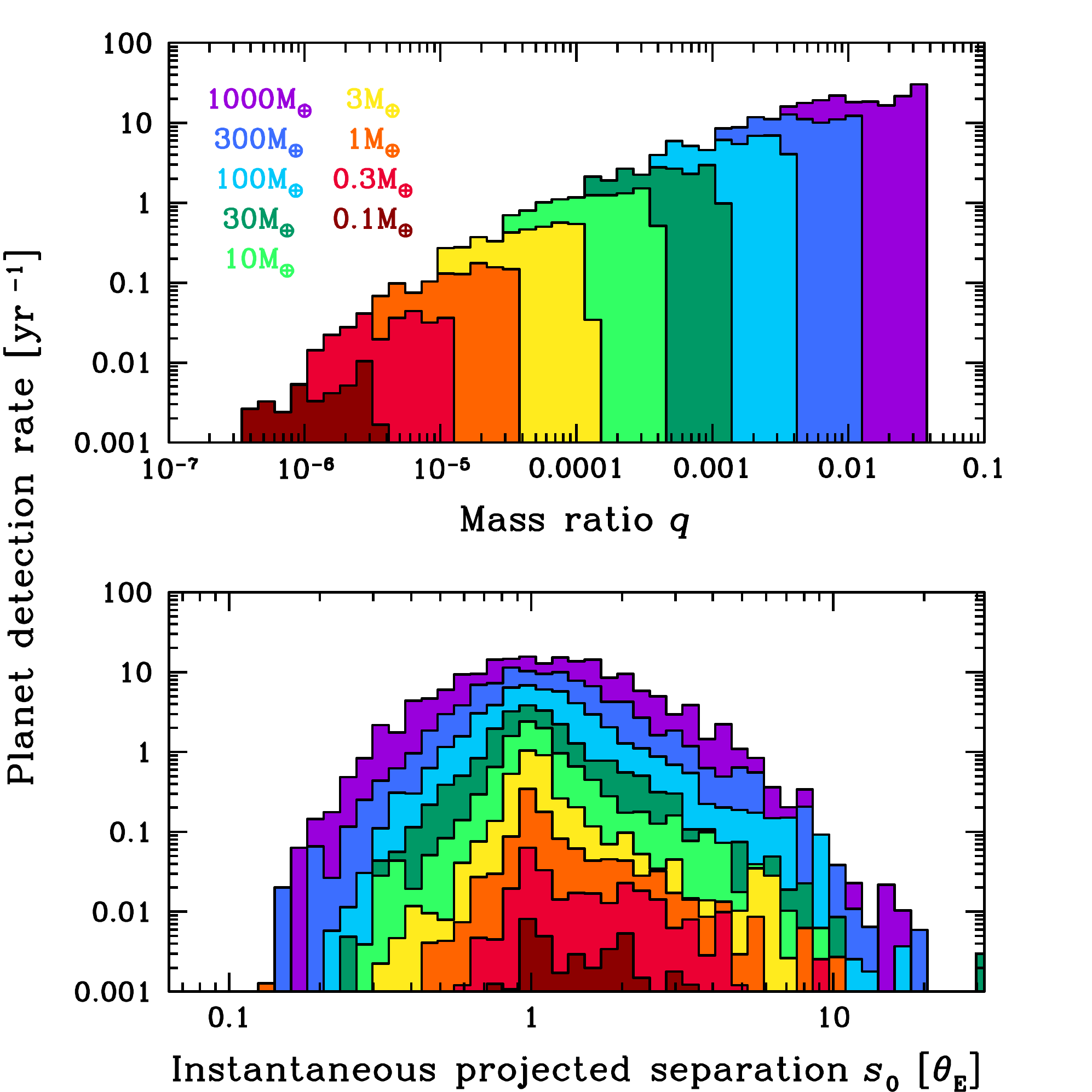}}
\caption{
\footnotesize{
Planet detection rate as a function of mass ratio $q$ and instantaneous projected separation $s_{\rm 0}$ for different planet masses for the full grid of our fiducial simulations, assuming one planet per dex$^{2}$ per star.
The distribution of {\snot} roughly follows a broken power-law with a change in slope where the instantaneous projected separation is approximately unity.
These scalings, $s_{0}^{3}$ for ${\snot} < 1$ and $s_{0}^{-2}$ for ${\snot} > 1$, arise from the dependence of the size of the planetary caustics on {\snot}, as shown in \citet{han06}.
This behavior changes for low-mass planets with mass ${\mpl} \lesssim 1{\rm M}_{\Earth}$.
While finite-source effects strongly suppress the detection rates for ${\snot} < 1$, they act to potentially enhance the perturbation for ${\snot} > 1$, enhancing the resulting detection rate (\citealt{bennett96,gould97}).
}
}
\label{fig:q_thetae_ndet}
\end{figure}

Figure \ref{fig:q_thetae_ndet} shows the planet detection rate as a function of $q$ and {\snot} for all planet masses.
For a given planet mass, the spread in $q$ is slightly larger than one dex, corresponding to the spread in the primary masses we consider.
We find that KMTNet will have moderate sensitivity down to mass ratios as low as $\sim$10$^{-6}$.
For planets with mass ${\mpl} \gtrsim 1{\rm M}_{\Earth}$, the distribution of {\snot} roughly follows a broken power-law with a change in slope where the instantaneous projected separation is approximately unity.
For ${\snot} < 1$ the detection rate is a relatively steep function of ${\snot}$, scaling roughly as $s_{0}^{3}$.
On the other hand for ${\snot} > 1$, the detection rate scaling is less steep, going roughly as $s_{0}^{-2}$.
The asymmetry in this distribution arises from the scaling of the size of the planetary caustic(s) with {\snot}.
As shown in \citet{han06}, for ${\snot} \ll 1$ each of the two planetary caustics expand as $s_{0}^{3}$ and for ${\snot} \gg 1$ the single planetary caustic shrinks as $s_{0}^{-2}$.

For low-mass planet with mass ${\mpl} \lesssim 1{\rm M}_{\Earth}$, the behavior with $\snot$ is qualitatively different from that for more massive planets.
In particular, for ${\snot} < 1$, the detection rate for low-mass planets is strongly suppressed.
This is a consequence of finite-source effects.
As shown in \citet{gould97}, when the source is larger than and fully encloses the two triangular-shaped planetary caustics for ${\snot}<1$, the fractional difference in the magnification from the single lens is zero, up to fourth order in the source size in units of the planetary Einstein ring radius, $\rho_p \equiv \rho q^{-1/2}$.
On the other hand, when the source size is larger than and fully encloses the single, diamond-shaped planetary caustic that exists for ${\snot} > 1$, the fractional difference in the magnification is $2\rho_p^{-2}$.
Therefore, in the presence of strong finite source effects, planetary perturbations for ${\snot}<1$ are much more strongly suppressed than for ${\snot}>1$ \citep{bennett96}.
Furthermore, for sufficient photometric precision and when $\rho_p \gg 1$, the detection rate for ${\snot}>1$ perturbations may even be enhanced by finite-source effects, because the cross section for detection is $\propto \rho$ rather than $\propto q^{1/2}$.
This is likely the cause of the more gradual fall-off of the detection rate for low-mass planets for ${\snot}>1$.

\begin{figure}
\centerline{
\includegraphics[width=9cm]{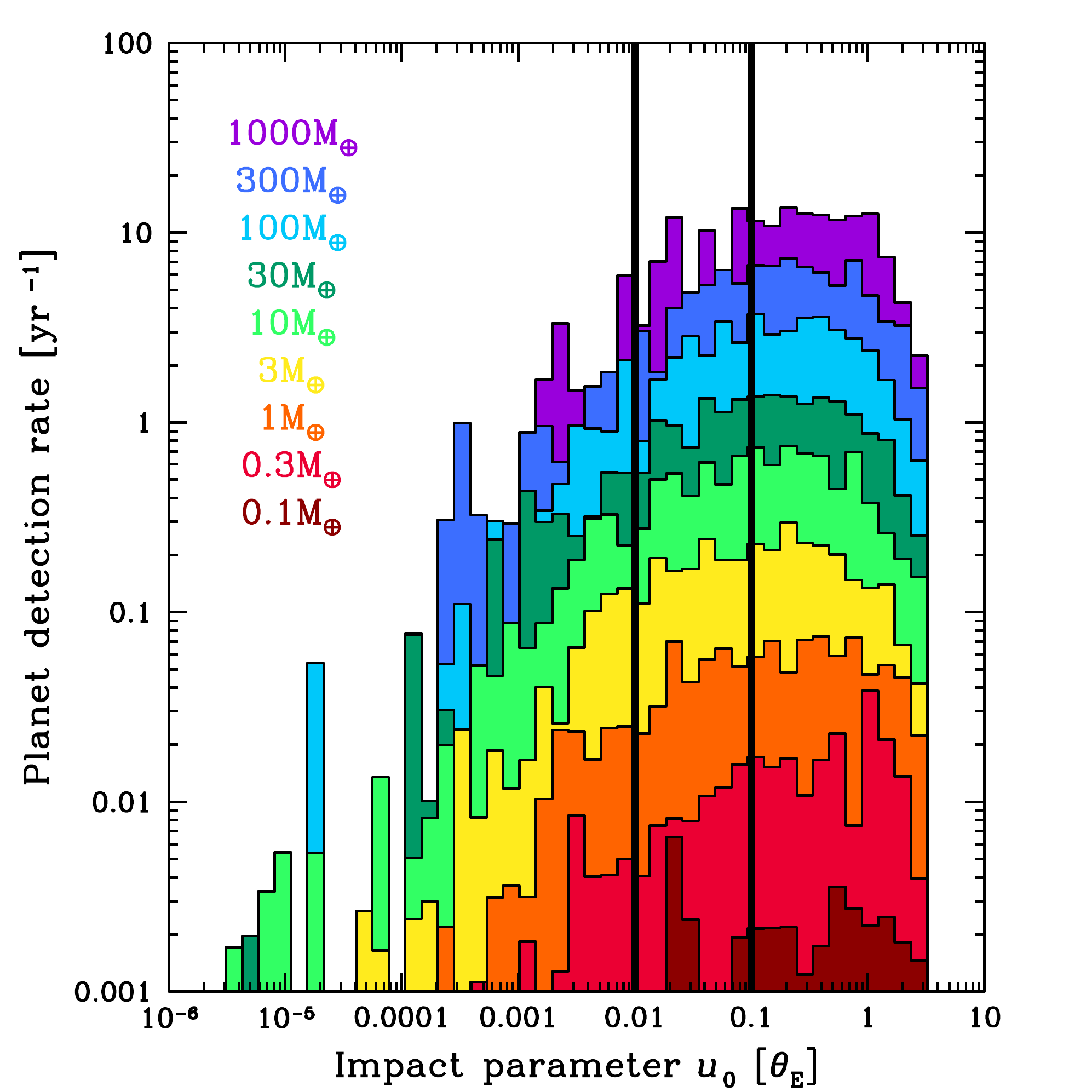}
}
\caption{
\footnotesize{
Planet detection rate as a function of impact parameter $u_{0}$ for different planet masses for the full grid of our fiducial simulations, assuming one planet per dex$^{2}$ per star.
The left-most solid black line marks a maximum magnification of $A_{max}\approx100$ while the right-most solid black line denotes a maximum magnification of $A_{max}\approx10$.
We find that 79$\%$ of events will have $A_{\rm max} \lesssim 10$ and 95$\%$ of events will have $A_{\rm max} \lesssim 100$.
Thus, the majority of detections arise from events with relatively low peak magnification.
}
}
\label{fig:u0_ndet}
\end{figure}

Figure \ref{fig:u0_ndet} shows the planet detection rate as a function of {\unot} for all planet masses.
Using {\unot} as a proxy for peak magnification and taking $A_{\rm max} \approx u_{\rm 0}^{-1}$, we find that 79$\%$ of events will have $A_{\rm max} \lesssim 10$ and 95$\%$ of events will have $A_{\rm max} \lesssim 100$.
Thus, the majority of detections arise from events with relatively low peak magnification, which has significant implications for the automated fitting routines that will be necessary to model such large numbers of events expediently.

\subsection{Free-floating Planets} \label{sec:ffps}

\begin{figure}
\centerline{
\includegraphics[width=9cm]{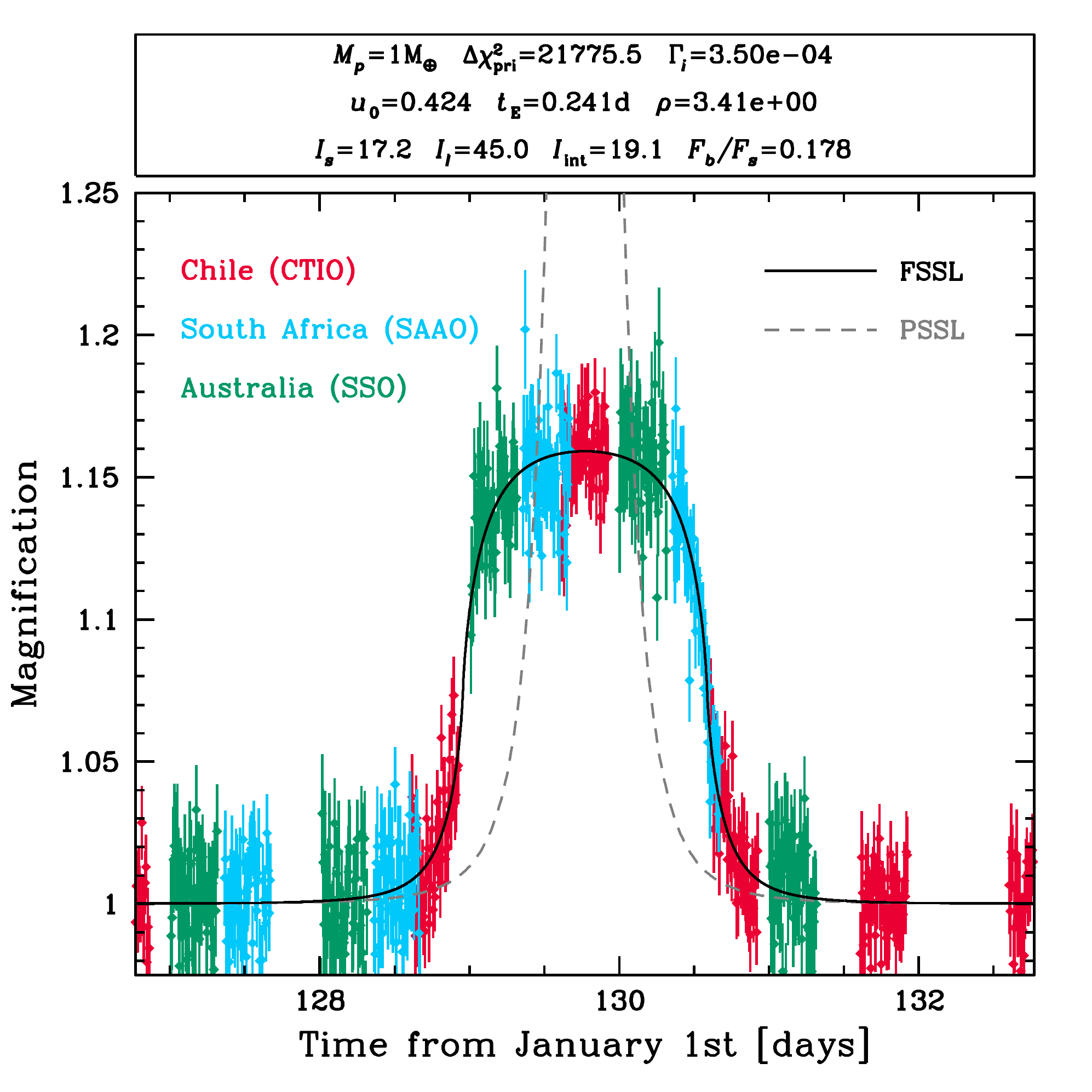}
}
\caption{
\footnotesize{
An example light curve for an Earth-mass FFP.
The top panel shows the physical, lensing, and observational parameters for this event.
We have included Gaussian scatter in the photometry solely for the purposes of visualizing the quality of data that KMTNet will actually obtain.
This demonstrates the ability of KMTNet to find FFPs down to the mass of Earth even for events with low peak magnifications and short time scales.
}
}
\label{fig:lc_ffp_earth}
\end{figure}

\begin{figure}
\centerline{
\includegraphics[width=9cm]{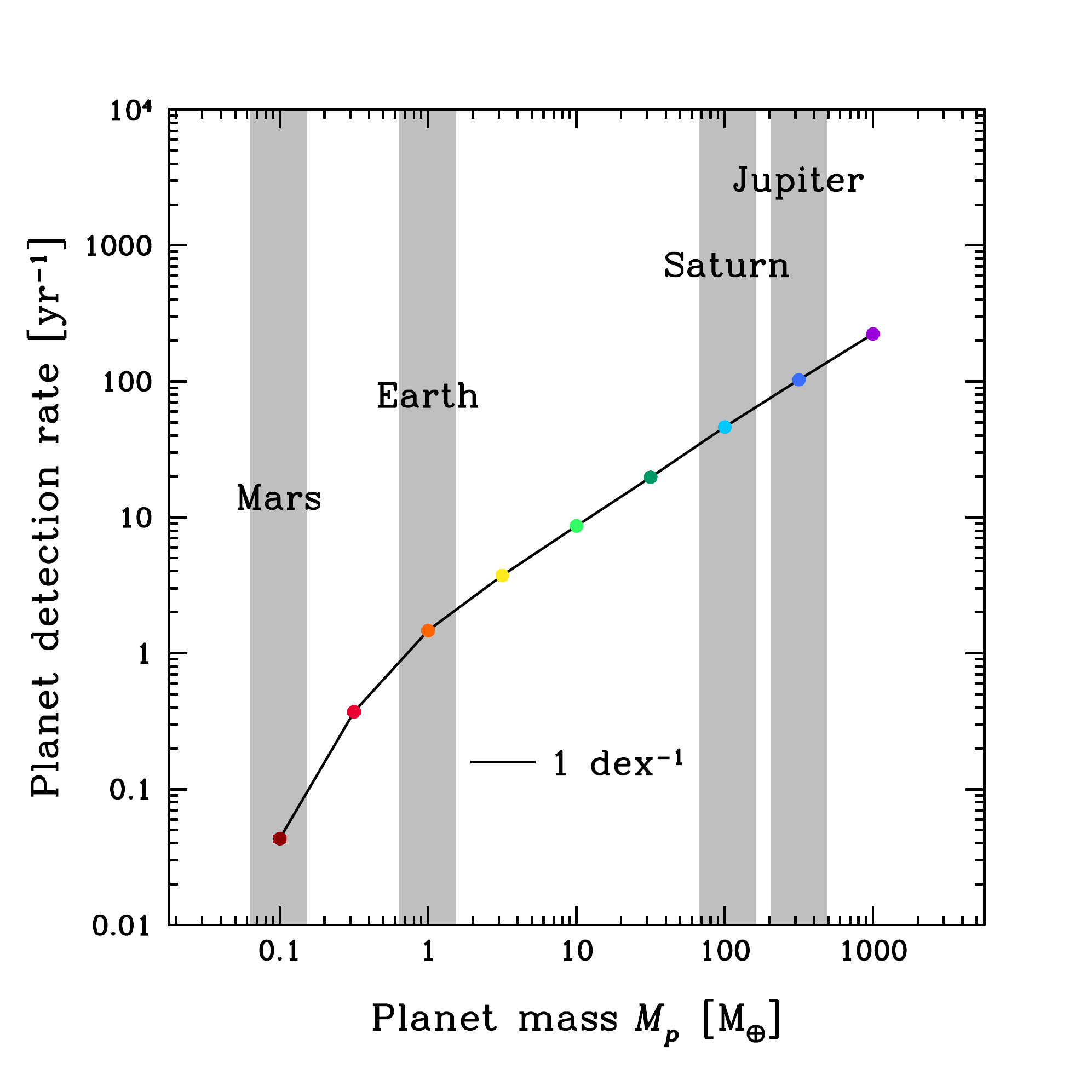}
}
\caption{
\footnotesize{
Planet detection rate for FFPs as a function of planet mass {\mpl}.
Here we have assumed an underlying FFP frequency of one such planet per star in the Galaxy.
The distribution roughly follows the same power-law as for bound planets, with a steeper slope for ${\mpl} \lesssim 1{\rm M}_{\Earth}$ due to the prominence of finite-source effects.
}
}
\label{fig:ndet_ffp}
\end{figure}

We also explore the detection rates KMTNet will obtain for FFPs.
We run a set of simulations across the same mass range as for bound planets, $0.1 \leq {\mplme} \leq 1000$, with the same spacing.
For these simulations, we assume the same velocity distributions as in \S \ref{sec:bulgedisk}, applying the lens velocities to the FFPs.
Our sole detection criterion is that the {\dchisq} of each FFP event from its flux-weighted mean brightness be greater than 500:
\begin{equation}
   \Delta\chi_{\rm FFP}^{2}\geq500.
\end{equation}
We choose a higher detection threshold for FFPs than for bound planets because a much larger sample of light curves must be searched over a longer time baseline for perturbations from FFPs than bound planets.
An example light curve for an Earth-mass FFP is shown in Figure \ref{fig:lc_ffp_earth}.
As with the light curves for bound planets, the scatter in the photometry is Gaussian and is solely for the purposes of visualizing the quality of data that KMTNet will actually obtain.
The high cadence of KMTNet produces densely sampled light curves even for time scales as short as ${\te} \approx 0.1$d.
Figure \ref{fig:ndet_ffp} shows the planet detection rates for FFPs as a function of {\mpl} for all planet masses, assuming an underlying planet frequency of one such planet per star in the Galaxy.
We find that the detection rate as a function of planet mass to have a slope and normalization similar to that for bound planets.
The detection rates are given in Table \ref{tab:rates_ffp}, and we predict that, assuming one FFP per star in the Galaxy, KMTNet will detect $\sim$100 with the mass of Jupiter per year and $\sim$1 with the mass of Earth.
\begin{deluxetable}{cc}
\tablecaption{Detection Rates for FFPs}
\tablewidth{0pt}
\tablehead{
\colhead{\mplme}   &
\colhead{$N_{\rm det}$}
}
\startdata
0.100   &   0.043 $\pm$ 0.002   \\
0.316   &   0.371 $\pm$ 0.009   \\
1.00    &   1.47 $\pm$ 0.02     \\
3.16    &   3.73 $\pm$ 0.05     \\
10.0    &   8.6 $\pm$ 0.1       \\
31.6    &   19.7 $\pm$ 0.2      \\
100.    &   46.2 $\pm$ 0.5      \\
316     &   103 $\pm$ 1         \\
1000    &   223 $\pm$ 3
\enddata
\tablecomments{Here we assume an underlying planet population of one planet per star in the Galaxy and the rates are per year.}
\label{tab:rates_ffp}
\end{deluxetable}

\section{Extrinsic Parameter Variation} \label{sec:extparm_vary}

While we have converged on an optimal set of observational parameters, discussed in \S \ref{sec:obsparm_opt}, there are other factors that will also impact KMTNet's detection rates.
Here we examine the dependence of the detection rates on the number of observatories, to understand the gain in detection rates as the three KMTNet telescopes successively come online, and the systematic error floor, ultimately set by the quality of the data and the photometry pipeline that KMTNet will employ.
Finally, we run a full grid of simulations with a more optimistic set of assumptions regarding seeing distributions, the systematic error floor, the photon rate normalization, and the photometric precision achieved.
We believe that this last set of simulations in conjunction with our fiducial results will bracket the detection rates that KMTNet will obtain.

\subsection{Observatory Chronology} \label{sec:ndet_obs}

\begin{figure}
\centerline{
\includegraphics[width=9cm]{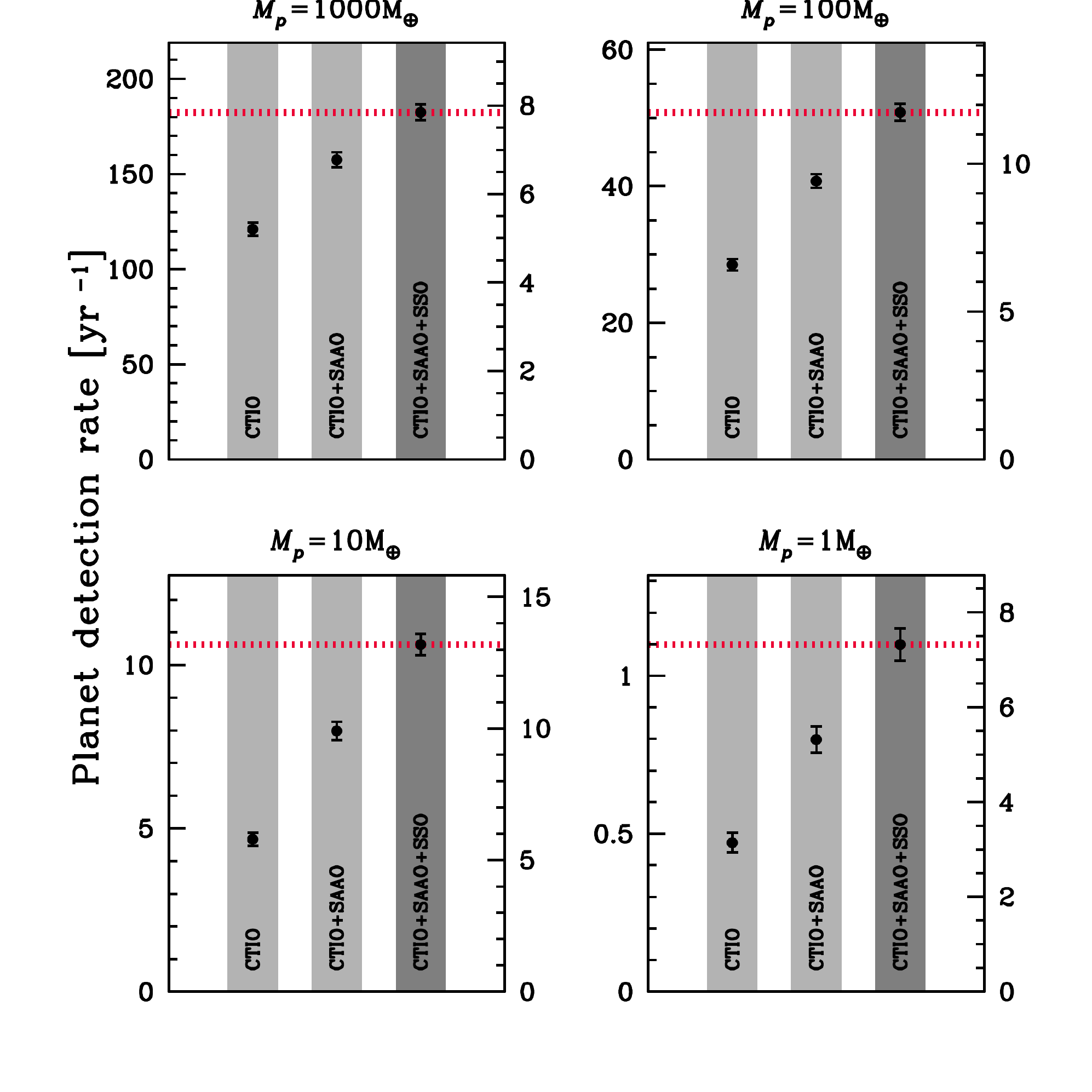}
}
\caption{
\footnotesize{
Planet detection rate as a function of the number of observatories for four different planet masses.
The rates on the left-hand side of each plot represent an assumed planet frequency of one planet per dex$^{2}$ per star.
The rates on the right-hand side of each plot assume the cool-planet mass function of \citet{cassan12}, where we have saturated it at ${\mpl} = 5{\rm M}_{\Earth}$ (corresponding to $\sim$2 planets per dex$^{2}$).
The red dashed line denotes the detection rates for our fiducial simulations, which includes all three observatories.
The fraction of planets that are detected by the first observatory and by the first and second observatories together decreases with the planet mass.
For ${\mpl} = 1000{\rm M}_{\Earth}$, 66$\%$ of the planet detections come from CTIO alone, and 86$\%$ come from CTIO in conjunction with SAAO.
For ${\mpl} = 1{\rm M}_{\Earth}$, only 43$\%$ of the detections come from CTIO alone, and the fraction of detections that arise from CTIO and SAAO working in concert has dropped to 73$\%$.
For high-mass planets with ${\mpl} \gtrsim 100{\rm M}_{\Earth}$, the planetary perturbation typically lasts substantially longer than a day (see, e.g., Figure \ref{fig:lc_jupiter}), and thus can be detected from a single site, given sufficient photometric precision and good weather.
In this regime, particularly given the cadence and photometric precision of KMTNet, adding additional observatories will only marginally increase the total number of detected planets.
On the other hand, planets with mass ${\mpl} \lesssim 1{\rm M}_{\Earth}$ create perturbations that typically last a day or less (see Figures \ref{fig:lc_earth} and \ref{fig:lc_mars}), and thus for such low-mass planets it is more probable for perturbations to be detected and observed by a single observatory, approaching the limit that each observatory contributes and equal fraction of the overall planet detection rate.
}
}
\label{fig:ndet_obs}
\end{figure}

Figure \ref{fig:ndet_obs} shows the planet detection rates as a function of the number of observatories, following the order in which the KMTNet telescopes will come online.
The rates have been summed across all semimajor axes.
The rate of increase of the number of detections shows a dependence on planet mass.
The fraction of planets that are detected by the first observatory and by the first and second observatories together decreases with the planet mass.
For ${\mpl} = 1000{\rm M}_{\Earth}$, 66$\%$ of the planet detections come from CTIO alone, and 86$\%$ come from CTIO in conjunction with SAAO.
For ${\mpl} = 1{\rm M}_{\Earth}$, only 43$\%$ of the detections come from CTIO alone, and the fraction of detections that arise from CTIO and SAAO working in concert has dropped to 73$\%$.
For high-mass planets with ${\mpl} \gtrsim 100{\rm M}_{\Earth}$, the planetary perturbation typically lasts substantially longer than a day (see, e.g., Figure \ref{fig:lc_jupiter}), and thus can be detected from a single site, given sufficient photometric precision and good weather.
In this regime, particularly given the cadence and photometric precision of KMTNet, adding additional observatories will only marginally increase the total number of detected planets.
On the other hand, planets with mass ${\mpl} \lesssim 1{\rm M}_{\Earth}$ create perturbations that typically last a day or less (see Figures \ref{fig:lc_earth} and \ref{fig:lc_mars}), and thus for such low-mass planets it is more probable for perturbations to be detected and observed by a single observatory, approaching the limit that each observatory contributes and equal fraction of the overall planet detection rate.
We conclude that, since the detection rates for the three observatories become more nearly independent as the planet mass decreases, having the full network is critical for maximizing the low-mass planet yield.

\subsection{Systematic Error Floor} \label{sec:ndet_sys}

\begin{figure}
\centerline{
\includegraphics[width=9cm]{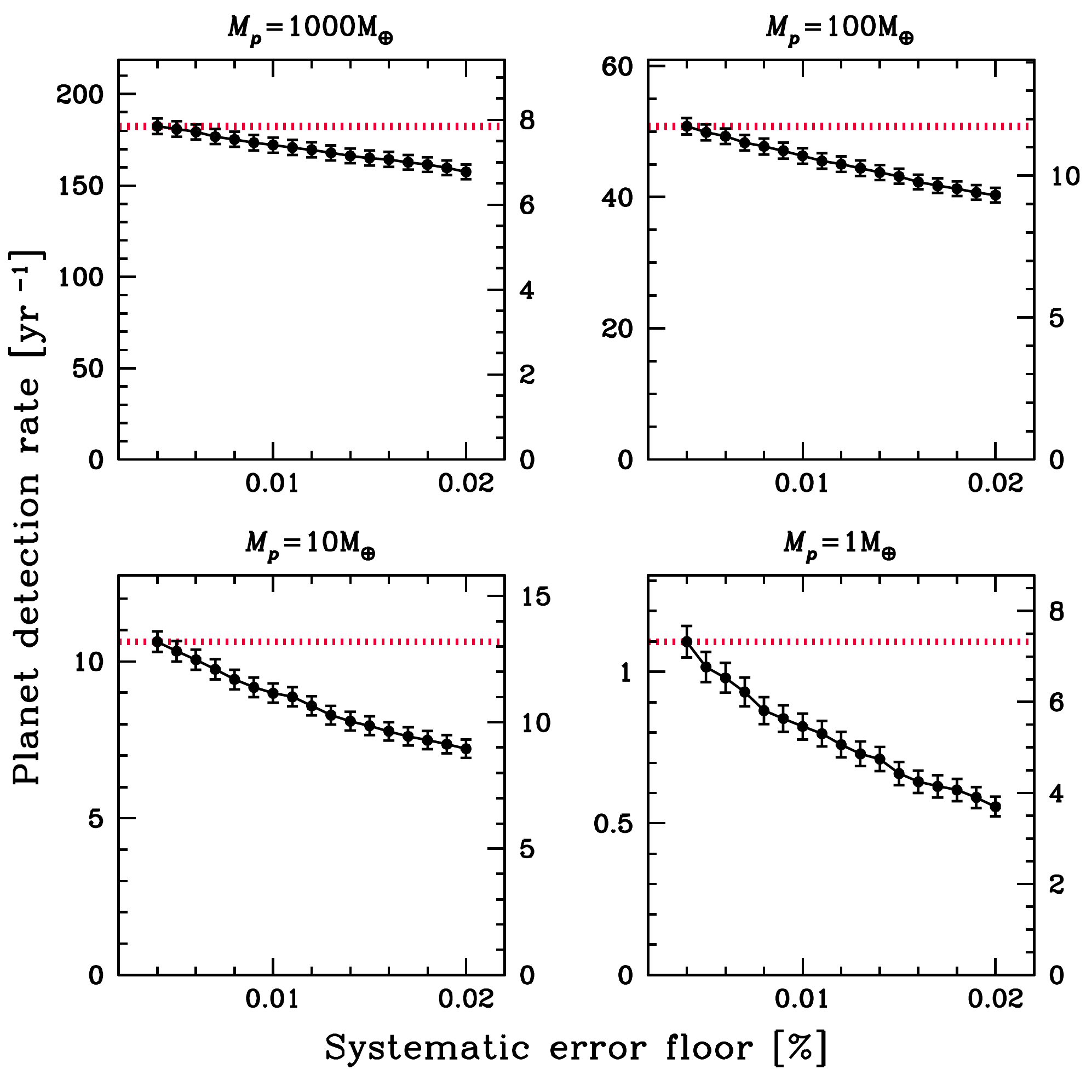}
}
\caption{
\footnotesize{
Planet detection rate as a function of the systematic error floor for four different planet masses.
The rates on the left-hand side of each plot represent an assumed planet frequency of one planet per dex$^{2}$ per star.
The rates on the right-hand side of each plot assume the cool-planet mass function of \citet{cassan12}, where we have saturated it at ${\mpl} = 5{\rm M}_{\Earth}$ (corresponding to $\sim$2 planets per dex$^{2}$).
The red dashed line denotes the detection rates for our fiducial simulations, for which we use a systematic error floor of 0.004~mag.
Here we see two trends.
First, the slope of the decrease in detection rate as a function of the photometric precision limit increases as planet mass decreases.
Secondly, the morphology of that dependence becomes more asymptotic and less linear as planet mass decreases.
}
}
\label{fig:ndet_sys}
\end{figure}

Figure \ref{fig:ndet_sys} shows the planet detection rate as a function of the systematic error floor, summed across semimajor axis.
We step through the range $0.004 \leq {\sigsys}/{\rm mag} \leq 0.02$ in steps of 0.001 mag.
The value adopted for our fiducial simulations is 0.004 mag, denoted by the dashed red line.
Here we see that the decrease in detections as a function of {\sigsys} depends on mass.
Over the range of systematic error floors we consider, we find that the trend is relatively weak and consistent with linear for planets with mass ${\mpl} \geq 100{\rm M}_{\Earth}$.
On the other hand, for lower mass planets the trend is much stronger and appears to exhibit an asymptotic behavior toward larger error floors.
In particular, the fraction of detections that remain even for the highest value (0.02 mag) of the systematic floor is 86$\%$ for ${\mpl} = 1000{\rm M}_{\Earth}$, but drops to 51$\%$ for ${\mpl} = 1{\rm M}_{\Earth}$.

This behavior is analogous to that seen in the distribution of $\Delta\chi^2$ values for the detected planets shown in Figure \ref{fig:chi2_apars}, where the low-mass planets produce perturbations with a steeper distribution of $\Delta\chi^2$ than high-mass planets.
Both trends are likely a consequence of the underlying spectrum of planetary perturbations that is being probed in the simulations.
In particular, the bulk of the detections from high-mass planets arise from perturbations that have lower amplitudes and longer time scales.
Such perturbations have very broad but shallow signals, and thus the detection rate is less sensitive to the precise value of systematic error floor or threshold $\Delta\chi^2$.
On the other hand, the detectable signals from smaller-mass planets are more often due to short-duration but high-amplitude perturbations.
The spectrum of such perturbations is narrow and steep, and thus the detection rate is more sensitive to the systematic error floor or detection threshold.

\subsection{Optimistic Simulation Run} \label{sec:rates_opt}

We also run a set of simulations across the same 9x17 grid of {\mpl} and $a$ and with the same spacing, this time using a set of more optimistic input assumptions to bracket the planet detection rates that KMTNet will obtain.
For these simulations we
\begin{itemize}
   \item{decrease the assumed value of the systematic error floor, from ${\sigsys} = 0.004$ mag to ${\sigsys} = 0.002$ mag,}

   \item{remove the extra component of the smooth background of $\mu_{I} = 18.8$ mag/$\square \arcsec$, discussed in \S \ref{sec:fluxdeter}, that we had previously included to match OGLE's photometric uncertainties, which were higher than expected due to additional noise of unknown origin,}

   \item{adopt seeing distributions that approximate the native seeing of each site, shown in Table \ref{tab:see_opt}, and}

   \item{implement a photon rate normalization, derived below, that yields a photon rate of $9.25~\dot{\gamma}$ for $I=22$.}
\end{itemize}

\begin{deluxetable}{cccc}
\tablecaption{Optimistic Site-dependent Seeing Distribution Parameters}
\tablewidth{0pt}
\tablehead{
\colhead{Site} &
\colhead{min.} &
\colhead{$\mu$} &
\colhead{$\sigma$}
}
\startdata
CTIO   &   0.5   &   0.80   &   0.16   \\
SAAO   &   0.6   &   0.92   &   0.20   \\
SSO    &   0.6   &   1.2    &   0.40
\enddata
\tablecomments{All values are in arcseconds.}
\label{tab:see_opt}
\end{deluxetable}

Here we derive the simplified theoretical photon rate normalization that we employ for the optimistic simulations.
We compute the rate of detected photons for a star with an apparent magnitude of $I=22$ as
\begin{equation}
   \dot{\gamma}_{I=22}=f_{I=22} \cdot E_{\lambda_{{\rm eff},I}} \cdot \Delta\lambda_{I} \cdot A_{\rm KMTNet} \cdot TE \cdot QE \cdot AE,
\end{equation}
where $f_{I=22}$ is the flux per unit wavelength of a 22nd magnitude star, $E_{\lambda_{{\rm eff},I}}$ is the energy of a photon with a wavelength equal to the effective wavelength $\lambda_{\rm eff}$ of the Cousins $I$-band filter, $\Delta\lambda_{I}$ is the width of the Cousins $I$-band filter, and $A_{\rm KMTNet}$ is the effective collecting area of KMTNet.
The last three terms encapsulate assumptions made about the photon loss rate from the top of the atmosphere to the production of a photoelectron.
Here $TE$ is the telescope efficiency, $QE$ is the $I$-band quantum efficiency of a given pixel on the KMTNet CCD, and $AE$ is the atmospheric extinction.

Using these and the flux received from a 0-magnitude star through the Cousins $I$-band filter, $f_{I=0} = 112.6 \cdot 10^{-11}~{\rm erg}~{\rm s}^{-1}~{\rm cm}^{-2}~\text{\AA}^{-1}$ \citep{bessell98}, we obtain the flux for a star with $I = 22$: $f_{I=22}=1.785 \cdot 10^{-18}~{\rm erg}~{\rm s}^{-1}~{\rm cm}^{-2}~\text{\AA}^{-1}$.
From \citet{bessell05} we obtain $\lambda_{{\rm eff},I}=7980\text{\AA}$, making $E_{\lambda_{{\rm eff},I}}=2.491\cdot10^{-12}~{\rm erg}~\gamma^{-1}$, and $\Delta\lambda_{I}=1540\text{\AA}$.
The effective diameter of the clear aperture of each KMTNet telescope is 1.6m, given in Table \ref{tab:kmt_scope_parms}, yielding $A_{\rm KMTNet}=20106.2~{\rm cm}^{2}$.
Finally, we adopt $TE=0.662$, given in Table \ref{tab:kmt_scope_parms}, and $QE=0.7$, listed in Table \ref{tab:kmt_cam_parms}.
While in reality $AE$ depends on the airmass of a given observation, we make the simplifying assumption that it is constant and use a slightly pessimistic value of $AE=0.9$.
Together these give a naive theoretical photon rate normalization of
\begin{equation}
   \dot{\gamma}=9.25~{\rm ph/s}\cdot10^{-0.4(I-22.0)},
\end{equation}
which is a factor of $\sim$2 higher than the empirical rate and which we use for our optimistic simulations.

\begin{figure}
\centerline{
\includegraphics[width=9cm]{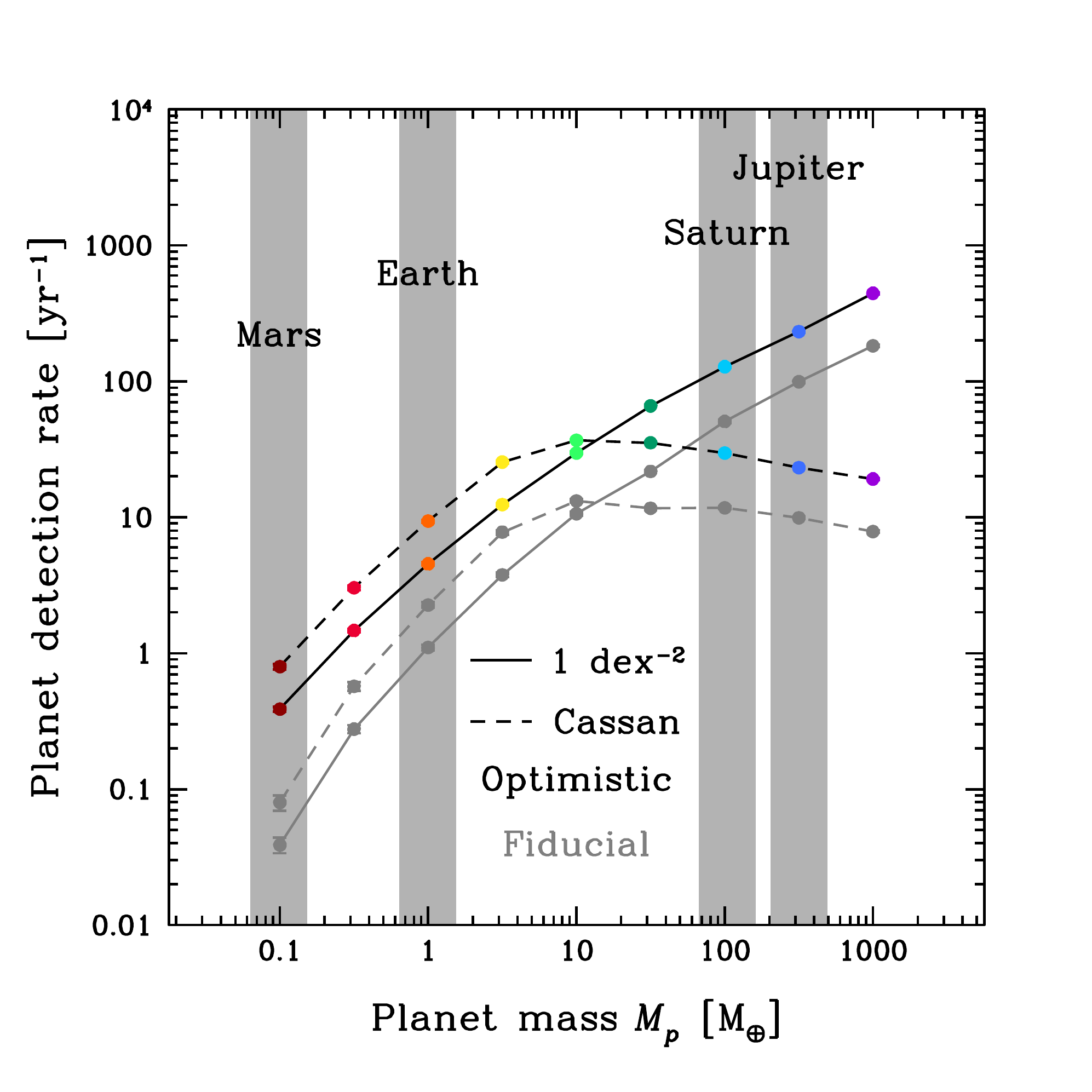}
}
\caption{
\footnotesize{
Planet detection rate as a function of planet mass {\mpl}, summed across semimajor axis $a$, for the full grid of our optimistic simulations.
The solid black line represents an assumed planet frequency of one planet per dex$^{2}$ per star.
The dashed black line represents our predictions for the detection rates after normalizing to the cool-planet mass function of \citet{cassan12}, where we have saturated it at ${\mpl} = 5{\rm M}_{\Earth}$ (corresponding to $\sim$2 planets per dex$^{2}$).
The grey points and lines represent the detection rates for our fiducial simulations, also shown in Figure \ref{fig:ndet_suma_pes}.
As with our fiducial detection rates, we see that the steep dependence that the optimistic detection rates have as a function of mass assuming one planet per dex$^{2}$ is nearly exactly canceled out by the increasing frequency of planets with decreasing planet mass according to the cool-planet mass function for ${\mpl} \geq 5{\rm M}_{\Earth}$.
For the highest-mass planets with ${\mpl} = 1000{\rm M}_{\Earth}$, the detection rate is roughly two times larger for the optimistic simulations than for our fiducial simulations.
This boost factor increases monotonically as planet mass decreases, resulting in a gain in detection rate by a factor of $\sim$10 for the lowest-mass planets with ${\mpl} = 0.1{\rm M}_{\Earth}$.
This trend is likely a result of the evolution of the dependence of the detection rate on the photometric precision limit shown in Figure \ref{fig:ndet_obs}, given that the optimistic simulations have a systematic error floor that is half of the value assumed for the fiducial simulations, coupled with the improved photometric precision arising from the removal of the additional smooth background component.
}
}
\label{fig:ndet_suma_pesopt}
\end{figure}

Figure \ref{fig:ndet_suma_pesopt} shows the planet detection rates as a function of planet mass for our optimistic simulations, summed across semimajor axis.
For ${\mpl} = 1000{\rm M}_{\Earth}$ our optimistic assumptions result in an increase in planet detection rates by a factor of $\sim$2 over the fiducial results.
This increase in the ratio of detection rates rises to $\sim$4 for ${\mpl} = 1{\rm M}_{\Earth}$ and $\sim$10 for ${\mpl} = 0.1{\rm M}_{\Earth}$.
These results show that even modest improvements in technology over that used by the extant generation of observational microlensing that boost the photon rate, decrease the systematics, and limit the background noise lead to an significant increase in KMTNet's planet detection rates, particularly for low-mass planets.

\section{Discussion} \label{sec:discussion}

If our fiducial and optimistic simulations reasonably bracket the expected KMTNet detection rates, we find that KMTNet will substantially increase the annual detection rates of exoplanets via gravitational microlensing.
Adopting the cool-planet mass function of \citet{cassan12} and leveling it off at ${\mpl} = 5{\rm M}_{\Earth}$ (corresponding to $\sim$2 planets per dex$^{2}$), we find that the slope of the mass function almost exactly cancels the power-law slope of detections as a function of mass that we find with our simulations.
For ${\mpl} \gtrsim 5{\rm M}_{\Earth}$, the regime in which we believe our rates to be robust and less sensitive to small number statistics, the detection rate is roughly flat, indicating that KMTNet will detect approximately $\sim$20 planets per year per dex in mass across this range.
For lower-mass planets with mass $0.1 \leq {\mplme} < 5$, we predict KMTNet will detect $\sim$10 planets per year.

\begin{figure}
\centerline{
\includegraphics[width=9cm]{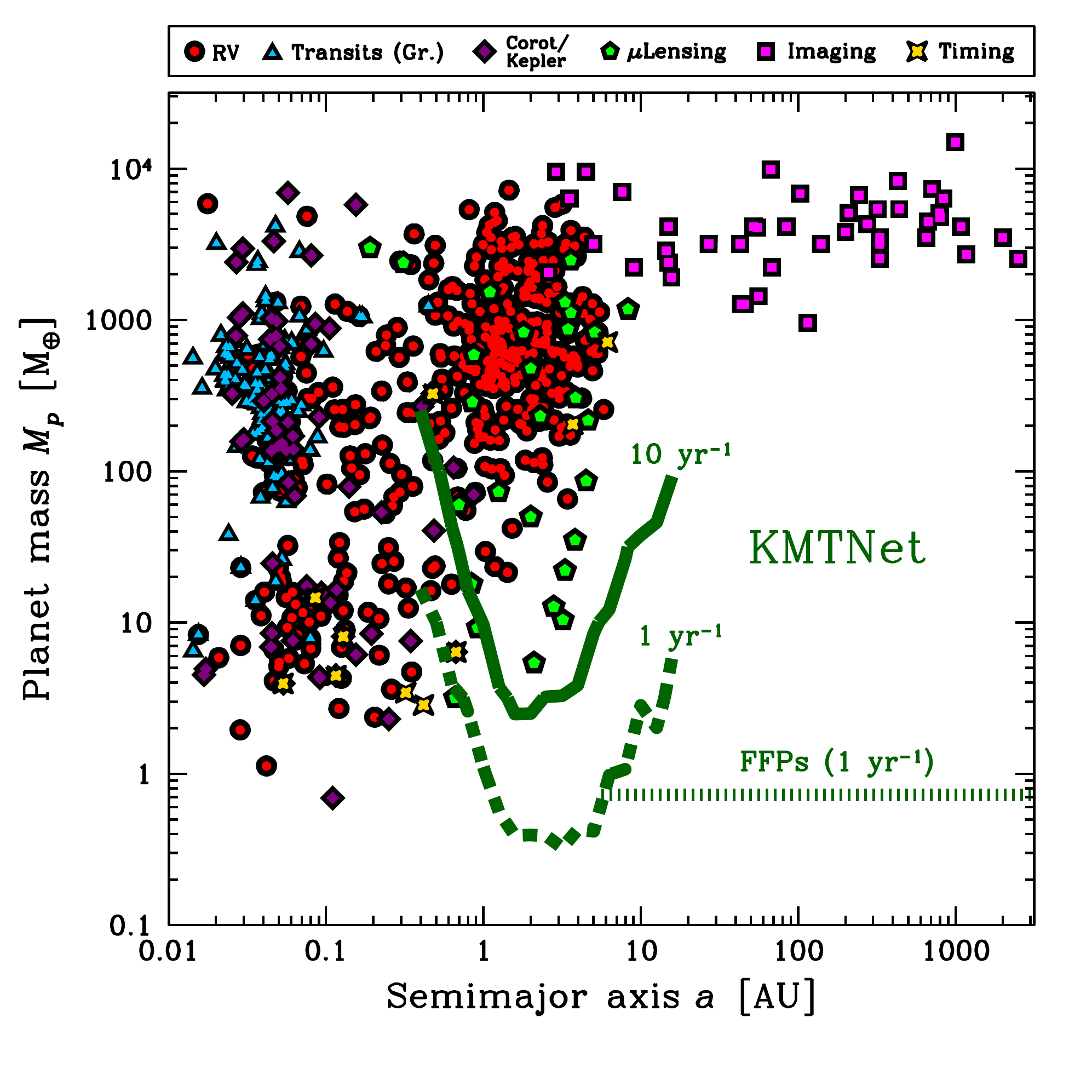}
}
\caption{
\footnotesize{
Planet mass {\mpl} as a function of semimajor axis $a$ for known exoplanets as of 29/May/2014.
The data for planets discovered via transits or RV come from http://exoplanets.org while the data for planets discovered via microlensing, imaging, or timing come from http://exoplanet.eu.
The thick solid green line marks the KMTNet detection contour of 10 planets per year while the thick dashed green line represents the KMTNet detection contour for one planet per year, both assuming that each lens star hosts exactly one planet, of the specified mass, at the specified separation.
The thick horizontal green dashed line denotes the mass at which KMTNet will detect one FFP per year, assuming one such planet per star in the Galaxy.
KMTNet will significantly augment the number of known exoplanets near and beyond the snow lines of their host stars, facilitating synoptic studies of exoplanet demographics across an unprecedented range in both planet mass and planet-star separation.
}
}
\label{fig:a_mass}
\end{figure}

Figure \ref{fig:a_mass} shows all known exoplanets\footnote{Data are taken from \url{http://exoplanets.org} and \url{http://exoplanet.eu} as of 29/May/2014}, color-coded by discovery technique.
We overlay contours for our fiducial KMTNet detection rates of one and 10 planets per year for bound planets, assuming that each lens star hosts exactly one planet, of the specified mass, at the specified separation, and show the planet mass at which the detection rate for FFPs is approximately unity, assuming one such planet per star in the Galaxy.
Even for our more conservative set of assumptions we find that KMTNet's annual detection rate will significantly augment the sample of known planets at planet-star distances near and beyond the snow line, and will do so even for planets less-massive than Earth.
This explosion in the microlensing detection rate will complement the \textit{Kepler} transiting planet population, specifically at semimajor axes of a few AU all the way down to Earth-mass planets, allowing for studies of the demographics of exoplanets across a range of over four dex in mass and three dex in separation.
Furthermore, we predict that KMTNet will have the ability to detect FFPs with mass below that of the Earth, providing a statistically large sample with which to improve our understanding of their prevalence within the Galaxy.

\acknowledgments
This material is based upon work supported by the National Science Foundation (NSF) Graduate Research Fellowship Program under Grant No.\ DGE-0822215 and an international travel allowance through the Graduate Research Opportunities Worldwide.
Any opinions, findings, and conclusions or recommendations expressed in this material are those of the authors and do not necessarily reflect the views of the NSF.
Work by B.~S.~G.\ was partially supported by NSF CAREER grant AST-1056524.
C.\ Han was supported by the Creative Research Initiative Program (2009-0081561) of the National Research Foundation of Korea.
J.~S.\ acknowledges support of the Space Exploration Research Fund of The Ohio State University, and also from the OGLE project's funding received from the European Research Council under the European Community's Seventh Framework Programme (FP7/2007--2013)/ERC grant agreement No.~246678.
Work by B.S.G.\ and A.G.\ was partially supported by NSF grant AST 1103471.

\begin{appendices}

\section{Optical Depth Comparison} \label{app:optical_depth}

\begin{figure}
\includegraphics[width=9cm]{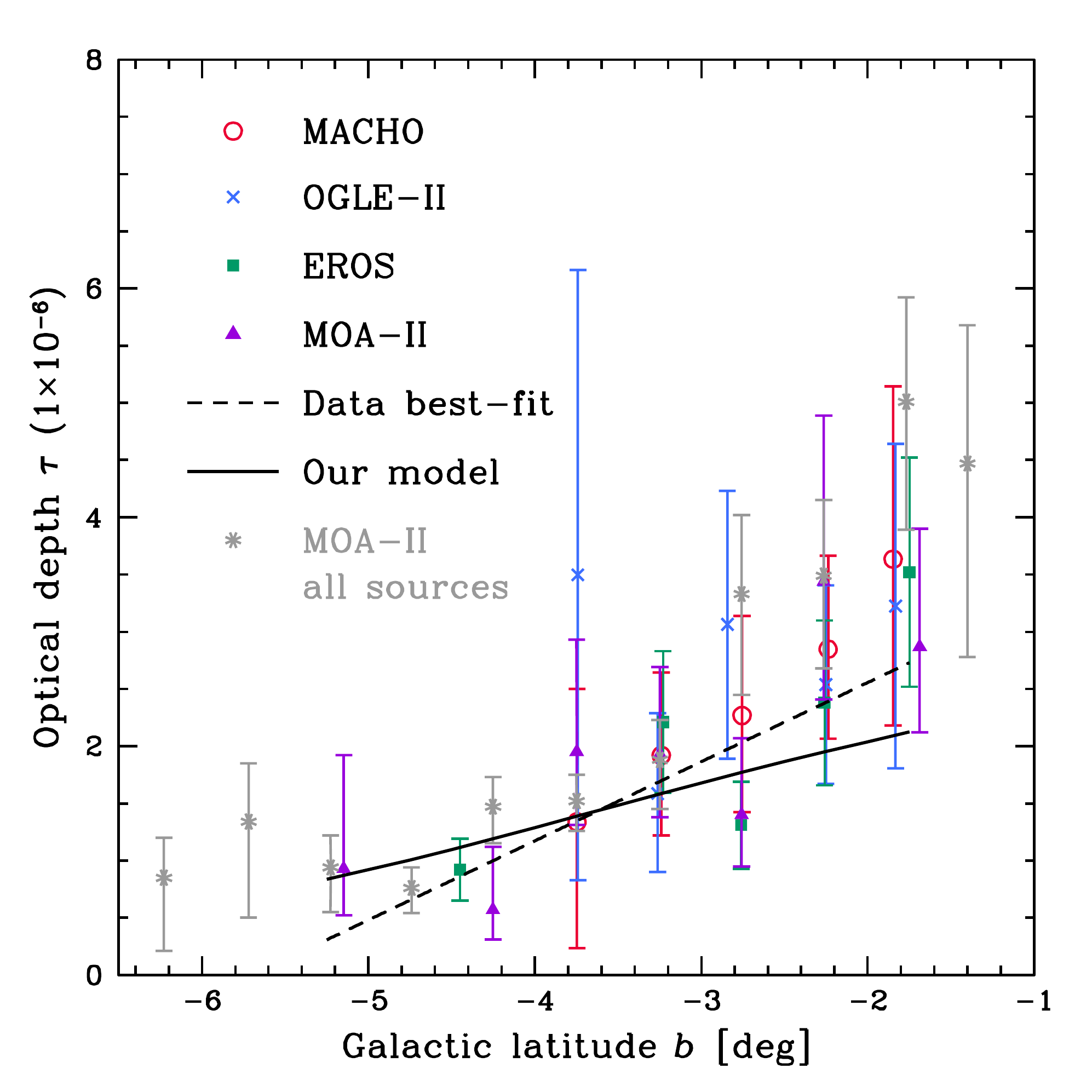}
\caption{
\footnotesize{
Optical depth as a function of absolute Galactic latitude.
All points except the grey ones represent RCG-derived measurements of $\tau$.
The dashed black line is a best-fit to the combined MACHO, OGLE, EROS, and MOA data.
The solid black line is the optical depth of our Galactic model, in steps of 0.25 degrees in $b$, averaged over $-1.0
\leq l/\mathrm{deg} \leq 4.0$, with 0.5 degree spacing.
The discrepancy between our model and the best-fit would correspondingly affect our event rates.
The grey points show the optical depth derived using all sources in the MOA-II catalog, and increases more steeply toward higher Galactic latitude.
}
}
\label{fig:optical_depth}
\end{figure}

In this appendix we compare the optical depth predicted by our Galactic model to that measured in recent microlensing studies.
Figure \ref{fig:optical_depth} shows the optical depth $\tau$ as a function of Galactic latitude as measured from RCG stars from the MACHO \citep{popowski05}, OGLE-II \citep{sumi06}, EROS \citep{hamadache06}, and MOA-II \citep{sumi13} surveys, as well as a simple linear fit to these combined data.
To compare to these results, we also show the optical depth produced by our model in the range $-5.25 \leq b/{\rm deg} \leq -1.75$ with spacing of 0.25 degrees, averaging $\tau$ across $-1.0 \leq l/{\rm deg} \leq 4.0$, with 0.5 degree spacing, at each value of $b$.
Our gradient of $\tau$ with $b$ is slightly less steep than that found by a simple linear fit to the RCG microlensing survey data.
For the inner survey fields of $b \ga -3$, which constitute the majority of the event rate, our model underestimates the optical depth as inferred from RCGs by up to $\sim 50\%$, and thus underestimates the event rate by a similar factor.
Also shown are the optical depths as measured using all sources from the MOA-II survey \citep{sumi13}.
At higher Galactic latitudes of $b \sim -1.5$, these optical depths are $\sim$30-70$\%$ larger than the RCG optical depths measured in the same fields, driven by the peak in the event rate per square degree per year that they find at $(l,b) \approx (1.0, -2.5)$ (see their Figures 3 and 12).
While we acknowledge that these results would affect our detection rates, we assert that it would not have an impact with regard to the lack of a clear preference
for the tiling of the observational fields, because the shifts between the fields are not substantial and the tilings generally cover similar regions of the inner Bulge.
On the other hand, it is possible that the steep gradient in the event rate implied by these results would have an appreciable effect on our chosen optimal number of fields.
However, we argue that, given the substantial uncertainties in the measured optical depths and event rates, the prudent strategy is to initially monitor the four fields we have advocated here.
The results of this initial survey can then be used to more accurately determine the event rate in these fields, and this information can then be used to further optimize a second phase of the KMTNet survey.
Indeed, it may be useful initially to monitor a dozen or so additional ``outrigger'' fields with a much lower cadence, in order to map the event rate over a much larger area of the bulge.

\section{Online Light Curve Atlas} \label{app:lc_atlas}
\begin{figure}
\includegraphics[width=9cm]{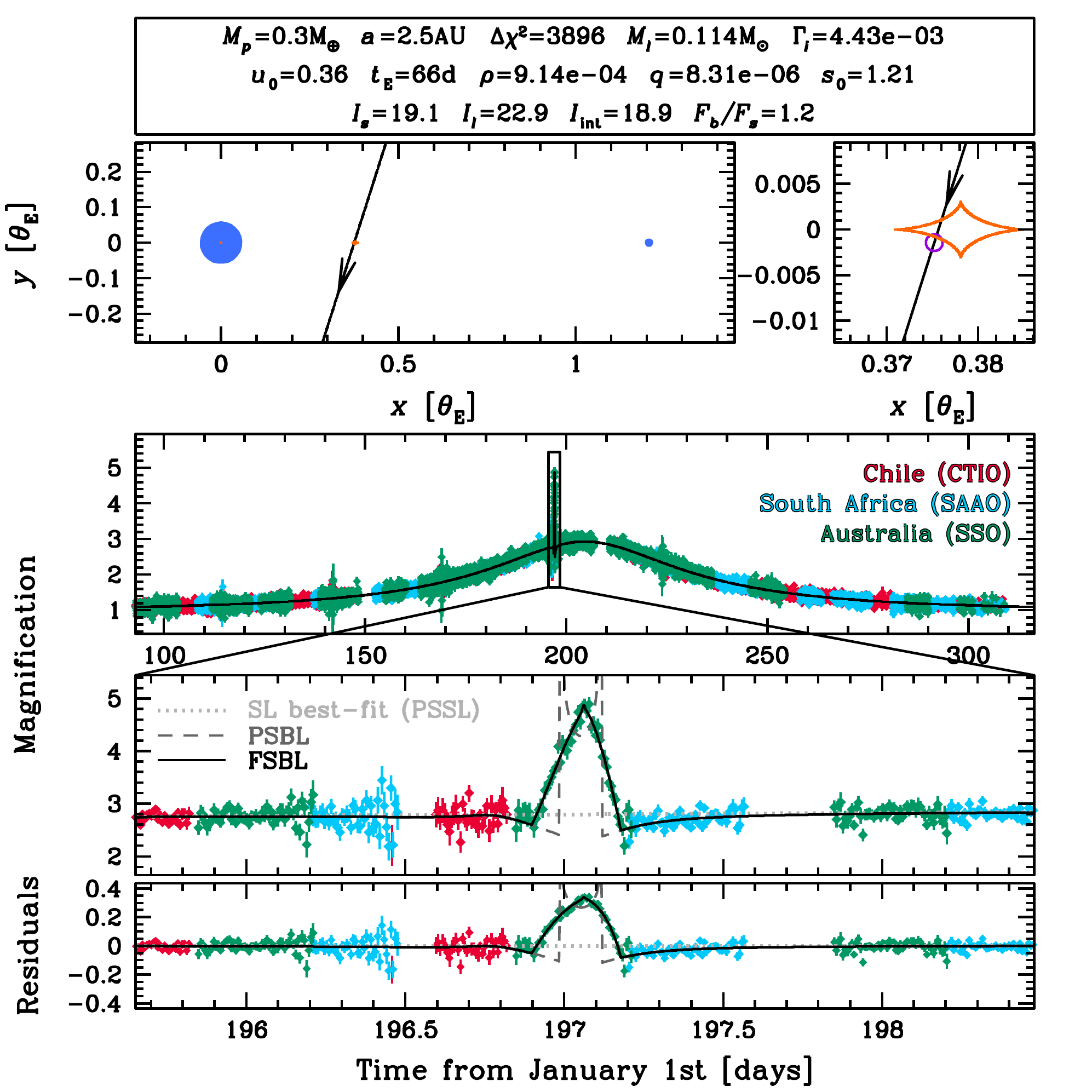}
\caption{
\footnotesize{
An example light curve from our online atlas\footnote{To view the light curve atlas in its entirety, please visit \url{http://www.astronomy.ohio-state.edu/\string~henderson/kmtnet\_lcatlas}}.
The middle panel contains an overview of the magnification of the event as a function of time, with the data color-coded to the respective observatories from which they were taken.
We have included Gaussian scatter in the photometry solely for the purposes of visualizing the quality of data that KMTNet will actually obtain.
The second panel from the bottom highlights the planetary perturbation in greater detail and also includes the curves for three different models.
The solid black line represents the FSBL model, about which the data are scattered.
The dashed dark grey line shows the PSBL model, indicating the effect of a finite source.
The dotted light grey line represents the best-fit single-lens (SL) model, either PSSL or FSSL as determined by our simulation in \S \ref{sec:finitesource}.
The lowest panel shows the residuals and models over the same temporal range as the second panel.
The fourth panel contains two plots that show an overview (left) and a zoomed-in version (right) of the geometry of the event.
Here the blue circles represent the primary lensing mass (left) and planetary companion (right).
The absolute sizes of the lensing masses are arbitrary but the radius of the primary is fixed while the radius of the planet scales as log($q$).
The origin is located at either the center of mass (if ${\snot} < 1$) or photocenter (if ${\snot} > 1$) of the lens system.
The orange curves represent the caustic(s) while the black line and arrow together specify the source trajectory.
In the right plot of the fourth panel the purple open circle is the source, scaled appropriately, centered on the time of the maximum deviation of the observed light curve.
The top panel shows the physical, lensing, and observational parameters for the event.
We have generated dozens of such figures for a random selection of planet detections for each grid point combination of planet mass {\mpl} and semimajor axis $a$.
}
}
\label{fig:lc_atlas_example}
\end{figure}

In this appendix we describe a large-scale atlas of light curves we have created to facilitate visualization of and improve intuition about the planetary systems that we predict KMTNet will detect.
For each grid point combination of planet mass {\mpl} and planet-star separation $a$ we have generated dozens of light curves from a random selection of the detections.
Figure \ref{fig:lc_atlas_example} shows an example for a detection with ${\mpl} = 0.5{\rm M}_{\Earth}$ and $a = 2.51$AU.
Each figure has five panels.
The middle panel contains an overview of the light curve that shows the overarching magnification structure of the primary event and, when of sufficient amplitude, the deviation due to the planet.
We have included Gaussian scatter in the photometry solely for the purposes of visualizing the quality of data that KMTNet will actually obtain.
The second panel from the bottom shows the planetary perturbation in greater detail with curves for three different models --- the FSBL model about which the data are scattered, the PSBL model to indicate the effect a finite source has, and the best-fit single-lens model, either PSSL or FSSL as determined by our simulation in \S \ref{sec:finitesource}.
The lowest panel shows the residuals and models over the same temporal range as the second panel.
The fourth panel contains two plots that show an overview and a detailed version of the geometry of the event, including the lensing masses, the caustic(s), and the source trajectory and size.
The top panel shows the physical, lensing, and observational parameters for the event.

From $I_{s}$, $I_{l}$, $I_{\rm int}$, and an estimate of the peak magnification from the bottom panel it is possible to estimate the maximum brightness of the event.
It is also possible to estimate by-eye from where the signal of the planet arises, both in time and in the source plane with respect to the caustics, and to estimate the {\dchisq}.
To view the light curve atlas in its entirety, please visit \url{http://www.astronomy.ohio-state.edu/\string~henderson/kmtnet\_lcatlas}.

\end{appendices}


\begin{thebibliography}{95}
\expandafter\ifx\csname natexlab\endcsname\relax\def\natexlab#1{#1}\fi

\bibitem[{{Alard}(2000)}]{alard00}
{Alard}, C. 2000, \aaps, 144, 363

\bibitem[{{Alard} \& {Lupton}(1998)}]{alard98}
{Alard}, C., \& {Lupton}, R.~H. 1998, \apj, 503, 325

\bibitem[{{Alexander} {et~al.}(2012){Alexander}, {Bowden}, {Fogel}, {Howard},
  {Herd}, \& {Nittler}}]{alexander12}
{Alexander}, C.~M.~O.~., {Bowden}, R., {Fogel}, M.~L., {Howard}, K.~T., {Herd},
  C.~D.~K., \& {Nittler}, L.~R. 2012, Science, 337, 721

\bibitem[{{Bahcall}(1986)}]{bahcall86}
{Bahcall}, J.~N. 1986, \araa, 24, 577

\bibitem[{{Bakos} {et~al.}(2009){Bakos}, {Howard}, {Noyes}, {Hartman},
  {Torres}, {Kov{\'a}cs}, {Fischer}, {Latham}, {Johnson}, {Marcy}, {Sasselov},
  {Stefanik}, {Sip{\H o}cz}, {Kov{\'a}cs}, {Esquerdo}, {P{\'a}l},
  {L{\'a}z{\'a}r}, {Papp}, \& {S{\'a}ri}}]{bakos09}
{Bakos}, G.~{\'A}., {et~al.} 2009, \apj, 707, 446

\bibitem[{{Baraffe} {et~al.}(1998){Baraffe}, {Chabrier}, {Allard}, \&
  {Hauschildt}}]{baraffe98}
{Baraffe}, I., {Chabrier}, G., {Allard}, F., \& {Hauschildt}, P.~H. 1998, \aap,
  337, 403

\bibitem[{{Baraffe} {et~al.}(2002){Baraffe}, {Chabrier}, {Allard}, \&
  {Hauschildt}}]{baraffe02}
---. 2002, \aap, 382, 563

\bibitem[{{Barclay} {et~al.}(2013){Barclay}, {Rowe}, {Lissauer}, {Huber},
  {Fressin}, {Howell}, {Bryson}, {Chaplin}, {D{\'e}sert}, {Lopez}, {Marcy},
  {Mullally}, {Ragozzine}, {Torres}, {Adams}, {Agol}, {Barrado}, {Basu},
  {Bedding}, {Buchhave}, {Charbonneau}, {Christiansen},
  {Christensen-Dalsgaard}, {Ciardi}, {Cochran}, {Dupree}, {Elsworth},
  {Everett}, {Fischer}, {Ford}, {Fortney}, {Geary}, {Haas}, {Handberg},
  {Hekker}, {Henze}, {Horch}, {Howard}, {Hunter}, {Isaacson}, {Jenkins},
  {Karoff}, {Kawaler}, {Kjeldsen}, {Klaus}, {Latham}, {Li}, {Lillo-Box},
  {Lund}, {Lundkvist}, {Metcalfe}, {Miglio}, {Morris}, {Quintana}, {Stello},
  {Smith}, {Still}, \& {Thompson}}]{barclay13}
{Barclay}, T., {et~al.} 2013, \nat, 494, 452

\bibitem[{{Beaulieu} {et~al.}(2006){Beaulieu}, {Bennett}, {Fouqu{\'e}},
  {Williams}, {Dominik}, {J{\o}rgensen}, {Kubas}, {Cassan}, {Coutures},
  {Greenhill}, {Hill}, {Menzies}, {Sackett}, {Albrow}, {Brillant}, {Caldwell},
  {Calitz}, {Cook}, {Corrales}, {Desort}, {Dieters}, {Dominis}, {Donatowicz},
  {Hoffman}, {Kane}, {Marquette}, {Martin}, {Meintjes}, {Pollard}, {Sahu},
  {Vinter}, {Wambsganss}, {Woller}, {Horne}, {Steele}, {Bramich}, {Burgdorf},
  {Snodgrass}, {Bode}, {Udalski}, {Szyma{\'n}ski}, {Kubiak}, {Wi{\c e}ckowski},
  {Pietrzy{\'n}ski}, {Soszy{\'n}ski}, {Szewczyk}, {Wyrzykowski},
  {Paczy{\'n}ski}, {Abe}, {Bond}, {Britton}, {Gilmore}, {Hearnshaw}, {Itow},
  {Kamiya}, {Kilmartin}, {Korpela}, {Masuda}, {Matsubara}, {Motomura},
  {Muraki}, {Nakamura}, {Okada}, {Ohnishi}, {Rattenbury}, {Sako}, {Sato},
  {Sasaki}, {Sekiguchi}, {Sullivan}, {Tristram}, {Yock}, \&
  {Yoshioka}}]{beaulieu06}
{Beaulieu}, J.-P., {et~al.} 2006, \nat, 439, 437

\bibitem[{{Bennett}(2010)}]{bennett10b}
{Bennett}, D.~P. 2010, \apj, 716, 1408

\bibitem[{{Bennett} \& {Rhie}(1996)}]{bennett96}
{Bennett}, D.~P., \& {Rhie}, S.~H. 1996, \apj, 472, 660

\bibitem[{{Bennett} {et~al.}(2003){Bennett}, {Bally}, {Bond}, {Cheng}, {Cook},
  {Deming}, {Garnavich}, {Griest}, {Jewitt}, {Kaiser}, {Lauer}, {Lunine},
  {Luppino}, {Mather}, {Minniti}, {Peale}, {Rhie}, {Rhodes}, {Schneider},
  {Sonneborn}, {Stevenson}, {Stubbs}, {Tenerelli}, {Woolf}, \&
  {Yock}}]{bennett02}
{Bennett}, D.~P., {et~al.} 2003, in Society of Photo-Optical Instrumentation
  Engineers (SPIE) Conference Series, Vol. 4854, Society of Photo-Optical
  Instrumentation Engineers (SPIE) Conference Series, ed. J.~C. {Blades} \&
  O.~H.~W. {Siegmund}, 141--157

\bibitem[{{Bennett} {et~al.}(2008){Bennett}, {Bond}, {Udalski}, {Sumi}, {Abe},
  {Fukui}, {Furusawa}, {Hearnshaw}, {Holderness}, {Itow}, {Kamiya}, {Korpela},
  {Kilmartin}, {Lin}, {Ling}, {Masuda}, {Matsubara}, {Miyake}, {Muraki},
  {Nagaya}, {Okumura}, {Ohnishi}, {Perrott}, {Rattenbury}, {Sako}, {Saito},
  {Sato}, {Skuljan}, {Sullivan}, {Sweatman}, {Tristram}, {Yock}, {Kubiak},
  {Szyma{\'n}ski}, {Pietrzy{\'n}ski}, {Soszy{\'n}ski}, {Szewczyk},
  {Wyrzykowski}, {Ulaczyk}, {Batista}, {Beaulieu}, {Brillant}, {Cassan},
  {Fouqu{\'e}}, {Kervella}, {Kubas}, \& {Marquette}}]{bennett08}
{Bennett}, D.~P., {et~al.} 2008, \apj, 684, 663

\bibitem[{{Bennett} {et~al.}(2010){Bennett}, {Rhie}, {Nikolaev}, {Gaudi},
  {Udalski}, {Gould}, {Christie}, {Maoz}, {Dong}, {McCormick}, {Szyma{\'n}ski},
  {Tristram}, {Macintosh}, {Cook}, {Kubiak}, {Pietrzy{\'n}ski},
  {Soszy{\'n}ski}, {Szewczyk}, {Ulaczyk}, {Wyrzykowski}, {OGLE Collaboration},
  {DePoy}, {Han}, {Kaspi}, {Lee}, {Mallia}, {Natusch}, {Park}, {Pogge},
  {Polishook}, {{$\mu$}FUN Collaboration}, {Abe}, {Bond}, {Botzler}, {Fukui},
  {Hearnshaw}, {Itow}, {Kamiya}, {Korpela}, {Kilmartin}, {Lin}, {Ling},
  {Masuda}, {Matsubara}, {Motomura}, {Muraki}, {Nakamura}, {Okumura},
  {Ohnishi}, {Perrott}, {Rattenbury}, {Sako}, {Saito}, {Sato}, {Skuljan},
  {Sullivan}, {Sumi}, {Sweatman}, {Yock}, {MOA Collaboration}, {Albrow},
  {Allan}, {Beaulieu}, {Bramich}, {Burgdorf}, {Coutures}, {Dominik}, {Dieters},
  {Fouqu{\'e}}, {Greenhill}, {Horne}, {Snodgrass}, {Steele}, {Tsapras},
  {PLANET}, {RoboNet Collaborations}, {Chaboyer}, {Crocker}, \&
  {Frank}}]{bennett10a}
---. 2010, \apj, 713, 837

\bibitem[{{Bessell}(2005)}]{bessell05}
{Bessell}, M.~S. 2005, \araa, 43, 293

\bibitem[{{Bessell} {et~al.}(1998){Bessell}, {Castelli}, \& {Plez}}]{bessell98}
{Bessell}, M.~S., {Castelli}, F., \& {Plez}, B. 1998, \aap, 333, 231

\bibitem[{{Bond} {et~al.}(2001){Bond}, {Abe}, {Dodd}, {Hearnshaw}, {Honda},
  {Jugaku}, {Kilmartin}, {Marles}, {Masuda}, {Matsubara}, {Muraki}, {Nakamura},
  {Nankivell}, {Noda}, {Noguchi}, {Ohnishi}, {Rattenbury}, {Reid}, {Saito},
  {Sato}, {Sekiguchi}, {Skuljan}, {Sullivan}, {Sumi}, {Takeuti}, {Watase},
  {Wilkinson}, {Yamada}, {Yanagisawa}, \& {Yock}}]{bond01}
{Bond}, I.~A., {et~al.} 2001, \mnras, 327, 868

\bibitem[{{Bond} {et~al.}(2004){Bond}, {Udalski}, {Jaroszy{\'n}ski},
  {Rattenbury}, {Paczy{\'n}ski}, {Soszy{\'n}ski}, {Wyrzykowski},
  {Szyma{\'n}ski}, {Kubiak}, {Szewczyk}, {{\.Z}ebru{\'n}}, {Pietrzy{\'n}ski},
  {Abe}, {Bennett}, {Eguchi}, {Furuta}, {Hearnshaw}, {Kamiya}, {Kilmartin},
  {Kurata}, {Masuda}, {Matsubara}, {Muraki}, {Noda}, {Okajima}, {Sako},
  {Sekiguchi}, {Sullivan}, {Sumi}, {Tristram}, {Yanagisawa}, {Yock}, \& {OGLE
  Collaboration}}]{bond04}
---. 2004, \apjl, 606, L155

\bibitem[{{Bonfils} {et~al.}(2013){Bonfils}, {Delfosse}, {Udry}, {Forveille},
  {Mayor}, {Perrier}, {Bouchy}, {Gillon}, {Lovis}, {Pepe}, {Queloz}, {Santos},
  {S{\'e}gransan}, \& {Bertaux}}]{bonfils13}
{Bonfils}, X., {et~al.} 2013, \aap, 549, A109

\bibitem[{{Butler} {et~al.}(1999){Butler}, {Marcy}, {Fischer}, {Brown},
  {Contos}, {Korzennik}, {Nisenson}, \& {Noyes}}]{butler99}
{Butler}, R.~P., {Marcy}, G.~W., {Fischer}, D.~A., {Brown}, T.~M., {Contos},
  A.~R., {Korzennik}, S.~G., {Nisenson}, P., \& {Noyes}, R.~W. 1999, \apj, 526,
  916

\bibitem[{{Cassan} {et~al.}(2012){Cassan}, {Kubas}, {Beaulieu}, {Dominik},
  {Horne}, {Greenhill}, {Wambsganss}, {Menzies}, {Williams}, {J{\o}rgensen},
  {Udalski}, {Bennett}, {Albrow}, {Batista}, {Brillant}, {Caldwell}, {Cole},
  {Coutures}, {Cook}, {Dieters}, {Prester}, {Donatowicz}, {Fouqu{\'e}}, {Hill},
  {Kains}, {Kane}, {Marquette}, {Martin}, {Pollard}, {Sahu}, {Vinter},
  {Warren}, {Watson}, {Zub}, {Sumi}, {Szyma{\'n}ski}, {Kubiak}, {Poleski},
  {Soszynski}, {Ulaczyk}, {Pietrzy{\'n}ski}, \& {Wyrzykowski}}]{cassan12}
{Cassan}, A., {et~al.} 2012, \nat, 481, 167

\bibitem[{{Charbonneau} {et~al.}(2009){Charbonneau}, {Berta}, {Irwin}, {Burke},
  {Nutzman}, {Buchhave}, {Lovis}, {Bonfils}, {Latham}, {Udry}, {Murray-Clay},
  {Holman}, {Falco}, {Winn}, {Queloz}, {Pepe}, {Mayor}, {Delfosse}, \&
  {Forveille}}]{charbonneau09}
{Charbonneau}, D., {et~al.} 2009, \nat, 462, 891

\bibitem[{{Cumming} {et~al.}(2008){Cumming}, {Butler}, {Marcy}, {Vogt},
  {Wright}, \& {Fischer}}]{cumming08}
{Cumming}, A., {Butler}, R.~P., {Marcy}, G.~W., {Vogt}, S.~S., {Wright}, J.~T.,
  \& {Fischer}, D.~A. 2008, \pasp, 120, 531

\bibitem[{{Cumming} {et~al.}(1999){Cumming}, {Marcy}, \& {Butler}}]{cumming99}
{Cumming}, A., {Marcy}, G.~W., \& {Butler}, R.~P. 1999, \apj, 526, 890

\bibitem[{{Dominik} {et~al.}(2010){Dominik}, {J{\o}rgensen}, {Rattenbury},
  {Mathiasen}, {Hinse}, {Calchi Novati}, {Harps{\o}e}, {Bozza}, {Anguita},
  {Burgdorf}, {Horne}, {Hundertmark}, {Kerins}, {Kj{\ae}rgaard}, {Liebig},
  {Mancini}, {Masi}, {Rahvar}, {Ricci}, {Scarpetta}, {Snodgrass}, {Southworth},
  {Street}, {Surdej}, {Th{\"o}ne}, {Tsapras}, {Wambsganss}, \&
  {Zub}}]{dominik10}
{Dominik}, M., {et~al.} 2010, Astronomische Nachrichten, 331, 671

\bibitem[{{Dong} \& {Zhu}(2013)}]{dong13}
{Dong}, S., \& {Zhu}, Z. 2013, \apj, 778, 53

\bibitem[{{Dressing} \& {Charbonneau}(2013)}]{dressing13}
{Dressing}, C.~D., \& {Charbonneau}, D. 2013, \apj, 767, 95

\bibitem[{{Dwek} {et~al.}(1995){Dwek}, {Arendt}, {Hauser}, {Kelsall}, {Lisse},
  {Moseley}, {Silverberg}, {Sodroski}, \& {Weiland}}]{dwek95}
{Dwek}, E., {et~al.} 1995, \apj, 445, 716

\bibitem[{{Gaudi}(2000)}]{gaudi00}
{Gaudi}, B.~S. 2000, \apjl, 539, L59

\bibitem[{{Gaudi}(2012)}]{gaudi12}
---. 2012, \araa, 50, 411

\bibitem[{{Gaudi} {et~al.}(2008){Gaudi}, {Bennett}, {Udalski}, {Gould},
  {Christie}, {Maoz}, {Dong}, {McCormick}, {Szyma{\'n}ski}, {Tristram},
  {Nikolaev}, {Paczy{\'n}ski}, {Kubiak}, {Pietrzy{\'n}ski}, {Soszy{\'n}ski},
  {Szewczyk}, {Ulaczyk}, {Wyrzykowski}, {OGLE Collaboration}, {DePoy}, {Han},
  {Kaspi}, {Lee}, {Mallia}, {Natusch}, {Pogge}, {Park}, {{$\mu$}-Fun
  Collabortion}, {Abe}, {Bond}, {Botzler}, {Fukui}, {Hearnshaw}, {Itow},
  {Kamiya}, {Korpela}, {Kilmartin}, {Lin}, {Masuda}, {Matsubara}, {Motomura},
  {Muraki}, {Nakamura}, {Okumura}, {Ohnishi}, {Rattenbury}, {Sako}, {Saito},
  {Sato}, {Skuljan}, {Sullivan}, {Sumi}, {Sweatman}, {Yock}, {MOA
  Collaboration}, {Albrow}, {Allan}, {Beaulieu}, {Burgdorf}, {Cook},
  {Coutures}, {Dominik}, {Dieters}, {Fouqu{\'e}}, {Greenhill}, {Horne},
  {Steele}, {Tsapras}, {Planet Collaboration}, {RoboNet Collaborations},
  {Chaboyer}, {Crocker}, {Frank}, \& {Macintosh}}]{gaudi08}
{Gaudi}, B.~S., {et~al.} 2008, Science, 319, 927

\bibitem[{{Girardi} {et~al.}(2000){Girardi}, {Bressan}, {Bertelli}, \&
  {Chiosi}}]{girardi00}
{Girardi}, L., {Bressan}, A., {Bertelli}, G., \& {Chiosi}, C. 2000, \aaps, 141,
  371

\bibitem[{{Gorbikov} {et~al.}(2010){Gorbikov}, {Brosch}, \&
  {Afonso}}]{gorbikov10}
{Gorbikov}, E., {Brosch}, N., \& {Afonso}, C. 2010, \apss, 326, 203

\bibitem[{{Gould}(2000)}]{gould00}
{Gould}, A. 2000, \apj, 535, 928

\bibitem[{{Gould}(2003)}]{gould03}
---. 2003, ArXiv Astrophysics e-prints

\bibitem[{{Gould}(2008)}]{gould08}
---. 2008, \apj, 681, 1593

\bibitem[{{Gould} \& {Gaucherel}(1997)}]{gould97}
{Gould}, A., \& {Gaucherel}, C. 1997, \apj, 477, 580

\bibitem[{{Gould} \& {Loeb}(1992)}]{gould92b}
{Gould}, A., \& {Loeb}, A. 1992, \apj, 396, 104

\bibitem[{{Gould} {et~al.}(2006){Gould}, {Udalski}, {An}, {Bennett}, {Zhou},
  {Dong}, {Rattenbury}, {Gaudi}, {Yock}, {Bond}, {Christie}, {Horne},
  {Anderson}, {Stanek}, {DePoy}, {Han}, {McCormick}, {Park}, {Pogge},
  {Poindexter}, {Soszy{\'n}ski}, {Szyma{\'n}ski}, {Kubiak}, {Pietrzy{\'n}ski},
  {Szewczyk}, {Wyrzykowski}, {Ulaczyk}, {Paczy{\'n}ski}, {Bramich},
  {Snodgrass}, {Steele}, {Burgdorf}, {Bode}, {Botzler}, {Mao}, \&
  {Swaving}}]{gould06}
{Gould}, A., {et~al.} 2006, \apjl, 644, L37

\bibitem[{{Gould} {et~al.}(2010){Gould}, {Dong}, {Gaudi}, {Udalski}, {Bond},
  {Greenhill}, {Street}, {Dominik}, {Sumi}, {Szyma{\'n}ski}, {Han}, {Allen},
  {Bolt}, {Bos}, {Christie}, {DePoy}, {Drummond}, {Eastman}, {Gal-Yam},
  {Higgins}, {Janczak}, {Kaspi}, {Koz{\l}owski}, {Lee}, {Mallia}, {Maury},
  {Maoz}, {McCormick}, {Monard}, {Moorhouse}, {Morgan}, {Natusch}, {Ofek},
  {Park}, {Pogge}, {Polishook}, {Santallo}, {Shporer}, {Spector}, {Thornley},
  {Yee}, {{$\mu$}FUN Collaboration}, {Kubiak}, {Pietrzy{\'n}ski},
  {Soszy{\'n}ski}, {Szewczyk}, {Wyrzykowski}, {Ulaczyk}, {Poleski}, {OGLE
  Collaboration}, {Abe}, {Bennett}, {Botzler}, {Douchin}, {Freeman}, {Fukui},
  {Furusawa}, {Hearnshaw}, {Hosaka}, {Itow}, {Kamiya}, {Kilmartin}, {Korpela},
  {Lin}, {Ling}, {Makita}, {Masuda}, {Matsubara}, {Miyake}, {Muraki}, {Nagaya},
  {Nishimoto}, {Ohnishi}, {Okumura}, {Perrott}, {Philpott}, {Rattenbury},
  {Saito}, {Sako}, {Sullivan}, {Sweatman}, {Tristram}, {von Seggern}, {Yock},
  {MOA Collaboration}, {Albrow}, {Batista}, {Beaulieu}, {Brillant}, {Caldwell},
  {Calitz}, {Cassan}, {Cole}, {Cook}, {Coutures}, {Dieters}, {Dominis Prester},
  {Donatowicz}, {Fouqu{\'e}}, {Hill}, {Hoffman}, {Jablonski}, {Kane}, {Kains},
  {Kubas}, {Marquette}, {Martin}, {Martioli}, {Meintjes}, {Menzies},
  {Pedretti}, {Pollard}, {Sahu}, {Vinter}, {Wambsganss}, {Watson}, {Williams},
  {Zub}, {PLANET Collaboration}, {Allan}, {Bode}, {Bramich}, {Burgdorf},
  {Clay}, {Fraser}, {Hawkins}, {Horne}, {Kerins}, {Lister}, {Mottram},
  {Saunders}, {Snodgrass}, {Steele}, {Tsapras}, {RoboNet Collaboration},
  {J{\o}rgensen}, {Anguita}, {Bozza}, {Calchi Novati}, {Harps{\o}e}, {Hinse},
  {Hundertmark}, {Kj{\ae}rgaard}, {Liebig}, {Mancini}, {Masi}, {Mathiasen},
  {Rahvar}, {Ricci}, {Scarpetta}, {Southworth}, {Surdej}, {Th{\"o}ne}, \&
  {MiNDSTEp Consortium}}]{gould10}
---. 2010, \apj, 720, 1073

\bibitem[{{Hamadache} {et~al.}(2006){Hamadache}, {Le Guillou}, {Tisserand},
  {Afonso}, {Albert}, {Andersen}, {Ansari}, {Aubourg}, {Bareyre}, {Beaulieu},
  {Charlot}, {Coutures}, {Ferlet}, {Fouqu{\'e}}, {Glicenstein}, {Goldman},
  {Gould}, {Graff}, {Gros}, {Haissinski}, {de Kat}, {Lesquoy}, {Loup},
  {Magneville}, {Marquette}, {Maurice}, {Maury}, {Milsztajn}, {Moniez},
  {Palanque-Delabrouille}, {Perdereau}, {Rahal}, {Rich}, {Spiro},
  {Vidal-Madjar}, {Vigroux}, \& {Zylberajch}}]{hamadache06}
{Hamadache}, C., {et~al.} 2006, \aap, 454, 185

\bibitem[{{Han}(2006)}]{han06}
{Han}, C. 2006, \apj, 638, 1080

\bibitem[{{Han} \& {Gould}(1995{\natexlab{a}})}]{han95a}
{Han}, C., \& {Gould}, A. 1995{\natexlab{a}}, \apj, 449, 521

\bibitem[{{Han} \& {Gould}(1995{\natexlab{b}})}]{han95b}
---. 1995{\natexlab{b}}, \apj, 447, 53

\bibitem[{{Han} \& {Gould}(2003)}]{han03}
---. 2003, \apj, 592, 172

\bibitem[{{Han} {et~al.}(2013){Han}, {Udalski}, {Choi}, {Yee}, {Gould},
  {Christie}, {Tan}, {Szyma{\'n}ski}, {Kubiak}, {Soszy{\'n}ski},
  {Pietrzy{\'n}ski}, {Poleski}, {Ulaczyk}, {Pietrukowicz}, {Koz{\l}owski},
  {Skowron}, {Wyrzykowski}, {OGLE Collaboration}, {Almeida}, {Batista},
  {Depoy}, {Dong}, {Drummond}, {Gaudi}, {Hwang}, {Jablonski}, {Jung}, {Lee},
  {Koo}, {McCormick}, {Monard}, {Natusch}, {Ngan}, {Park}, {Pogge}, {Porritt},
  {Shin}, \& {{$\mu$}FUN Collaboration}}]{han13}
{Han}, C., {et~al.} 2013, \apjl, 762, L28

\bibitem[{{Hartman} {et~al.}(2004){Hartman}, {Bakos}, {Stanek}, \&
  {Noyes}}]{hartman04}
{Hartman}, J.~D., {Bakos}, G., {Stanek}, K.~Z., \& {Noyes}, R.~W. 2004, \aj,
  128, 1761

\bibitem[{{Holtzman} {et~al.}(1998){Holtzman}, {Watson}, {Baum}, {Grillmair},
  {Groth}, {Light}, {Lynds}, \& {O'Neil}}]{holtzman98}
{Holtzman}, J.~A., {Watson}, A.~M., {Baum}, W.~A., {Grillmair}, C.~J., {Groth},
  E.~J., {Light}, R.~M., {Lynds}, R., \& {O'Neil}, Jr., E.~J. 1998, \aj, 115,
  1946

\bibitem[{{Howard} {et~al.}(2012){Howard}, {Marcy}, {Bryson}, {Jenkins},
  {Rowe}, {Batalha}, {Borucki}, {Koch}, {Dunham}, {Gautier}, {Van Cleve},
  {Cochran}, {Latham}, {Lissauer}, {Torres}, {Brown}, {Gilliland}, {Buchhave},
  {Caldwell}, {Christensen-Dalsgaard}, {Ciardi}, {Fressin}, {Haas}, {Howell},
  {Kjeldsen}, {Seager}, {Rogers}, {Sasselov}, {Steffen}, {Basri},
  {Charbonneau}, {Christiansen}, {Clarke}, {Dupree}, {Fabrycky}, {Fischer},
  {Ford}, {Fortney}, {Tarter}, {Girouard}, {Holman}, {Johnson}, {Klaus},
  {Machalek}, {Moorhead}, {Morehead}, {Ragozzine}, {Tenenbaum}, {Twicken},
  {Quinn}, {Isaacson}, {Shporer}, {Lucas}, {Walkowicz}, {Welsh}, {Boss},
  {Devore}, {Gould}, {Smith}, {Morris}, {Prsa}, {Morton}, {Still}, {Thompson},
  {Mullally}, {Endl}, \& {MacQueen}}]{howard12}
{Howard}, A.~W., {et~al.} 2012, \apjs, 201, 15

\bibitem[{{Ida} \& {Lin}(2005)}]{ida05}
{Ida}, S., \& {Lin}, D.~N.~C. 2005, \apj, 626, 1045

\bibitem[{{Jacquet} \& {Robert}(2013)}]{jacquet13}
{Jacquet}, E., \& {Robert}, F. 2013, Icarus, 223, 722

\bibitem[{{Kalas} {et~al.}(2008){Kalas}, {Graham}, {Chiang}, {Fitzgerald},
  {Clampin}, {Kite}, {Stapelfeldt}, {Marois}, \& {Krist}}]{kalas08}
{Kalas}, P., {et~al.} 2008, Science, 322, 1345

\bibitem[{{Kennedy} \& {Kenyon}(2008)}]{kennedy08}
{Kennedy}, G.~M., \& {Kenyon}, S.~J. 2008, \apj, 673, 502

\bibitem[{{King}(1983)}]{king83}
{King}, I.~R. 1983, \pasp, 95, 163

\bibitem[{{Krisciunas} \& {Schaefer}(1991)}]{krisciunas91}
{Krisciunas}, K., \& {Schaefer}, B.~E. 1991, \pasp, 103, 1033

\bibitem[{{Lagrange} {et~al.}(2010){Lagrange}, {Bonnefoy}, {Chauvin}, {Apai},
  {Ehrenreich}, {Boccaletti}, {Gratadour}, {Rouan}, {Mouillet}, {Lacour}, \&
  {Kasper}}]{lagrange10}
{Lagrange}, A.-M., {et~al.} 2010, Science, 329, 57

\bibitem[{{Lissauer}(1987)}]{lissauer87}
{Lissauer}, J.~J. 1987, Icarus, 69, 249

\bibitem[{{Lissauer} {et~al.}(2011){Lissauer}, {Fabrycky}, {Ford}, {Borucki},
  {Fressin}, {Marcy}, {Orosz}, {Rowe}, {Torres}, {Welsh}, {Batalha}, {Bryson},
  {Buchhave}, {Caldwell}, {Carter}, {Charbonneau}, {Christiansen}, {Cochran},
  {Desert}, {Dunham}, {Fanelli}, {Fortney}, {Gautier}, {Geary}, {Gilliland},
  {Haas}, {Hall}, {Holman}, {Koch}, {Latham}, {Lopez}, {McCauliff}, {Miller},
  {Morehead}, {Quintana}, {Ragozzine}, {Sasselov}, {Short}, \&
  {Steffen}}]{lissauer11}
{Lissauer}, J.~J., {et~al.} 2011, \nat, 470, 53

\bibitem[{{Lovis} {et~al.}(2006){Lovis}, {Mayor}, {Pepe}, {Alibert}, {Benz},
  {Bouchy}, {Correia}, {Laskar}, {Mordasini}, {Queloz}, {Santos}, {Udry},
  {Bertaux}, \& {Sivan}}]{lovis06}
{Lovis}, C., {et~al.} 2006, \nat, 441, 305

\bibitem[{{Majewski} {et~al.}(2011){Majewski}, {Zasowski}, \&
  {Nidever}}]{majewski11}
{Majewski}, S.~R., {Zasowski}, G., \& {Nidever}, D.~L. 2011, \apj, 739, 25

\bibitem[{{Mao} \& {Paczynski}(1991)}]{mao91}
{Mao}, S., \& {Paczynski}, B. 1991, \apjl, 374, L37

\bibitem[{{Marois} {et~al.}(2008){Marois}, {Macintosh}, {Barman}, {Zuckerman},
  {Song}, {Patience}, {Lafreni{\`e}re}, \& {Doyon}}]{marois08}
{Marois}, C., {Macintosh}, B., {Barman}, T., {Zuckerman}, B., {Song}, I.,
  {Patience}, J., {Lafreni{\`e}re}, D., \& {Doyon}, R. 2008, Science, 322, 1348

\bibitem[{{Marshall} {et~al.}(2006){Marshall}, {Robin}, {Reyl{\'e}},
  {Schultheis}, \& {Picaud}}]{marshall06}
{Marshall}, D.~J., {Robin}, A.~C., {Reyl{\'e}}, C., {Schultheis}, M., \&
  {Picaud}, S. 2006, \aap, 453, 635

\bibitem[{{Mayor} \& {Queloz}(1995)}]{mayor95}
{Mayor}, M., \& {Queloz}, D. 1995, \nat, 378, 355

\bibitem[{{Mayor} {et~al.}(2009){Mayor}, {Udry}, {Lovis}, {Pepe}, {Queloz},
  {Benz}, {Bertaux}, {Bouchy}, {Mordasini}, \& {Segransan}}]{mayor09}
{Mayor}, M., {et~al.} 2009, \aap, 493, 639

\bibitem[{{Moffat}(1969)}]{moffat69}
{Moffat}, A.~F.~J. 1969, \aap, 3, 455

\bibitem[{{Morton} \& {Swift}(2013)}]{morton13}
{Morton}, T.~D., \& {Swift}, J.~J. 2013, ArXiv e-prints

\bibitem[{{Nataf} {et~al.}(2013){Nataf}, {Gould}, {Fouqu{\'e}}, {Gonzalez},
  {Johnson}, {Skowron}, {Udalski}, {Szyma{\'n}ski}, {Kubiak},
  {Pietrzy{\'n}ski}, {Soszy{\'n}ski}, {Ulaczyk}, {Wyrzykowski}, \&
  {Poleski}}]{nataf13}
{Nataf}, D.~M., {et~al.} 2013, \apj, 769, 88

\bibitem[{{Nidever} {et~al.}(2012){Nidever}, {Zasowski}, \&
  {Majewski}}]{nidever12}
{Nidever}, D.~L., {Zasowski}, G., \& {Majewski}, S.~R. 2012, \apjs, 201, 35

\bibitem[{{Orosz} {et~al.}(2012){Orosz}, {Welsh}, {Carter}, {Fabrycky},
  {Cochran}, {Endl}, {Ford}, {Haghighipour}, {MacQueen}, {Mazeh},
  {Sanchis-Ojeda}, {Short}, {Torres}, {Agol}, {Buchhave}, {Doyle}, {Isaacson},
  {Lissauer}, {Marcy}, {Shporer}, {Windmiller}, {Barclay}, {Boss}, {Clarke},
  {Fortney}, {Geary}, {Holman}, {Huber}, {Jenkins}, {Kinemuchi}, {Kruse},
  {Ragozzine}, {Sasselov}, {Still}, {Tenenbaum}, {Uddin}, {Winn}, {Koch}, \&
  {Borucki}}]{orosz12}
{Orosz}, J.~A., {et~al.} 2012, Science, 337, 1511

\bibitem[{{Paczynski}(1986)}]{paczynski86}
{Paczynski}, B. 1986, \apj, 304, 1

\bibitem[{{Peale}(1997)}]{peale97}
{Peale}, S.~J. 1997, Icarus, 127, 269

\bibitem[{{Peale}(1998)}]{peale98}
---. 1998, \apj, 509, 177

\bibitem[{{Pejcha} \& {Heyrovsk{\'y}}(2009)}]{pejcha09}
{Pejcha}, O., \& {Heyrovsk{\'y}}, D. 2009, \apj, 690, 1772

\bibitem[{{Penny}(2013)}]{penny13a}
{Penny}, M.~T. 2013, ArXiv e-prints

\bibitem[{{Petigura} {et~al.}(2013){Petigura}, {Howard}, \&
  {Marcy}}]{petigura13}
{Petigura}, E.~A., {Howard}, A.~W., \& {Marcy}, G.~W. 2013, Proceedings of the
  National Academy of Science, 110, 19273

\bibitem[{{Popowski} {et~al.}(2005){Popowski}, {Griest}, {Thomas}, {Cook},
  {Bennett}, {Becker}, {Alves}, {Minniti}, {Drake}, {Alcock}, {Allsman},
  {Axelrod}, {Freeman}, {Geha}, {Lehner}, {Marshall}, {Nelson}, {Peterson},
  {Quinn}, {Stubbs}, {Sutherland}, {Vandehei}, {Welch}, \& {MACHO
  Collaboration}}]{popowski05}
{Popowski}, P., {et~al.} 2005, \apj, 631, 879

\bibitem[{{Press} {et~al.}(1992){Press}, {Teukolsky}, {Vetterling}, \&
  {Flannery}}]{press92}
{Press}, W.~H., {Teukolsky}, S.~A., {Vetterling}, W.~T., \& {Flannery}, B.~P.
  1992, {Numerical recipes in C. The art of scientific computing}, ed. {Press,
  W.~H., Teukolsky, S.~A., Vetterling, W.~T., \& Flannery, B.~P. }

\bibitem[{{Raymond} {et~al.}(2004){Raymond}, {Quinn}, \& {Lunine}}]{raymond04}
{Raymond}, S.~N., {Quinn}, T., \& {Lunine}, J.~I. 2004, Icarus, 168, 1

\bibitem[{{Shvartzvald} \& {Maoz}(2012)}]{shvartzvald12}
{Shvartzvald}, Y., \& {Maoz}, D. 2012, \mnras, 419, 3631

\bibitem[{{Skowron} \& {Gould}(2012)}]{skowron12}
{Skowron}, J., \& {Gould}, A. 2012, ArXiv e-prints

\bibitem[{{Sumi} {et~al.}(2003){Sumi}, {Abe}, {Bond}, {Dodd}, {Hearnshaw},
  {Honda}, {Honma}, {Kan-ya}, {Kilmartin}, {Masuda}, {Matsubara}, {Muraki},
  {Nakamura}, {Nishi}, {Noda}, {Ohnishi}, {Petterson}, {Rattenbury}, {Reid},
  {Saito}, {Saito}, {Sato}, {Sekiguchi}, {Skuljan}, {Sullivan}, {Takeuti},
  {Tristram}, {Wilkinson}, {Yanagisawa}, \& {Yock}}]{sumi03}
{Sumi}, T., {et~al.} 2003, \apj, 591, 204

\bibitem[{{Sumi} {et~al.}(2006){Sumi}, {Wo{\'z}niak}, {Udalski},
  {Szyma{\'n}ski}, {Kubiak}, {Pietrzy{\'n}ski}, {Soszy{\'n}ski},
  {{\.Z}ebru{\'n}}, {Szewczyk}, {Wyrzykowski}, \& {Paczy{\'n}ski}}]{sumi06}
---. 2006, \apj, 636, 240

\bibitem[{{Sumi} {et~al.}(2010){Sumi}, {Bennett}, {Bond}, {Udalski}, {Batista},
  {Dominik}, {Fouqu{\'e}}, {Kubas}, {Gould}, {Macintosh}, {Cook}, {Dong},
  {Skuljan}, {Cassan}, {Abe}, {Botzler}, {Fukui}, {Furusawa}, {Hearnshaw},
  {Itow}, {Kamiya}, {Kilmartin}, {Korpela}, {Lin}, {Ling}, {Masuda},
  {Matsubara}, {Miyake}, {Muraki}, {Nagaya}, {Nagayama}, {Ohnishi}, {Okumura},
  {Perrott}, {Rattenbury}, {Saito}, {Sako}, {Sullivan}, {Sweatman}, {Tristram},
  {Yock}, {MOA Collaboration}, {Beaulieu}, {Cole}, {Coutures}, {Duran},
  {Greenhill}, {Jablonski}, {Marboeuf}, {Martioli}, {Pedretti}, {Pejcha},
  {Rojo}, {Albrow}, {Brillant}, {Bode}, {Bramich}, {Burgdorf}, {Caldwell},
  {Calitz}, {Corrales}, {Dieters}, {Dominis Prester}, {Donatowicz}, {Hill},
  {Hoffman}, {Horne}, {J{\o}rgensen}, {Kains}, {Kane}, {Marquette}, {Martin},
  {Meintjes}, {Menzies}, {Pollard}, {Sahu}, {Snodgrass}, {Steele}, {Street},
  {Tsapras}, {Wambsganss}, {Williams}, {Zub}, {PLANET Collaboration},
  {Szyma{\'n}ski}, {Kubiak}, {Pietrzy{\'n}ski}, {Soszy{\'n}ski}, {Szewczyk},
  {Wyrzykowski}, {Ulaczyk}, {OGLE Collaboration}, {Allen}, {Christie}, {DePoy},
  {Gaudi}, {Han}, {Janczak}, {Lee}, {McCormick}, {Mallia}, {Monard}, {Natusch},
  {Park}, {Pogge}, {Santallo}, \& {{$\mu$}FUN Collaboration}}]{sumi10}
---. 2010, \apj, 710, 1641

\bibitem[{{Sumi} {et~al.}(2011){Sumi}, {Kamiya}, {Bennett}, {Bond}, {Abe},
  {Botzler}, {Fukui}, {Furusawa}, {Hearnshaw}, {Itow}, {Kilmartin}, {Korpela},
  {Lin}, {Ling}, {Masuda}, {Matsubara}, {Miyake}, {Motomura}, {Muraki},
  {Nagaya}, {Nakamura}, {Ohnishi}, {Okumura}, {Perrott}, {Rattenbury}, {Saito},
  {Sako}, {Sullivan}, {Sweatman}, {Tristram}, {Udalski}, {Szyma{\'n}ski},
  {Kubiak}, {Pietrzy{\'n}ski}, {Poleski}, {Soszy{\'n}ski}, {Wyrzykowski},
  {Ulaczyk}, \& {Microlensing Observations in Astrophysics (MOA)
  Collaboration}}]{sumi11}
---. 2011, \nat, 473, 349

\bibitem[{{Sumi} {et~al.}(2013){Sumi}, {Bennett}, {Bond}, {Abe}, {Botzler},
  {Fukui}, {Furusawa}, {Itow}, {Ling}, {Masuda}, {Matsubara}, {Muraki},
  {Ohnishi}, {Rattenbury}, {Saito}, {Sullivan}, {Suzuki}, {Sweatman}, {P.},
  {Tristram}, {Wada}, \& {Yock}}]{sumi13}
---. 2013, ArXiv e-prints

\bibitem[{{Tsapras} {et~al.}(2009){Tsapras}, {Street}, {Horne}, {Snodgrass},
  {Dominik}, {Allan}, {Steele}, {Bramich}, {Saunders}, {Rattenbury}, {Mottram},
  {Fraser}, {Clay}, {Burgdorf}, {Bode}, {Lister}, {Hawkins}, {Beaulieu},
  {Fouqu{\'e}}, {Albrow}, {Menzies}, {Cassan}, \&
  {Dominis-Prester}}]{tsapras09}
{Tsapras}, Y., {et~al.} 2009, Astronomische Nachrichten, 330, 4

\bibitem[{{Udalski}(2003)}]{udalski03}
{Udalski}, A. 2003, AcA, 53, 291

\bibitem[{{Udry} {et~al.}(2003){Udry}, {Mayor}, \& {Santos}}]{udry03}
{Udry}, S., {Mayor}, M., \& {Santos}, N.~C. 2003, \aap, 407, 369

\bibitem[{{Williams} \& {Cieza}(2011)}]{williams11}
{Williams}, J.~P., \& {Cieza}, L.~A. 2011, \araa, 49, 67

\bibitem[{{Witt} \& {Mao}(1994)}]{witt94}
{Witt}, H.~J., \& {Mao}, S. 1994, \apj, 430, 505

\bibitem[{{Witt} \& {Mao}(1995)}]{witt95}
---. 1995, \apjl, 447, L105

\bibitem[{{Woolf}(1982)}]{woolf82}
{Woolf}, N.~J. 1982, \araa, 20, 367

\bibitem[{{Yee} {et~al.}(2013){Yee}, {Hung}, {Bond}, {Allen}, {Monard},
  {Albrow}, {Fouqu{\'e}}, {Dominik}, {Tsapras}, {Udalski}, {Gould}, {Zellem},
  {Bos}, {Christie}, {DePoy}, {Dong}, {Drummond}, {Gaudi}, {Gorbikov}, {Han},
  {Kaspi}, {Klein}, {Lee}, {Maoz}, {McCormick}, {Moorhouse}, {Natusch}, {Nola},
  {Park}, {Pogge}, {Polishook}, {Shporer}, {Shvartzvald}, {Skowron},
  {Thornley}, {The {$\mu$}FUN Collaboration}, {Abe}, {Bennett}, {Botzler},
  {Chote}, {Freeman}, {Fukui}, {Furusawa}, {Harris}, {Itow}, {Ling}, {Masuda},
  {Matsubara}, {Miyake}, {Ohnishi}, {Rattenbury}, {Saito}, {Sullivan}, {Sumi},
  {Suzuki}, {Sweatman}, {Tristram}, {Wada}, {Yock}, {The MOA Collaboration},
  {Szyma{\'n}ski}, {Soszy{\'n}ski}, {Kubiak}, {Poleski}, {Ulaczyk},
  {Pietrzy{\'n}ski}, {Wyrzykowski}, {The OGLE Collaboration}, {Bachelet},
  {Batista}, {Beatty}, {Beaulieu}, {Bennett}, {Bowens-Rubin}, {Brillant},
  {Caldwell}, {Cassan}, {Cole}, {Corrales}, {Coutures}, {Dieters}, {Dominis
  Prester}, {Donatowicz}, {Greenhill}, {Henderson}, {Kubas}, {Marquette},
  {Martin}, {Menzies}, {Shappee}, {Williams}, {Wouters}, {van Saders}, {Zub},
  {The PLANET Collaboration}, {Street}, {Horne}, {Bramich}, {Steele}, {The
  RoboNet Collaboration}, {Alsubai}, {Bozza}, {Browne}, {Burgdorf}, {Calchi
  Novati}, {Dodds}, {Finet}, {Gerner}, {Hardis}, {Harps{\o}e}, {Hessman},
  {Hinse}, {Hundertmark}, {J{\o}rgensen}, {Kains}, {Kerins}, {Liebig},
  {Mancini}, {Mathiasen}, {Penny}, {Proft}, {Rahvar}, {Ricci}, {Sahu},
  {Scarpetta}, {Sch{\"a}fer}, {Sch{\"o}nebeck}, {Snodgrass}, {Southworth},
  {Surdej}, {Wambsganss}, \& {MiNDSTEp Consortium}}]{yee13}
{Yee}, J.~C., {et~al.} 2013, \apj, 769, 77

\bibitem[{{Youdin}(2011)}]{youdin11}
{Youdin}, A.~N. 2011, \apj, 742, 38

\end{thebibliography}
\end{document}